\documentclass[manuscript]{aastex}

\newcommand{\hi}{\ion{H}{1}}

\newcommand{\cden}{cm$^{-2}$}

\newcommand{\msun}{M$_{\sun}$}
\newcommand{\kms}{km~s$^{-1}$}
\newcommand{\sfr}{$M_{\sun}$\,yr$^{-1}$}
\newcommand{\hissc}{_{\textrm{H}\textsc{i}}}

\shorttitle{VLA-ANGST}
\shortauthors{Ott et al.}

\begin{document}

\title{VLA-ANGST: A high-resolution \ion{H}{1} Survey of Nearby Dwarf
  Galaxies}

\author{J\"urgen Ott}
\affil{National Radio Astronomy Observatory, P.O. Box O, 1003
  Lopezville Road, Socorro, NM 87801, USA}
\email{jott@nrao.edu}

\author{Adrienne M. Stilp}
\affil{Department of Astronomy, Box 351580, University of Washington, Seattle, WA 98195,
USA}
\email{adrienne@astro.washington.edu}

\author{Steven R. Warren}
\affil{Minnesota Institute for Astrophysics, University of Minnesota, 116 Church St. SE,
Minneapolis, MN 55455, USA}
\email{warren@astro.umn.edu}

\author{Evan D. Skillman}
\affil{Minnesota Institute for Astrophysics, University of Minnesota, 116 Church St. SE,
Minneapolis, MN 55455, USA}
\email{skillman@astro.umn.edu}

\author{Julianne J. Dalcanton}
\affil{Department of Astronomy, Box 351580, University of Washington, Seattle, WA 98195,
USA}
\email{jd@astro.washington.edu}

\author{Fabian Walter}
\affil{Max-Planck-Institut f\"ur Astronomie, K\"onigstuhl 17, D-69117 Heidelberg, Germany}
\email{walter@mpia.de}

\author{W.J.G. de Blok}
\affil{Netherlands Institute for Radio Astronomy, Oude Hoogeveensedijk
  4, 7991 PD Dwingeloo, The Netherlands}
\email{blok@astron.nl}

\author{B\"arbel Koribalski}
\affil{Australia Telescope National Facility, CSIRO Astronomy and Space Science, PO Box 76,
Epping NSW 1710, Australia}
\email{Baerbel.Koribalski@csiro.au}

\and

\author{Andrew A. West}
\affil{Department of Astronomy, Boston University, 725 Commonwealth Avenue, Boston, MA
02215, USA}
\email{aawest@bu.edu}

\begin{abstract}

  We present the ``Very Large Array survey of Advanced Camera for
  Surveys Nearby Galaxy Survey Treasury galaxies (VLA-ANGST).''
  VLA-ANGST is a National Radio Astronomy Observatory Large Program
  consisting of high spectral ($0.6-2.6$\,\kms) and spatial ($\sim
  6\arcsec$) resolution observations of neutral, atomic hydrogen (\hi)
  emission toward 35 nearby dwarf galaxies from the ANGST survey.
  ANGST is a systematic Hubble Space Telescope survey to establish a
  legacy of uniform multi-color photometry of resolved stars for a
  volume-limited sample of nearby galaxies (D $\lesssim$ 4 Mpc).
  VLA-ANGST provides VLA \hi\ observations of the sub-sample of ANGST
  galaxies with recent star formation that are observable from the
  northern hemisphere and that were not observed in the ``The \hi\
  Nearby Galaxy Survey'' (THINGS). The overarching scientific goal of
  VLA-ANGST is to investigate fundamental characteristics of the
  neutral interstellar medium (ISM) of dwarf galaxies. Here we
  describe the VLA observations, the data reduction, and the final
  VLA-ANGST data products. We present an atlas of the integrated \hi\
  maps, the intensity-weighted velocity fields, the second moment maps
  as a measure for the velocity dispersion of the \hi, individual
  channel maps, and integrated \hi\ spectra for each VLA-ANGST
  galaxy. We closely follow the observational setup and data reduction
  of THINGS to achieve comparable sensitivity and angular
  resolution. A major difference between VLA-ANGST and THINGS,
  however, is the high velocity resolution of the VLA-ANGST
  observations (0.65 and 1.3\,\kms\ for the majority of the galaxies).
  The VLA-ANGST data products are made publicly available through a
  dedicated
  webpage\footnote{\url{https://science.nrao.edu/science/surveys/vla-angst}}.
  With available star formation histories from resolved stellar
  populations and lower resolution ancillary observations from the far
  infrared to the ultraviolet, VLA-ANGST will enable detailed studies
  of the relationship between the ISM and star formation in dwarf
  galaxies on a $\sim 100$\,pc scale.
\end{abstract}


\object{NGC 247}
\object{DDO 6}
\object{NGC 404}
\object{KKH 37}
\object{UGC 4483}
\object{KK 77}
\object{BK 3N}
\object{AO 0952+69}
\object{Sextans B}
\object{NGC 3109}
\object{Antlia}
\object{KDG 63}
\object{Sextans A}
\object{HS 117}
\object{DDO 82}
\object{KDG 73}
\object{NGC 3741}
\object{DDO 99}
\object{NGC 4163}
\object{NGC 4190}
\object{DDO 113}
\object{MCG +09-20-131}
\object{DDO 125}
\object{UGCA 292}
\object{GR 8}
\object{UGC 8508}
\object{DDO 181}
\object{DDO 183}
\object{KKH 86}
\object{UGC 8833}
\object{KK 230}
\object{DDO 187}
\object{DDO 190}
\object{KKR 25}
\object{KKH 98}

\section{Introduction}

\label{sec:intro}

Star formation is driven by complicated interactions between gas and
stars.  Untangling the interplay between these processes is difficult,
because in most cases, the events that trigger star formation are not
obvious, nor are those that shape the structure and dynamics of the
ISM. For a full understanding it is therefore indispensable to obtain
a comprehensive view of all processes that come together to form
stars, stellar associations, and stellar clusters.  Observationally,
one requires knowledge of the gas distribution and kinematics as well
as the stellar energy input into the ISM over time.  

In recent years, large systematic surveys have made superb progress on
the first of these requirements.  The number of nearby galaxies for
which high-quality \hi\ data is available has dramatically increased
in the last few years, including campaigns such as THINGS \citep[``The
\ion{H}{1} Nearby Galaxy Survey'';][]{wal08}, FIGGS \citep[``Faint
Irregular Galaxies GMRT Survey'';][]{beg08}, LITTLE-THINGS
\citep[``LITTLE: Local Irregulars That Trace Luminosity
Extremes'';][]{hun07}, SHIELD \citep[``Survey of \hi\ in Extremely
Low-mass Dwarfs'';][]{can11a}, LVHIS \citep[``The Local Volume
\ion{H}{1} Survey'';][]{kor08}, WHISP \citep[``Westerbork observations
of neutral Hydrogen in Irregular and Spiral galaxies'';][]{vdh01}, and
HALOGAS \citep[``The Westerbork Hydrogen Accretion in Local
Galaxies'';][]{hea11}. The difficult work of mapping the molecular
medium in the brighter galaxies has begun as well, e.g. in BIMA SONG
\citep[``The BIMA Survey of Nearby Galaxies''][]{hel03}, HERACLES,
\citep[``The HERA CO Line Extragalactic Survey'';][]{ler09}, and STING
\citep[``CARMA Survey Toward Infrared-bright Nearby
Galaxies'';][]{rah12}.

Unfortunately, the needed measurements of time-resolved stellar energy
input are more difficult to acquire.  Large systematic surveys in the
far-ultraviolet (e.g., obtained with the {\it GALEX} telescope) and the
far-infrared (e.g., the Local Volume Legacy survey
\citep[LVL;][]{dale09}, SINGS \citep[``Spitzer Infrared Nearby
Galaxies Survey'';][]{ken03}, Kingfish \citep[``Key Insights on Nearby
Galaxies: a Far-Infrared Survey with Herschel'';] []{ken11}, etc) have
made excellent progress in measuring the recent star formation rate
(SFR), while surveys like 11HUGS \citep[``The 11\,Mpc H$\alpha$ UV
Galaxy Survey'';][]{lee04} have provided the H$\alpha$ mapping needed
to trace star formation (SF) on much shorter timescales.  However,
these approaches to measuring star formation lack all but the broadest
time resolution, with the different tracers being sensitive to SF over 
different timescales. H$\alpha$ is emitted on the timescale of O-star
lifetimes ($\lesssim 5$\,Myr) and the far-ultraviolet on the timescale of
A-star lifetimes ($\lesssim 100$\,Myr). The far-infrared is sensitive to 
timescales similar to the $\lesssim 100$\,Myr of far-ultraviolet
heating. These timescales are not necessarily well-matched to the
relevant energy input timescales for the gas.

For the survey presented here, we take a different approach, and focus
\hi\ observations on galaxies which are sufficiently close that their
stellar populations can be resolved with the {\it Hubble Space
  Telescope} (HST).  The resulting color--magnitude diagrams (CMDs)
allow one to construct spatially-resolved star formation histories
(SFHs) via analyses of galaxies' stellar populations, and thus reveal
the time-resolved SFR of these nearby galaxies over timescales of
several hundred Myr at $\sim5-10\arcsec$ spatial resolution
\citep[e.g.,][]{doh02,can03,wei08,wil10,mcq10,crn11,can11b,wei11}.
With nearly 300 orbits of {\it HST} time, one of the most ambitious
programs to obtain spatially-resolved SFHs is the ACS Nearby Galaxy
Survey Treasury program \citep[ANGST;][]{dal09}. The ANGST
observations enable to map spatially-resolved SFHs for a
volume-limited sample of 69 nearby ($<4$\,Mpc) galaxies, probing both
group and field environments. These data provide an entirely new,
statistically significant view on the SFH of the local universe dwarf
galaxy population.

The survey presented here, VLA-ANGST (``Very Large Array survey of ACS
Nearby Galaxy Survey Treasury galaxies''), complements ANGST with high
spatial and spectral resolution data cubes of the atomic gas traced by
the 21\,cm line of neutral atomic hydrogen (\hi). VLA-ANGST was
designed to aim for the best available resolution and sensitivity, 
using the NRAO Very Large Array (VLA) in multiple configurations in a Large Program
worth $\sim 500$\,hours of observing time to achieve that
goal. VLA-ANGST is designed to match the \hi\ spatial resolution
($\sim 6\arcsec$) to the cell sizes over which the SFHs can be
determined. Furthermore, the majority of the VLA-ANGST galaxies were
observed at very high spectral resolution of $0.6-1.3$\,\kms\ which is
important for detailed ISM dynamic modeling studies of the rather
low-mass objects which dominate the galaxy population within $4$\,Mpc.

The VLA-ANGST survey has a number of features that make it a valuable
addition to the many other \hi\ surveys, beyond the existence
of resolved stellar population data.  First, the galaxies in VLA-ANGST
are all quite close, which ensures high linear resolution for studying
small-scale features in the \hi\ distribution.  Second, because care
was taken to match the observational setups of THINGS, the VLA-ANGST
survey can be readily combined with the THINGS survey, giving much broader
coverage towards low galaxy masses.  When further combined with surveys like
LITTLE-THINGS and SHIELD, which have used a similar observational
strategy,
we will have a relatively uniform database of \ion{H}{1} toward $>100$ objects
spanning a large variety of galaxy types that is comparable in terms
of sensitivity, angular, and spectral resolution.

The variety of galaxy types in the sample
allows the study of (1) the response of gas and star formation to the
propagation of spiral arms and to interactions, as seen in
massive spirals and starburst galaxies (2) star formation propagation
in the absence of strong perturbations of the gas density; gas rich
dwarves are ideal for such a study due to their lack of internal shear
and spiral density waves (3) star formation triggered in unusual
kinematic environments such as in tidal dwarfs, and (4) dIrr/dSph
transition-type galaxies, whose lack of current SF but ample gas
reservoirs allow studies of galaxies that possess the raw material for
star formation, but somehow remain dormant.

In the following we present the data of the VLA-ANGST survey, and
in-depth scientific analyses will follow in subsequent
publications. First analyses based on VLA-ANGST data cubes are
provided by \citet{war11} on the energy requirements of large \hi\
holes, by \citet{war12} on the distribution of cold and warm \hi, and by
\citet{sti12} on the global \hi\ velocity dispersion profiles
correlated to the properties of the galaxies.
Section\,\ref{sec:sample} describes the selection of the targets,
followed by the observational setup and data reduction
(Section\,\ref{sec:reduction}). Our data products are presented in
Section\,\ref{sec:dataProducts} and a summary of the VLA-ANGST given
in Section\,\ref{sec:summary}.

\section{Target Selection} 

\label{sec:sample}
The ANGST survey targeted a complete volume-limited sample of 69
galaxies. The volume consists of those galaxies above a Galactic
latitude of $\mid b \mid $ $>$ 20$^{\circ}$, outside the Local Group
but within 3.5 Mpc, with additional cones out to 4 Mpc in the
directions of the M\,81 and Sculptor groups.  This volume provides a
wide variety of gas-rich galaxies of all morphological types (Sb,
dIrr, dSph/dE types, and tidal dwarfs), spanning a range of 10
magnitudes in luminosity, 10$^4$ in current star formation rate (SFR),
and 1.3 dex in metallicity.

To complement the ANGST HST data with interferometric \hi\
observations, for VLA-ANGST we selected a sub-sample of ANGST targets
that comprises galaxies with known \hi\ content and galaxies with
indications of recent star formation, even if \hi\ was not previously
detected by single dish observations. We excluded galaxies that are
too far south for the VLA ($\delta<-30^{\degr}$) and galaxies that were
previously covered by THINGS.

Due to the volume limited nature of ANGST, most of the objects that
fit our selection criteria are rather low-mass, low-luminosity dwarf
galaxies. An exception is the ANGST galaxy NGC\,253, a massive barred
starburst galaxy. The \hi\ properties of NGC\,253, in particular its
large linewidth, are very different to the rest of our sample and can
only be adequately observed with the new correlator capabilities of
the upgraded Karl G. Jansky Very Large Array\footnote{The ``Very Large
  Array'' was recently renamed to ``The Karl G. Jansky Very Large
  Array (VLA)'' to mark the end of the construction phase of the VLA
  upgrade. During construction, the project carried the name ``Expanded
  Very Large Array'' (EVLA).}. Thus, we excluded the massive
starburst galaxy NGC\,253 from our sample.  In total, the VLA-ANGST
sample amounts to 35 galaxies, approximately half the objects that
comprise the entire ANGST {\it HST} survey.

We list all VLA-ANGST galaxies and their basic properties in
Table\,\ref{tab:sample}. Column (1) contains the galaxy names, and
column (2) a range of alternative names as found on the ``NASA/IPAC
Extragalactic Database''
(NED)\footnote{\url{http://ned.ipac.caltech.edu/}}. Columns (3) and
(4) are the central equatorial J2000 coordinates, followed by the
distance $D$ of the galaxies in column (derived using the tip of the
red giant branch [TRGB] method) (5). The optical diameters at
25\,mag\,$\arcsec^{2}$ (D$_{25}$) surface brightness and the absolute
$B$ magnitudes $M_{B}$ are listed in columns (6) and (7),
respectively. Columns (8) contains the galaxies' 3.6\,$\mu$m infrared
luminosities, which is a rough measure for the old, unobscured stellar
population, followed by the morphological types in numerical code
according to \citet{dev91} taken from \citet{kar04} in column
(9). Finally, we list the UV-based star formation rate in column (10).

In the VLA-ANGST sample, 29 galaxies are classified as T-type = 10, 2
as T-type = 9, 1 as T-type = 7, 1 as T-type = $-$3, and 2 as T-type =
$-$3, where negative T-types are early type galaxies and positive
T-types are late-type galaxies. Gas-rich irregulars are at the upper
end of the -10 to 10 scale.  Six galaxies were not detected in our
\hi\ observations.  The two galaxies classified as T-type = $-$3 were
non-detections, and the other 4 non-detections were T-type =
10. DDO\,82 was a single dish \hi\ non-detection but we detected the
atomic hydrogen gas in VLA-ANGST.

\begin{figure}
\includegraphics[angle=0,scale=0.8]{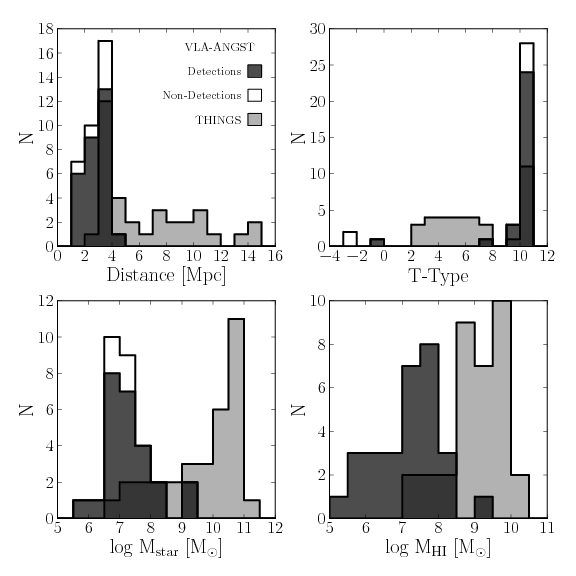}
\caption{VLA-ANGST galaxy distribution across distance (upper left),
  T-Type (upper right), logarithmic stellar mass (lower left), and
  logarithmic \hi\ mass (lower right). Galaxies that are detected in
  VLA-ANGST are shown in medium gray bins and non-detections, colored
  white, add to the distribution. THINGS galaxies are shown in light gray
  and the VLA-ANGST and THINGS overlap areas in dark gray
  color.  \label{fig:histo}}
\end{figure}

\begin{figure}
\includegraphics[angle=0,scale=0.4]{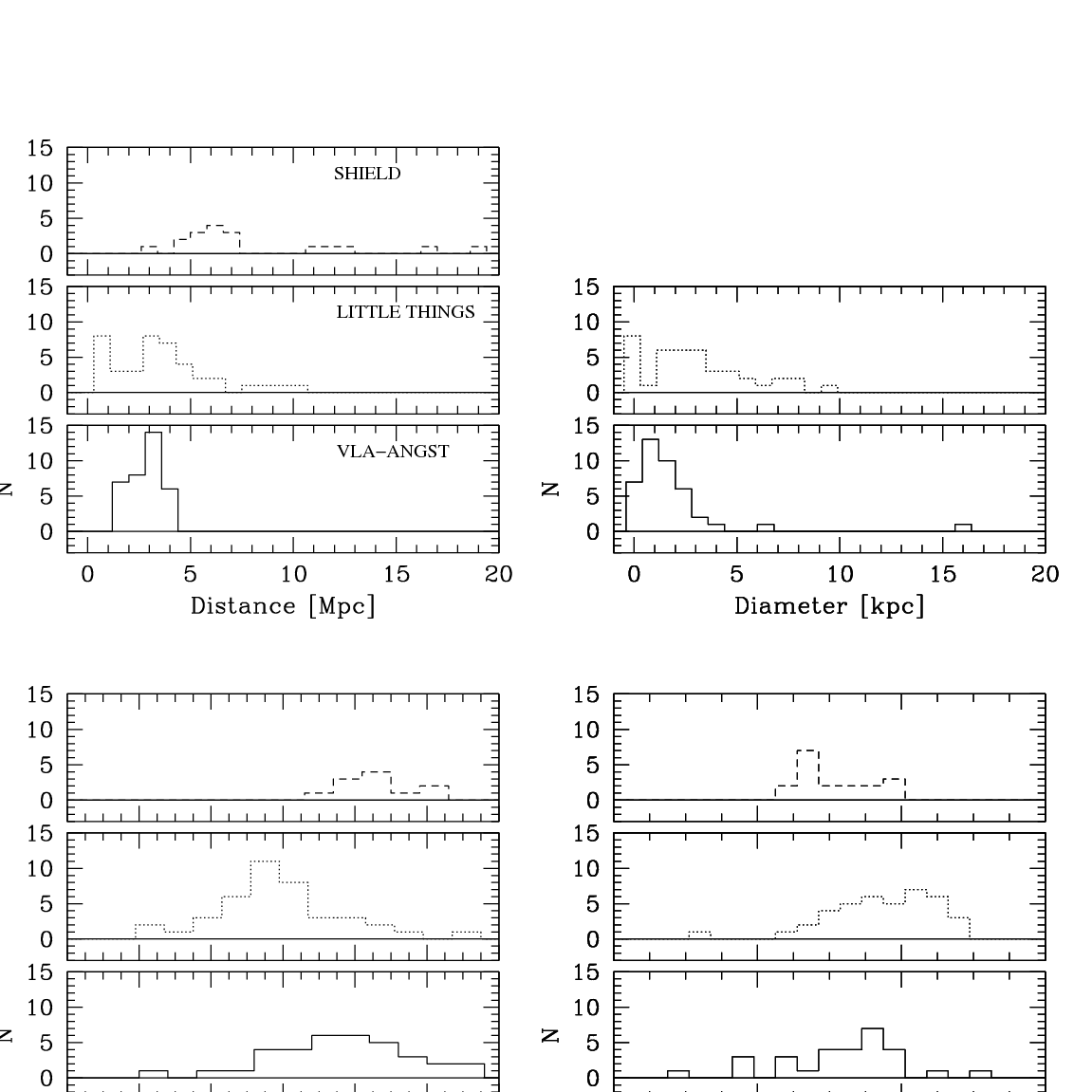}
\caption{Comparison of VLA-ANGST galaxy properties ({\it solid}) with the
  LITTLE THINGS ({\it dotted}) and SHIELD ({\it dashed})
  samples. Clockwise from the upper left: Distance, Diameter (D$_{25}$
  for VLA-ANGST, Holmberg diameter for LITTLE THINGS, no information
  on SHIELD targets), logarithmic \hi\ masses, and optical absolute
  $B$ magnitudes.  \label{fig:surveys}}
\end{figure}

Fig.\,\ref{fig:histo} shows the distributions of VLA-ANGST galaxies as
a function of distance, T-type, logarithmic stellar mass (based on the
$3.6$\,$\mu$m emission), and logarithmic \hi\ mass, as compared to the
THINGS sample. VLA-ANGST galaxies are on average more nearby, late
type, and low mass galaxies, both in terms of stellar and \hi\
mass. The VLA-ANGST sample is thus much more dominated by low-mass
dwarf irregulars and thus provides a complementary sample to
THINGS. Comparison with LITTLE THINGS and SHIELD are shown in
Fig.\,\ref{fig:surveys}. Given the volume filling sample selection
constraints, the VLA-ANGST galaxies are much more tightly located at
distances of $\sim 2-4$\,Mpc, whereas LITTLE THINGS and SHIELD have a
larger distance spread. Galaxy sizes are also broader distributed in
LITTLE THINGS as compred to VLA-ANGST. LITTLE THINGS galaxies are on
average brighter and SHIELD galaxies are at the faint end of optical
absolute $B$ magnitudes when compared to the VLA-ANGST sample. LITTLE
THINGS contains more \hi-massive galaxies than VLA-ANGST and SHIELD is
again at the fainter \hi\ end. We include the SHIELD VLA Pilot
galaxies in this plot, which are brighter than the proper SHIELD
targets \citep[see][]{can11a}. The high mass end of the final SHIELD
\hi\ mass distribution is thus somewhat fainter than depicted in
Fig.\,\ref{fig:surveys}.

\section{Observations and Data Reduction}
\label{sec:reduction}

The vast majority of the observations in VLA-ANGST are new.  In a few
cases, archival data were used in place of obtaining new observations
to improve efficiency.  Here we describe both the new and archival
observations.

\subsection{Description of Observations}
\label{ssec:obssetup}

The parameters for our new observations and the subsequent data
reduction strategy closely followed the design of the THINGS survey
\citep{wal08}, with the goal of obtaining comparable sensitivity and
spatial resolution. A significant difference between the two surveys,
however, is the $\sim5$ times better velocity resolution of
VLA-ANGST. Each of the VLA-ANGST galaxies was observed with the NRAO's
VLA in the B- (9\,h total observing time per galaxy), C- (3\,h), and
D- (3\,h) array configurations (under project code AO215). The compact
D-configuration exhibits the largest number of short baselines with
antenna separations down to 35\,m. D-configuration is thus the most
sensitive antenna configuration to image spatial scales of up to $\sim
16\arcmin$, which is the maximum for the VLA at 1.4\,GHz. The addition of
B-configuration observations yield spatial resolutions $\sim6\arcsec$
or $\sim 100$\,pc for the nearby ($D\sim3$\,Mpc) objects. This scale
is necessary to compare star forming regions with their \hi\
counterparts and matches well the resolution of surveys at other
wavelengths such as LVL (SPITZER) and 11HUGS (GALEX).  The southern
sources NGC\,3109, NGC\,247, Antlia, and DDO\,6 were observed mostly
in the hybrid BnA-, CnB-, and DnC-array configurations that feature
elongated placements of the antennas along the northern arm. The
projected baselines of these antenna configurations produce a more
circular beam toward southern targets.

At the time of the observations, 2007 October to 2008 August, the VLA
was in the process of being upgraded EVLA. This transition implied
that the observations were taken with a mix of already converted EVLA
and original VLA antennas. The conversion period mostly affected the
signal distribution from the front-ends to the correlator. At the
time, the old VLA correlator was still in use, and we configured it in
the 2AC or 2AD modes to capture both RR and LL polarization
products. Doppler tracking in the transition phase would have
introduced phase jumps on baselines involving EVLA
antennas. Consequently, we calculated and fixed the appropriate
observing frequency for each observing run such that the \hi\ emission
line was well placed within a VLA bandpass at the start of an
observation (Doppler setting). Line shifts
during a single observation are minimal ($<0.5$\,\kms) and are
corrected in post-processing (\S.\,\ref{ssec:calibration}).

The observational parameters for each galaxy were based on its \hi\
linewidth as taken from single dish \hi\ spectra, plus $\sim20$\%
additional line free channels to enable a good continuum
subtraction. To reach our goal of the best possible velocity
resolution, we used a correlator mode with 0.78\,MHz bandwidth and 256
channels for 15 galaxies in our sample. This corresponds to a channel
width of $\sim 0.65$\,\kms\ over a total velocity range of $\sim
120$\,\kms\ after cropping about 20\% edge channels, 10\% on each
side.  Eight galaxies had wider \hi\ linewidths and were therefore
observed with a bandwidth of 1.56\,MHz and 256 channels ($\sim
300$\,\kms\ velocity range and $\sim 1.3$\,\kms\ channel width; see
Table\,\ref{tab:dataproperties}). The systemic velocity of KK\,77 is
unknown. To maximize on the range of velocities for this source,
we used the 4IF mode for this galaxy, a correlator mode that enables
observing with two simultaneous frequencies. NGC\,247 was also observed in
the 4IF mode because of its extreme \hi\ linewidth of $\sim
200$\,\kms, by far the widest line of our sample. In the presence of
extremely strong, narrow line features, the response of the VLA can
include signs of the Gibbs phenomenon, a {\it sinc} function that
oscillates between channels. This oscillation can be suppressed by
``Hanning'' smoothing the data with a triangular smoothing kernel.  We
decided not to use online Hanning smoothing as the lines are too weak
to show any signs of the Gibbs phenomenon. Some of the archival
observations that we added in, however, were observed with online
Hanning smoothing turned on. We also applied a 25\,MHz frontend filter
to reduce the impact on radio frequency interference (RFI) in our
data.

For flux and bandpass calibration purposes, we observed the VLA
standard flux calibrators 3C286 and/or 3C48 (with fluxes of $\sim
15.0$\,Jy and $\sim 16.5$\,Jy, respectively) depending on their
visibility at the time of the observations. Typical integration times
were 12\,minutes on the flux calibrator, split between the beginning
and end of a track. We observed our target galaxies in 40\,minute
intervals alternating with 3\,minutes on a nearby phase/complex gain
calibrator. The complex gain calibrators were chosen to be nearby
point sources with a minimum flux of $\sim 1$\,Jy.  B-configuration
observations were obtained in single 9\,hour programs, or, in
the case of southern galaxies, two 4.5\,hour observations. In C- and
D-configurations we combined sources into a few observations to reduce
overhead and to obtain an improved {\it uv-}coverage.

During the VLA to EVLA transition time, the EVLA-EVLA baselines showed
considerable aliasing and were largely unusable. Since the
D-configuration observations had the largest number of EVLA antennas
and therefore EVLA-EVLA baselines, we obtained an additional hour of
observing time per object to reach our original sensitivity goals.
Overall we obtained a total of 3\,h in D-configuration per galaxy. The
aliasing is a monotonic function that affects the first $\sim
0.5$\,MHz (or $\sim 100$\,\kms\ at the \hi\ frequency), decreasing
from the bandpass edge toward the center\footnote{see
  \url{http://www.vla.nrao.edu/astro/guides/evlareturn/aliasing.shtml}}.
In addition to the make-up time and using mostly baselines with old
VLA antennas, we placed the \hi\ lines away from the affected
frequency ranges but could not fully eliminate the effects. As a
result, the noise levels in our data products are not entirely uniform
across all channels and increase toward lower frequencies where
aliasing is strongest. Across the full width of our most narrow
0.78\,MHz bandwidths the noise can vary up to 35\%. For the smaller
width of the spectral line feature, the effect is smaller. In addition,
for part of the observations, we tried to place the \hi\ signal away
from the most affected frequency ranges and for most galaxies the
noise level changes could be constrained to an rms variation of
$\lesssim 15$\,\% over the width of the \hi\ line. For the two
galaxies that were observed in the 4IF mode, KK\,77 and NGC\,247, the
full bandwidths were required to cover the requested velocity
ranges. The aliased signal appears in each IF and, consequently, the
full noise variations of up to 35\% are visible across the combined
spectrum.  Overall, however, this will not have a huge
impact on most of the data analysis; KK\,77 is a non-detection, and
NGC\,247 is one of our brightest objects such that the signal-to-noise
ratio is only moderately affected.

A few galaxies (marked ``a'' in Table\,\ref{tab:obs}) had \hi\
emission whose velocity range overlapped with Galactic \hi\ line
features. We addressed this problem by observing the bandpass
calibrator at two different frequencies, offset by $\pm4$\,MHz from
the source frequency, in order to interpolate the calibration to the
source frequency.

In addition to the new observations, in preparing for the program, we
identified a few archival VLA observations which would be of use to
the program. Most archival data, however, were observed at the
relatively lower spectral resolution of $2.6$\,\kms\ with online
Hanning smoothing applied. To avoid interpolation, we kept the lower
velocity resolution of the archival data and rebinned our new
VLA-ANGST observations after calibration for the final cube.

Table\,\ref{tab:obs} lists all of our observational setups. Column (1)
lists the galaxy names followed by the array configurations of the
observations and the project codes in columns (2) and (3). AO215 is
the genuine VLA-ANGST project, while other project codes refer to
archival data. The observing dates are listed in column (4). Asterisks
denote observations that span across midnight, for which we refer to the
start dates. The equatorial pointing positions are provided in columns
(5) and (6) and the appropriate equinox is tabulated in column (7).
Column (8) gives the phase calibrators that were used for the
observations. The correlator setups are displayed in columns (9)
through (14) with (9) the correlator modes, (10) the bandwidths, (11)
the number of channels, and (12) the channel widths in \kms\ (for the
rest frequency of \hi\ at 1.420405752\,GHz). VLA-ANGST data were taken
at the fixed sky frequencies listed in column (13) whereas many
archival observations were Doppler tracked at the velocities listed in
column (14). Finally, column (15) provides the number of converted
EVLA antennas in the dataset with the remaining antennas being old,
not yet converted, VLA antennas at the time.

\subsection{Data Calibration}
\label{ssec:calibration}

Data calibration was performed using the {\sc AIPS}\footnote{The
  Astronomical Image Processing System (AIPS) has been developed by
  the NRAO.} package and deviated from the ``standard'' VLA data
reduction procedure due to the effects from the VLA to EVLA
transition period. As mentioned above, EVLA-EVLA baselines showed
considerable aliasing that affected primarily the narrow bandwidth
observations. Our reduction scheme followed the following steps.

To avoid low signal-to-noise solutions during calibration, we started
by eliminating edge channels. About 20\% of the entire bandwidths were
cropped, $\sim 10$\% at the upper and $\sim 10$\% at the lower
frequency ends (task {\sc UVCOP}). After correcting the absolute
antenna positions ({\sc VLANT}), the absolute flux scale of the
primary flux calibrators 3C286 and 3C48 (see \S.\,\ref{ssec:obssetup})
were calculated by the task {\sc SETJY} using NRAO-provided models and
``Perley-Taylor 99'' flux scales.  We manually inspected all
calibrator data in each array configuration for bad visibilities due
to RFI or cross-talk between antennas (AIPS tasks {\sc TVFLAG, SPFLG,
  UVFLG, WIPER}). If calibrators showed higher than expected signal on
short baselines due to solar interference, we excluded baselines with
a {\it uv} range between 0 and 1 k$\lambda$ in the calibration. Solar
RFI was usually well removed by continuum subtraction in the source
visibilities, so we performed no additional source flagging due to
solar interference.

The data were then bandpass-corrected by deriving a channel-based,
normalized gain solution from the flux calibrator data via {\sc
  BPASS}. We produced a new broad band ``channel zero'' ({\sc CH0})
map from these bandpass corrected data ({\sc AVSPC}), and utilized the
new {\sc CH0} map in all subsequent calibration steps. Next, we
calculated the complex antenna gain as a function of time for all
calibrators ({\sc CALIB}). The complex gain is a solution for both
gain and phase, and for all solutions we assumed that the complex gain
calibrators are point-like. With {\sc CLCAL} we linearly interpolated
all phase/gain solutions for across all time intervals. {\sc GETJY}
transferred the flux calibration to the complex gain calibrator. At
this stage, we manually inspected the quality of the calibration and
repeated the above procedure if RFI corrupted the solution and further
flagging was required. Finally, we applied all calibration solutions
to the target galaxies. We estimate the calibration flux uncertainties
to be $\sim 5$\%.

For galaxies whose \hi\ emission is within the velocity range of the
foreground Galactic emission (marked ``a'' in Table\,\ref{tab:obs}),
the flux and bandpass calibrators were observed at offsets of
$\pm4$\,MHz relative to the frequency of the targets. The calibration
of these data followed the same steps as above with each observed
frequency calibrated separately. We then obtained an interpolated flux
and bandpass solution calibration from a linear interpolation across
these offset frequencies. The EVLA uses different internal local
oscillator settings than the VLA. For our observations during the VLA
to EVLA transition phase, this difference had the unfortunate effect
that, for some observations, the 25\,MHz frontend filter overlapped in
frequency with the upper frequency offset bandpass observations. This
resulted in an extreme phase gradient across that particular offset
frequency and the affected observations were unusable.  In these
cases, we calibrated the source data by extrapolating the calibration
from the single offset frequency that most closely matched the
bandpass derived from the phase calibrator.  This method provided
bandpasses that are accurate to few percent.  For observations where
the {\it complex gain} calibrator was contaminated by Galactic \hi, we
simply flagged the affected channels before calculating a new {\sc
  CH0} to be used for the gain and phase solutions.

After applying all calibration tables to the source, the galaxy data
were separated ({\sc SPLIT}) from the flux and phase calibrator data
for further processing. Both NGC\,247 and KK\,77 were observed in the
4IF mode. We calibrated each of the 2IF windows separately as
described above, and ``stitched'' the data together, averaging
overlapping velocity ranges in the process ({\sc UJOIN}) prior to
continuum subtraction.

To determine the continuum level, we fit a linear function to the
line-free channels, typically 20 channels on each side of the
spectrum. The fit was then subtracted from the complex $uv$-data (task
{\sc UVLSF}). Since the frequencies for each observing set up were
fixed (see \S.\ref{ssec:obssetup}), we regridded the \hi\ data to a
common heliocentric, optical velocity system ({\sc CVEL}).  Finally,
data from all observations and array configurations were combined
({\sc DBCON}) into a single data set for imaging.  We produced dirty
images with {\sc IMAGR} and inspected the cubes for artifacts. If
further flagging was required, we went back to the original source
data, flagged, and reapplied the calibration.

\subsection{Mapping and Deconvolution}\label{ssec:mapping}

After satisfactory calibration and source editing, we used the AIPS
task \textsc{IMAGR} to generate data cubes and final data products. We
followed the THINGS protocol when possible so that the two data sets
could be easily compared. For each VLA-ANGST galaxy, we imaged the
visibilities with two different weighting schemes: one using natural
weighting and one using the ``robust'' weighting \citep[originally
described by][with small modifications as described in the AIPS
help files]{bri95}. Natural weighting yields high sensitivity at moderate
resolution (typically $\sim 6-12\arcsec$ for our galaxies, or about
$\sim 90-170$\,pc for a distance of 3\,Mpc), while robust weighting
decreases the size of the synthesized beam at the cost of reduced
surface brightness sensitivity. We applied a robust parameter of 0.5,
which was found to be a good compromise between resolution and
sensitivity and matches the maps generated by THINGS. When compared to
the naturally-weighted cubes, the noise in the robust-weighted cubes
is typically $\sim 20$\% higher and the beam size $\sim 40$\% smaller.
Depending on the angular extent of each galaxy's \hi\ emission, the
cubes were imaged with either 1024$^2$ pixels at 1.5\arcsec{} per
pixel or 2048$^2$ pixels at 1.0\arcsec{} per pixel. The cubes were
deconvolved using the Clark {\sc Clean} deconvolution algorithm
\citep{cla80}, stopping at a residual flux threshold of 2.5 times the
noise level as measured in the cubes. Finally, we produced primary
beam corrected data cubes that were later used in the moment map
analysis (\S\,\ref{sec:dataProducts}).

The properties of all data cubes are listed in
Table\,\ref{tab:dataproperties} where column (1) lists the galaxy
names followed by columns (2), (3), (4), and (5) that contain the
weighting algorithms and the resulting beam major and minor axes sizes
as well as the position angles of the deconvolved data. The average
root mean squared noise per channel is shown in column (6) and the
channel width in (7), the number of pixels in each plane in column (8)
followed by the pixel size in column (9).

The most important difference between the VLA-ANGST datacubes and the
THINGS datacubes is the higher velocity resolution in the majority of
the VLA-ANGST datacubes which was possible because of the overall
smaller range in velocity of detectable \hi\ in the VLA-ANGST sample
of galaxies. The sensitive and spatial resolution of THINGS and
VLA-ANGST are very similar. The spatial resolution of LITTLE THINGS
and SHIELD are similar to VLA-ANGST, too, since all surveys share the
same VLA array configuration setups. Given the wider lines of
the somwehat more massive galaxies in LITTLE THINGS (see
Fig.\,\ref{fig:surveys}), the bandwidth had to be slightly increased
for this survey and their velocity resolution hovers around the upper
limit of ours at $\sim 2$\,\kms. On the other hand, LITTLE THINGS has
about 1/3 longer integration times than VLA-ANGST. The two Westerbork
surveys HALOGAS and WHISP exhibit 2 to 5 times lower spatial
resolution than the VLA surveys but their sensitivity limits are
comparable, observed at a lower velocity resolution of $\sim
5$\,\kms. FIGGS has a velocity resolution of $\sim 1.5$\,\kms\ and a
beam 2-8 times larger than VLA-ANGST.

To ensure that our datacubes would be as directly comparable to the
THINGS datacubes as possible, we reduced THINGS observations using our
calibration, mapping, and deconvolution protocols.  Comparisons of our
reductions of THINGS observations with the publicly available THINGS
datacubes showed no significant differences.

\subsection{Mask Generation}

To suppress noise for the production of moment maps, we generated
image cube masks that defined regions containing detectable \hi{}
emission from the galaxies. To do so, we convolved the
natural-weighted images to twice the original beam major axis and
applied spectral Hanning smoothing with a three channel wide kernel.
New ``mask'' cubes containing predominantly \hi\ signal were
constructed by keeping all regions corresponding to \hi\ emission
above the $1\sigma$ noise level and blanking all other regions.  To
remove the effects of sidelobes, noise spikes, or other spurious
signals from the masks, any individual regions with an area smaller
than the beam size were automatically removed. We then eliminated any
remaining non-emission regions by eye inspection.

The result is a single mask cube per galaxy that we applied prior to
the generation of the integrated spectra and moment maps. The same
mask is used for both the natural and robust data cubes.  New cubes
containing predominantly \hi\ signal were constructed by keeping all
\hi\ emission corresponding to unblanked regions in the mask cube.
For our data, this method discriminated well between significant, low
level emission and pure noise.  Note that this process produces a
lower limit to the total \hi\ emission.  Some very low surface
brightness \hi\ may be have been eliminated from these data cubes.

Note that mask generation is not entirely automatic, and, therefore,
our mask generation cannot be said to be strictly following THINGS
protocols.  This is exacerbated by the differences in typical channel
widths, which leads to differences between VLA-ANGST and THINGS in
average noise levels in the individual channels.  Nonetheless, we have
followed the intentions of the THINGS project to use masking to
suppress the noise and to provide optimal moment maps.  However, one
should be aware that small differences in masks are possible and that
the resulting moment maps have a small inherent systematic
uncertainty.

\subsection{Flux Densities}\label{ssec:fluxDensities}

Recovery of the total \hi\ flux from each channel and the resulting
\hi\ spectra is more complicated than simply summing up the total
emission. \citet{jor95} have shown that standard {\sc Clean} maps do
not in actuality yield correct flux measurements. Maps in AIPS are
created by summing the {\sc Clean} components, convolved with the {\sc
  Clean} beam, to the signal that is still present in the residuals.
While both maps are purportedly measured in
Jansky\,(beam~area)$^{-1}$, the relevant beam is different in each
map: the convolved {\sc Clean} component map is measured in Jy\,({\sc
  Clean}~beam~area)$^{-1}$ while the residual map has units of
Jy\,(dirty~beam~area)$^{-1}$.  Because the areas of the {\sc Clean}
beam and the dirty beam are different, the flux in the {\sc Clean}
components and in the residuals must be corrected to obtain the 
proper \hi\ flux. A full discussion of the following correction
technique is given in \citet{jor95}. The corrected flux of a channel
is given by:
\begin{equation}
  G = C + \epsilon \times R
\end{equation}
where $G$ is the corrected flux, $C$ is the flux in the convolved {\sc
  Clean} components [with units of Jy\,({\sc Clean}~beam)$^{-1}$], $R$
is the flux in the residual map [with units of
Jy\,(dirty~beam)$^{-1}$], and $\epsilon$ is the correction factor that
takes into account the ratio of the dirty beam area to the {\sc Clean}
beam area.  \textsc{IMAGR} provides a method to automatically apply
this correction and, following the THINGS protocol, we determined
$\epsilon$ within the inner $50\times50$ pixels of the dirty and {\sc
  Clean} beams. This produces a set of two new natural and robust
weighted cubes with the above correction applied in addition to our
standard cubes. When the residuals are scaled by $\epsilon$, the
noise in the corrected cubes is artificially suppressed. We thus
produced two sets of data cubes for different analyses:

\begin{enumerate}
\item ``Standard'' cube: the standard output from our pipeline, with
  uncorrected \hi{} fluxes but uniform noise properties. No primary
  beam correction is applied to these data. This cube should be used
  for any analysis that requires uniform noise properties or uses
  selection based on noise (e.g., fitting of individual
  profiles in order to construct velocity fields or measure profile
  shapes).

\item ``Rescaled'' cube: the cube with the flux correction applied, to
  be used only in regions with genuine \hi\ emission. The flux values
  in this cube are correct, and therefore any analysis that requires
  selection based on \hi\ fluxes should use this cube (e.g., mass and
  column density measurements). The ``rescaled'' cube is corrected for
  the attenuation from the primary beam.

\end{enumerate}

The data products that we make available follow this recipe; all
global \hi\ spectra (\S~\ref{ssec:spectra}) and moment maps
(\S~\ref{sec:dataProducts}) were derived using the masked, ``rescaled
cubes''. The online data cubes themselves are the ``standard'' cubes,
without primary beam attenuation or flux corrections.

\subsection{Global \hi\ Spectra and Masses}\label{ssec:spectra}

Global \hi\ spectra are derived from the masked, rescaled data cubes.
The spectra are used to derive velocity widths at 20\% ($w_{\rm 20}$)
and 50\% ($w_{\rm 50}$) of the peak. The central \hi{} velocity of
each galaxy is calculated by taking the mid-point of the $w_{\rm 20}$
boundaries.

We also use the integrated \hi\ spectra to calculate the total \hi{}
masses of our galaxies using the following equation:
\begin{equation}
  M\hissc \left[ M_{\odot}\right] = 
  2.36 \times 10^5 \; D^2 \times \sum_i S_i \Delta v
\end{equation}
where $D$ is the distance to the galaxy in Mpc (as given in
Table~\ref{tab:sample}) and $S_i \Delta v$ is the total flux of a
single channel in Jy\,km\,s$^{-1}$ \citep[e.g.][]{roh04}. This formula assumes
that the \hi\ emission is optically thin, an assumption that is valid
over a large flux range and may begin to fail at very extreme column
densities of $\gtrsim 10^{22}$\,cm$^{-2}$ \citep[e.g.][]{all12,bra12}. At
our spatial resolution, we do not observe column densities of
this magnitude.

In Table\,\ref{tab:hiproperties}, we present all of the derived \hi\
parameters starting with the galaxy names in column (1), followed by
the integrated \hi\ flux densities $S_{\rm HI}$ and the derived \hi\
masses in columns (2) and (3). For comparison, we compiled single dish
fluxes $S_{\rm HI}^{\rm SD}$ from the literature and list them in
column (4). The $w_{\rm 20}$ and $w_{\rm 50}$ values as well as the
central velocities are shown in columns (5), (6), and (7),
respectively, followed by the peak \hi\ column density
(\S\,\ref{ssec:m0maps}) taken from the natural-weighted map in the
final column (8). To derive upper limits for the non-detections, we
assume an \hi\ disk the same size as the optical diameter $D_{25}$ and
a hypothetical linewidth of 20\,\kms. Galaxies typically exhibit \hi\
dispersions of 5-10\,\kms\ and a linewidth of 20\,\kms\ thus implies
little rotation or face-on orientation.

In $\sim 70$\% of all cases the single dish fluxes are somewhat larger
than the interferometric VLA flux measurements. This difference is
expected to some level given that the VLA can only image structures
with an extent of up to $\sim 16\arcmin$ in D-configuration at
1.4\,GHz (limited by the minimum distance between two
antennas). Missing flux may therefore only be a significant issue for
the most extended objects in our sample. Some single dish flux
measurements deviate substantially from the trend of being slightly
larger than the VLA fluxes. The deviations can be either way: galaxies
like DDO\,6, UGC\,4483, DDO\,113, and KK\,230 have much larger single
dish measurements whereas others like BK\,3N, AO\,0952+69, Sextans\,B,
DDO\,82, and DDO\,190, have smaller single dish fluxes. Such
discrepancies may be explained by difficulties in single dish baseline
subtraction or by the larger single dish beam that tends to pick up
larger fractions of Galactic \hi\ emission as well as flux from nearby
objects.

\section{Data Products}\label{sec:dataProducts}

\subsection{\hi\ Spectra}

We used the naturally-weighted data cubes to derive \hi\ spectra
(\S\,\ref{ssec:spectra}) for our galaxies given their higher surface
brightness sensitivity compared to the robust-weighted data. This
approach captures as much low-level, extended emission as is possible
with interferometric VLA data. All fluxes are calculated from the
rescaled cubes described in $\S$\ref{ssec:fluxDensities} and are
presented in Fig.\,\ref{fig:spectra}. In the case of NGC\,247,
velocities around 108\,\kms{} were strongly contaminated with RFI; to
estimate the \hi{} flux in this channel, we interpolated the emission
from the adjacent channels.

Since our sample is composed primarily of dwarf galaxies, the galaxy
spectra typically show narrow, singly-peaked profiles. Extreme cases
like KDG\,73 and KKH\,86 exhibit linewidths of $<10$\,\kms, which
implies very little velocity dispersion, maybe due to low signal to
noise in the line wings. On the other end of the mass spectrum, a few
galaxies (NGC\,404 NGC\,3741, DDO\,190, Sextans\,A, DDO\,181,
NGC\,3109, NGC\,247) exhibit hints of the more familiar double-horned
profile expected from larger spiral disks. The maximum linewidth is
observed in NGC\,247 with $w_{20}\sim200$\,\kms.

\subsection{Channel Maps}\label{ssec:channelMaps}

Channel maps of the galaxies are presented in Figs.\,\ref{fig:n247}
through \ref{fig:kkh98} (natural weighting). Given the high spectral
resolution of our data, there is typically only little flux in each
velocity bin. Some galaxies, mainly the more massive ones such as
NGC\,247 or NGC\,3109 show the classic ``butterfly'' pattern of a
spiral galaxy, a tell-tale sign for a flat rotation curve. The bulk of
galaxies exhibit rotation despite the fact that the dispersion adds a
stochastic component to the velocity structure. NGC\,247 also features
a \hi\ absorption feature along the line of sight to the background
quasar NVSS J004713-205114 at RA\,(J2000)$=00^{h}:47^{m}:13.6^{s}$ and
DEC\,(J2000)$=-20^{\circ}:51\arcmin:15\arcsec$ \citep[e.g.,][]{dic92}. Some
data cubes are contaminated by Galactic foreground emission, but only
for NGC\,404 is the Galactic \hi\ close to the systemic velocity of
the source. Other data cubes, such as that for AO\,0952+69, contain
emission from a nearby object. AO\,0952+69, in fact, is likely not a
real galaxy but might be a feature within a spatially coincident
spiral arm that belongs to the massive M\,81 galaxy.

\subsection{Moment Maps}\label{ssec:momentmaps}

We used the AIPS task \textsc{XMOM} to generate moment maps from the
masked, flux-corrected cubes. For all calculations we require that
each pixel in a moment map is calculated from at least four unmasked
channels; pixels with fewer channels are masked in all moment maps.

\subsubsection{Integrated HI Maps}\label{ssec:m0maps}

Integrated \hi{} column density maps are created from the masked,
rescaled data cube by integrating along the velocity axis to generate
the moment 0 map: \begin{equation}
  I\hissc = \sum_i S_i \times \Delta v
\end{equation}
where $i$ is the channel, $S_i$ is the flux density in the $i$th
channel in Jy\,beam$^{-1}$, and $\Delta v$ is the channel spacing in
km~s$^{-1}$. We then convert the moment maps to column density with
\begin{equation}
  N\hissc = 1.104 \times 10^{24} \frac{1}{b_{\textrm{maj}}
  \,\,b_{\textrm{min}}} \sum_i S_i \Delta v
\end{equation}
where $b_{\textrm{maj}}$ and $b_{\textrm{min}}$ are the beam major and
minor axes in arcseconds and $\sum_i S_i \Delta v$ is the value of
moment 0 map at each pixel in units of Jy\,beam$^{-1}$\,\kms. In
Figs.\,\ref{fig:n247} to \ref{fig:kkh98} we show column density maps
for all galaxies with detected \hi\ (upper left panels on the second
page of each figure) We also show column density contours overlaid on
optical images for each galaxy. On these maps (upper right), we placed
the footprints of the HST observations that are available through
ANGST.

The maps exhibit resolved \hi\ structures comparable to their beam
sizes. Some galaxies, like KK\,230, or NGC\,404 have low columns with
peak values of a few times $10^{20}$\,\cden. Other galaxies like
DDO\,190, or UGCA\,292 reach columns of a few times $10^{21}$\,\cden\
(or $\sim 10$\,\,\msun\,pc$^{-2}$), which is a canonical threshold
for star formation \citep[e.g.,][]{ski87,ken89,big08,ler08}.

\subsubsection{Intensity-weighted Velocity Field Maps}\label{ssec:m1maps}

The \hi{} intensity-weighted velocity fields (moment 1) maps are
calculated using
\begin{equation}
  \langle v \rangle = \frac{\sum_i S_i \times v_i}{\sum_i S_i}.
\end{equation}

For well-behaved disks this equation gives a good indication of the
average velocity of gas in a given pixel. However, bulk motions,
outflows, and other non-circular motions can shift the derived
velocity to unexpected values. Therefore, profile fitting in order to
determine the velocity of the peak of the emission is a more reliable
method for finding the average rotational velocity of the gas at a
given location in the galaxy. While the velocity fields of lower mass
dwarfs are less ordered than those of their larger disky counterparts,
most still show velocity gradients across their disks that are
indicative of rotation, which is typical for \hi\ in dwarf galaxies
\citep{beg08,wal08}. The \hi\ intensity-weighted velocity maps are
shown in the lower right panels of the second page of for each galaxy
in Figs.\,\ref{fig:n247} to \ref{fig:kkh98}.

\subsubsection{Second Moment Maps}\label{ssec:m2maps}

The linewidth of \hi\ emission can be characterized by the
    intensity-weighted second velocity moment as given by:

\begin{equation}
  \sigma = \sqrt{ 
    \frac{ \sum_i S_i \times \left( v_i - \langle v \rangle \right)^2 }
	 { \sum_i S_i } }
\end{equation}
where $\langle v \rangle$ indicates the intensity-weighted velocity
derived in the first moment map. While the second moment can be
indicative of the turbulence of the ISM, it also reflects the
influence of large scale gas flows such as expanding shells or tidal
material. At lower resolution, the velocity dispersion can be
artificially inflated by beam smearing over the gradient in the
velocity field, especially towards the centers of the galaxies, where
this gradient is steepest. Overall, the velocity dispersion values
fall in a relatively narrow range of 5-15\,\kms, as seen in the lower
right panel on the second page of Figs.\,\ref{fig:n247} to
\ref{fig:kkh98}).

Pixels which yield first velocity moments outside
the velocity range of the data cube are blanked in all moment
maps. The first and second moment maps generated from the robust data
cubes are noisier than those from natural-weighted cubes and
occasionally have pixels with unrealistic values in low column density
regions. To counter this problem, we blanked all pixels with
column densities $N_{\rm HI} <3 \times 10^{19}$\,cm$^{-2}$ in the
robust moment maps.

\section{Summary}
\label{sec:summary}

Here we present the sample selection criteria, observational
parameters, data reduction procedures, and data product description of
the VLA-ANGST survey, a Large VLA project that targets nearby, mostly
dwarf irregular galaxies. Of the 35 galaxies in the survey, we detect
\hi\ in 29. The calibrated VLA data for these objects are publicly
available at
\url{https://science.nrao.edu/science/surveys/vla-angst}. This leads
to the following data products:

\begin{itemize}

\item Global \hi\ spectra for all galaxies, derived from the masked,
  flux-corrected, primary beam corrected, natural-weighted data cubes. 

\item \hi\ data cubes of both natural and robust weighting. The
  cubes are not primary beam or flux corrected and they are unmasked.

\item Integrated intensity maps (moment 0 maps) in units of
  Jy\,beam$^{-1}$\,\kms\ as well as converted to \hi\ column
  densities. These maps were derived from the masked, flux-corrected,
  primary beam corrected data cubes. We offer both, natural and
  robust-weighted maps for download.

\item The intensity-weighted velocity field maps (moment 1 maps), derived
  from the same data products as the integrated intensity maps.

\item Second moment maps which give a measure for the velocity
  dispersion of the gas; derived from the same data products as the
  integrated intensity maps.

\end{itemize}

This paper presents the observations; scientific analyses will follow
in subsequent publications. A study that compares the \hi\ kinematics
of large \hi\ shells to the supernovae and stellar wind output of the
underlying stellar populations is presented in \citet{war11}.
Detection and characterization of narrow \hi\ components that
presumably trace cold \hi\ are provided in \citet{war12}.
\citet{sti12} correlate averaged \hi\ dispersion values to the
physical properties of the host galaxies. The true value of our \hi\
survey is further unlocked by the extensive, multi-wavelength
ancillary data that is available for many of our objects. The
VLA-ANGST, THINGS, LITTLE THINGS, and SHIELD data products are
furthermore similar in sensitivity, spatial and spectral resolution
and provide \hi\ data cubes for $>100$ galaxies.


\begin{deluxetable}{llrrrrrrrc}

\tablewidth{0pt}
 \tabletypesize{\tiny}%
\rotate
\tablecaption{General properties of the VLA-ANGST galaxy sample.\label{tab:sample}}
\tablehead{
\colhead{(1)}           & 
\colhead{(2)}      &
\colhead{(3)}          & 
\colhead{(4)}  &
\colhead{(5)}    &
\colhead{(6)}  &
\colhead{(7)}  &
\colhead{(8)} & 
\colhead{(9)} &
\colhead{(10)}\\

\colhead{Name}           & 
\colhead{Alt. Name}      &
\colhead{RA}          & 
\colhead{DEC}  &
\colhead{D}    &
\colhead{$D_{\rm 25}$}  &
\colhead{$M_{\rm B}$}  &
\colhead{$\nu L_{\nu}$} & 
\colhead{Type}&
\colhead{$SFR$}\\

\colhead{}           & 
\colhead{}      &
\colhead{(J2000)}          & 
\colhead{(J2000)}  &
\colhead{}    &
\colhead{}  &
\colhead{}  &
\colhead{($3.6$\,$\mu$m)} & 
\colhead{}&
\colhead{(UV)}\\

\colhead{}           & 
\colhead{}      &
\colhead{[$^{h}$:$^{m}$:$^{s}$]}          & 
\colhead{[$\degr$:$\arcmin$:$\arcsec$]}  &
\colhead{[Mpc]}    &
\colhead{[kpc]}  &
\colhead{[mag]}  &
\colhead{[$10^{6}$\,L$_{\sun}$]} & 
\colhead{}&
\colhead{[$10^{-3}$\,\sfr]}

}

\startdata
NGC\,247              & ESO\,540-G022, UGCA\,11      & 00:47:08.3 & -20:45:36 & 3.50 & 15.7   & -17.86  &  270.52   & 7   &  229  \\
DDO\,6                & ESO\,540-G031, UGCA\,15      & 00:49:49.3 & -21:00:58 & 3.31 & 1.6    & -12.41  &  \nodata  & 10  &  1.4    \\
NGC\,404              & UGC\,718                     & 01:09:26.9 & 35:43:03  & 3.05 & 2.2    & -16.21  &  \nodata  & -1  &  6.6    \\
KKH\,37               & LEDA\,95597                  & 06:47:45.8 & 80:07:26  & 3.26 & 1.1\tablenotemark{a}    & -11.17  &  0.69     & 10  &  0.3    \\
UGC\,4483             & CGCG\,331-051                & 08:37:03.0 & 69:46:31  & 3.41 & 1.2    & -12.71  &  0.57     & 10  &  4.0    \\
KK\,77                & LEDA\,166101                 & 09:50:10.0 & 67:30:24  & 3.55 & 2.5\tablenotemark{a}    & -11.45  &  \nodata  & -3  &  \nodata \\
BK3N                  & PGC\,28529                   & 09:53:48.5 & 68:58:09  & 3.86 & 0.6    & -9.15   &  $<$0.03  & 10  &  0.3    \\
AO\,0952+69\tablenotemark{b}           & Arp's Loop                   & 09:57:29.0 & 69:16:20  & 3.78 & 2.0    & -11.09  &  \nodata  & 10  &  1.1    \\
Sextans\,B            & UGC\,5373, DDO\,70           & 10:00:00.1 & 05:19:56  & 1.39 & 2.1    & -13.87  &  2.48     & 10  &  4.5    \\
NGC\,3109             & ESO\,499-G036, DDO\,236      & 10:03:07.2 & -26:09:36 & 1.26 & 6.2    & -15.11  &  12.40    & 10  &  28.7  \\
Antlia                & PGC\,29194                   & 10:04:04.0 & -27:19:55 & 1.29 & 0.8    & -9.36   &  \nodata  & 10  &  1.8    \\
KDG\,63               & UGC\,5428\,DDO71             & 10:05:07.3 & 66:33:18  & 3.53 & 1.7    & -11.73  &  1.36     & -3  &  \nodata \\
Sextans\,A            & UGCA 205, DDO 075            & 10:11:00.8 & -04:41:34 & 1.38 & 2.2    & -13.84  &  1.82     & 10  &  12.3    \\
HS\,117               & \nodata                      & 10:21:25.2 & 71:06:58  & 3.82 & 1.7    & -11.41  &  1.08     & 10  &  \nodata \\
DDO\,82               & UGC\,5692                    & 10:30:35.0 & 70:37:10  & 3.80 & 3.8    & -14.33  &  \nodata  & 9   &  2.6    \\
KDG\,73               & PGC\,32667                   & 10:52:55.3 & 69:32:45  & 4.03 & 0.7    & -10.94  &  0.81     & 10  &  0.3    \\
NGC\,3741             & UGC\,6572                    & 11:36:06.4 & 45:17:07  & 3.24 & 1.9    & -13.17  &  1.29     & 10  &  6.2    \\
DDO\,99               & UGC\,6817                    & 11:50:53.0 & 38:52:50  & 2.59 & 3.1    & -13.37  &  1.43     & 10  &  5.5    \\
NGC\,4163             & NGC\,4167, UGC\,7199         & 12:12:08.9 & 36:10:10  & 2.86 & 1.6    & -13.65  &  3.66     & 10  &  4.0    \\
NGC\,4190             & UGC\,07232                   & 12:13:44.6 & 36:38:00  & 3.50\tablenotemark{c} & 1.7    & -14.20  &  5.80     & 10  &  10.1    \\
DDO\,113              & UGCA\,276                    & 12:14:57.9 & 36:13:08  & 2.95 & 1.3    & -11.65  &  0.61     & 10  &  \nodata \\
MCG\,+09-20-131       & CGCG\,269-049                & 12:15:46.7 & 52:23:15  & 1.60\tablenotemark{d} & 0.6    & -10.72  &  \nodata  & 10  &  0.4   \\
DDO\,125              & UGC\,7577                    & 12:27:41.8 & 43:29:38  & 2.58 & 3.2    & -14.11  &  4.15     & 10  &  5.3    \\
UGCA\,292             & PGC\,42275                   & 12:38:40.0 & 32:46:00  & 3.62 & 1.1    & -11.69  &  0.48     & 10  &  2.9    \\
GR\,8                 & UGC\,8091, DDO\,155          & 12:58:40.4 & 14:13:03  & 2.08 & 0.7    & -11.98  &  0.34     & 10  &  2.4    \\
UGC\,8508             & I\,Zw\,60                    & 13:30:44.4 & 54:54:36  & 2.58 & 1.3    & -12.94  &  1.40     & 10  &  \nodata \\
DDO\,181              & UGC\,8651                    & 13:39:53.8 & 40:44:21  & 3.14 & 2.1    & -13.03  &  1.51     & 10  &  3.8    \\
DDO\,183              & UGC\,8760                    & 13:50:51.1 & 38:01:16  & 3.22 & 2.1    & -13.09  &  1.63     & 9   &  3.2    \\
KKH\,86               & LEDA\,2807150                & 13:54:33.6 & 04:14:35  & 2.59 & 0.5\tablenotemark{a}    & -10.19  &  0.22     & 10  &  0.1    \\
UGC\,8833             & PGC\,49452                   & 13:54:48.7 & 35:50:15  & 3.08 & 0.8    & -12.29  &  0.60     & 10  &  1.4    \\
KK\,230               & KKR\,3                       & 14:07:10.7 & 35:03:37  & 1.97 & 0.3\tablenotemark{a}    & -8.57   &  0.05     & 10  &  0.2    \\
DDO\,187              & UGC\,9128                    & 14:15:56.5 & 23:03:19  & 2.21 & 1.1    & -12.34  &  0.39     & 10  &  1.1    \\
DDO\,190              & UGC\,9240                    & 14:24:43.5 & 44:31:33  & 2.79 & 1.5    & -14.13  &  3.12     & 10  &  6.1    \\
KKR\,25               & LEDA\,2801026                & 16:13:47.6 & 54:22:16  & 1.93 & 0.6\tablenotemark{a}    & -9.98   &  $<$0.02  & 10  &  \nodata \\
KKH\,98               & LEDA\,2807157                & 23:45:34.0 & 38:43:04  & 2.54 & 0.8\tablenotemark{a}    & -10.32  &  0.40     & 10  &  0.6    \\
\enddata

\tablerefs{(5) tip of the red giant branch distances from
  \citet{dal09}; (6) taken from \citet{dal09} and converted to
  physical diameters; (7) apparent blue magnitudes from \citet{kar04}
  and converted to absolute blue magnitudes; (8) converted from
  infrared fluxes given by \citet{dale09}; (9) T-type form
  \citet{dal09}; (10) converted from GALEX FUV asymptotic magnitudes
  given by \citet{lee11} and using
  $SFR=1.4\times10^{-28}\,L_{\nu}$(erg\,s$^{-1}$\,Hz$^{-1}$)  \citep{ken98}}

\tablenotetext{a}{for the KK-listed objects, the diameters are taken
  at a 26.5
  mag\,arcsec$^{-2}$ surface brightness level}
\tablenotetext{b}{object might be a feature in the spiral arm of M\,81 rather than a galaxy}
\tablenotetext{c}{TRGB distance from \citet{kar04}}
\tablenotetext{d}{the TRGB branch was not unambiguously identified in \citet{dal09}}

\end{deluxetable}

\clearpage
\begin{deluxetable}{lcccccccccccccc}
\tablewidth{0pt}
\tabletypesize{\scriptsize}
\rotate

\tablecaption{List of Observations \label{tab:obs}}
\tablehead{
  \colhead{(1)} &
  \colhead{(2)} &
  \colhead{(3)} &
  \colhead{(4)} &
  \colhead{(5)} &
  \colhead{(6)} &
  \colhead{(7)} &
  \colhead{(8)} &
  \colhead{(9)} &
  \colhead{(10)} &
  \colhead{(11)} &
  \colhead{(12)} &
  \colhead{(13)} &
  \colhead{(14)} &
  \colhead{(15)} \\

  \colhead{Galaxy} &           
  \colhead{Conf.} &            
  \colhead{Project} &          
  \colhead{Date} &             
  \colhead{RA} &                        
  \colhead{Dec} &                       
  \colhead{Equ.} &                      
  \colhead{Cal} &                       
  \colhead{Mode} &                      
  \colhead{BW} &                        
  \colhead{Chan} &                      
 \colhead{$\Delta$ v} &
  \colhead{$\nu_{\mathrm{obs}1}$} &     
  \colhead{v$_{\mathrm{obs}1}$} &       
  \colhead{N$_{EVLA}$} \\

  \colhead{} &
  \colhead{} &
  \colhead{} &
  \colhead{[yyyy-mm-dd]} &
  \colhead{[$^{h}$:$^{m}$:$^{s}$]} &
  \colhead{[$\degr$:$\arcmin$:$\arcsec$]} &
  \colhead{} &
  \colhead{} &
  \colhead{} &
  \colhead{[MHz]} &
  \colhead{\#} &
  \colhead{[km s$^{-1}$]} &
  \colhead{[MHz]} &
  \colhead{[km s$^{-1}$]} &
  \colhead{} 

}

\startdata

NGC\,247        & BnA & AO215 & 2007-10-10  & 00:47:08.5 & -20:45:37 & 2000 & 0110-076 & 4   & 1.56 & 128 &  2.6 & 1419.098\tablenotemark{a} & \nodata & 12 \\   
                &     &       &             &            &           &      &          &     &      &     &      & 1420.222\tablenotemark{a} & \nodata &    \\   
NGC\,247        & BnA & AO215 & 2007-10-11  & 00:47:08.5 & -20:45:37 & 2000 & 0110-076 & 4   & 1.56 & 128 &  2.6 & 1419.098\tablenotemark{a} & \nodata & 12 \\   
                &     &       &             &            &           &      &          &     &      &     &      & 1420.222\tablenotemark{a} & \nodata &    \\   
NGC\,247        & B   & AO215 & 2008-01-11  & 00:47:08.5 & -20:45:37 & 2000 & 0110-076 & 4   & 1.56 & 128 &  2.6 & 1419.008\tablenotemark{a} & \nodata & 13 \\   
                &     &       &             &            &           &      &          &     &      &     &      & 1420.132\tablenotemark{a} & \nodata &    \\   
NGC\,247        & B   & AO215 & 2008-01-12  & 00:47:08.5 & -20:45:37 & 2000 & 0110-076 & 4   & 1.56 & 128 &  2.6 & 1419.008\tablenotemark{a} & \nodata & 13 \\   
                &     &       &             &            &           &      &          &     &      &     &      & 1420.132\tablenotemark{a} & \nodata &    \\   
NGC\,247        & CnB & AO215 & 2008-02-17  & 00:47:08.5 & -20:45:37 & 2000 & 0110-076 & 4   & 1.56 & 128 &  2.6 & 1419.059\tablenotemark{a} & \nodata & 13 \\   
                &     &       &             &            &           &      &          &     &      &     &      & 1420.182\tablenotemark{a} & \nodata &    \\   
NGC\,247        & DnC & AO215 & 2008-06-12  & 00:47:08.5 & -20:45:37 & 2000 & 0116-208 & 4   & 1.56 & 128 &  2.6 & 1419.233\tablenotemark{a} & \nodata & 15 \\   
\\
DDO\,6          & BnA & AO215 & 2007-10-05  & 00:49:49.2 & -21:00:54 & 2000 & 0145-275 & 2AD & 0.78 & 256 & 0.6  & 1419.010                  & \nodata & 12 \\ 
DDO\,6          & BnA & AO215 & 2007-10-07  & 00:49:49.2 & -21:00:54 & 2000 & 0145-275 & 2AD & 0.78 & 256 & 0.6  & 1419.010                  & \nodata & 12 \\ 
DDO\,6          & CnB & AO215 & 2008-02-16  & 00:49:49.2 & -21:00:54 & 2000 & 0145-275 & 2AC & 0.78 & 256 & 0.6  & 1418.953                  & \nodata & 13 \\ 
DDO\,6          & DnC & AO215 & 2008-06-12  & 00:49:49.2 & -21:00:54 & 2000 & 0116-208 & 2AC & 0.78 & 256 & 0.6  & 1419.158                  & \nodata & 15 \\ 
DDO\,6          & DnC & AO215 & 2008-07-11  & 00:49:49.2 & -21:00:54 & 2000 & 0116-208 & 2AC & 0.78 & 256 & 0.6  & 1419.140                  & \nodata & 16 \\ 
\\
NGC\,404        & B   & AO215 & 2007-11-13  & 01:09:27.0 & +35:43:04 & 2000 & 0119+321 & 2AD & 0.78 & 256 & 0.6  & 1420.598\tablenotemark{a} & \nodata & 12 \\ 
NGC\,404        & C   & AC459 & 1996-01-01  & 01:06:39.0 & +35:28:00 & 1950 & 0116+319 & 2AD & 1.56 & 128 &  2.6 & \nodata                   & -56.0   &  0 \\ 
NGC\,404        & D   & AO215 & 2008-08-21  & 01:09:27.0 & +35:43:04 & 2000 & 0119+321 & 2AC & 0.78 & 256 & 0.6  & 1420.753\tablenotemark{a} & \nodata & 17 \\ 
NGC\,404        & D   & AC459 & 1996-07-16  & 01:06:39.0 & +35:28:00 & 1950 & 0116+319 & 2AD & 1.56 & 128 &  2.6 & \nodata                   & -56.0   &  0 \\ 
\\                   
KKH\,37         & B   & AO215 & 2007-12-15  & 06:47:45.8 & +80:07:26 & 2000 & 0410+769 & 2AD & 0.78 & 256 & 0.6  & 1421.121\tablenotemark{a} & \nodata & 12 \\ 
KKH\,37         & C   & AO215 & 2008-04-11  & 06:47:45.8 & +80:07:26 & 2000 & 0410+769 & 2AC & 0.78 & 256 & 0.6  & 1421.033\tablenotemark{a} & \nodata & 15 \\ 
KKH\,37         & D   & AO215 & 2008-07-12  & 06:47:45.8 & +80:07:26 & 2000 & 0410+769 & 2AC & 0.78 & 256 & 0.6  & 1421.150\tablenotemark{a} & \nodata & 16 \\ 
KKH\,37         & D   & AO215 & 2008-08-11  & 06:47:45.8 & +80:07:26 & 2000 & 0410+769 & 2AC & 0.78 & 256 & 0.6  & 1421.161\tablenotemark{a} & \nodata & 16 \\ 
\\                   
UGC\,4483       & B   & AO215 & 2007-12-22  & 08:37:03.0 & +69:46:31 & 2000 & 0834+555 & 2AD & 1.56 & 256 &  1.3 & 1419.698\tablenotemark{a} & \nodata & 12 \\
UGC\,4483       & B   & AZ090 & 1997-04-01  & 08:32:06.0 & +69:58:00 & 1950 & 0831+557 & 2AD & 1.56 & 128 &  2.6 & \nodata                   & 180.0   &  0 \\
UGC\,4483       & B   & AZ090 & 1997-04-11  & 08:32:06.0 & +69:58:00 & 1950 & 0831+557 & 2AD & 1.56 & 128 &  2.6 & \nodata                   & 180.0   &  0 \\
UGC\,4483       & B   & AZ090 & 1997-04-12  & 08:32:06.0 & +69:58:00 & 1950 & 0831+557 & 2AD & 1.56 & 128 &  2.6 & \nodata                   & 180.0   &  0 \\
UGC\,4483       & C   & AZ090 & 1997-06-28  & 08:32:06.0 & +69:58:00 & 1950 & 0831+557 & 2AD & 1.56 & 128 &  2.6 & \nodata                   & 180.0   &  0 \\
UGC\,4483       & C   & AZ090 & 1997-08-14  & 08:32:06.0 & +69:58:00 & 1950 & 0831+557 & 2AD & 1.56 & 128 &  2.6 & \nodata                   & 180.0   &  0 \\
UGC\,4483       & D   & AO215 & 2008-07-10  & 08:37:03.0 & +69:46:31 & 2000 & 0834+555 & 2AC & 1.56 & 256 &  1.3 & 1419.700\tablenotemark{a} & \nodata & 16 \\
UGC\,4483       & D   & AO215 & 2008-08-16  & 08:37:03.0 & +69:46:31 & 2000 & 0834+555 & 2AC & 1.56 & 256 &  1.3 & 1419.200\tablenotemark{a} & \nodata & 17 \\
\\                   
KK\,77          & B   & AO215 & 2007-12-07  & 09:50:10.5 & +67:30:24 & 2000 & 1035+564 & 4   & 1.56 & 128 &  2.2 & 1419.174\tablenotemark{a} & \nodata & 12 \\ 
                &     &       &             &            &           &      &          &     &      &     &      & 1420.298\tablenotemark{a} & \nodata &    \\ 
KK\,77          & C   & AO215 & 2008-03-31  & 09:50:10.5 & +67:30:24 & 2000 & 1035+564 & 4   & 1.56 & 128 &  2.6 & 1419.025\tablenotemark{a} & \nodata & 14 \\ 
                &     &       &             &            &           &      &          &     &      &     &      & 1420.148\tablenotemark{a} & \nodata &    \\ 
KK\,77          & D   & AO215 & 2008-08-10  & 09:50:10.5 & +67:30:24 & 2000 & 1035+564 & 4   & 1.56 & 128 &  2.6 & 1419.139\tablenotemark{a} & \nodata & 16 \\ 
                &     &       &             &            &           &      &          &     &      &     &      & 1420.262\tablenotemark{a} & \nodata &    \\ 
KK\,77          & D   & AO215 & 2008-08-19  & 09:50:10.5 & +67:30:24 & 2000 & 1035+564 & 4   & 1.56 & 128 &  2.6 & 1419.118\tablenotemark{a} & \nodata & 17 \\ 
                &     &       &             &            &           &      &          &     &      &     &      & 1420.242\tablenotemark{a} & \nodata &    \\ 
\\                   
BK3N            & B   & AO215 & 2007-12-18  & 09:53:48.5 & +68:58:08 & 2000 & 1035+564 & 2AD & 0.78 & 256 & 0.6  & 1420.644\tablenotemark{a} & \nodata & 12 \\ 
BK3N            & C   & AO215 & 2008-03-16  & 09:53:48.5 & +68:58:08 & 2000 & 1035+564 & 2AC & 0.78 & 256 & 0.6  & 1420.522\tablenotemark{a} & \nodata & 14 \\ 
BK3N            & D   & AO215 & 2008-07-18  & 09:53:48.5 & +68:58:08 & 2000 & 1035+564 & 2AC & 0.78 & 256 & 0.6  & 1420.590\tablenotemark{a} & \nodata & 16 \\ 
BK3N            & D   & AO215 & 2008-08-15  & 09:53:48.5 & +68:58:08 & 2000 & 1035+564 & 2AC & 0.78 & 256 & 0.6  & 1420.631\tablenotemark{a} & \nodata & 17 \\ 
\\                   
AO\,0952+69     & B   & AO215 & 2007-12-09  & 09:57:31.0 & +69:16:60 & 2000 & 1035+564 & 2AD & 1.56 & 256 &  1.3 & 1419.996\tablenotemark{a} & \nodata & 12 \\ 
AO\,0952+69     & C   & AO215 & 2008-03-31  & 09:57:31.0 & +69:16:60 & 2000 & 1035+564 & 2AC & 1.56 & 256 &  1.3 & 1419.855\tablenotemark{a} & \nodata & 14 \\ 
AO\,0952+69     & D   & AO215 & 2008-08-11  & 09:57:31.0 & +69:16:60 & 2000 & 1035+564 & 2AC & 1.56 & 256 &  1.3 & 1419.967\tablenotemark{a} & \nodata & 16 \\ 
\\                   
Sextans\,B      & B   & AO215 & 2007-11-16  & 10:00:00.1 & +05:19:56 & 2000 & 1024-008 & 2AD & 1.56 & 256 &  1.3 & 1419.103                  & \nodata & 12 \\
Sextans\,B      & C   & AM561 & 1997-08-02  & 09:59:59.9 & +05:19:43 & 2000 & 1008+075 & 2AD & 0.78 & 128 &  1.3 & \nodata                   & 301.0   &  0 \\
Sextans\,B      & D   & AO215 & 2008-08-03  & 10:00:00.1 & +05:19:56 & 2000 & 0943-083 & 2AC & 1.56 & 256 &  1.3 & 1418.890                  & \nodata & 16 \\
\\
NGC\,3109       & BnA & AO215 & 2007-10-07  & 10:03:06.9 & -26:09:34 & 2000 & 0921-263 & 2AD & 1.56 & 256 &  1.3 & 1418.570                  & \nodata & 12 \\
NGC\,3109       & BnA & AO215 & 2007-10-08  & 10:03:06.9 & -26:09:34 & 2000 & 0921-263 & 2AD & 1.56 & 256 &  1.3 & 1418.570                  & \nodata & 12 \\
NGC\,3109       & CnB & AO215 & 2008-02-26  & 10:03:06.9 & -26:09:34 & 2000 & 0921-263 & 2AC & 1.56 & 256 &  1.3 & 1418.526                  & \nodata & 14 \\
NGC\,3109       & DnC & AO215 & 2008-06-15  & 10:03:06.9 & -26:09:34 & 2000 & 0921-263 & 2AC & 1.56 & 256 &  1.3 & 1418.402                  & \nodata & 15 \\
NGC\,3109       & DnC & AO215 & 2008-07-12  & 10:03:06.9 & -26:09:34 & 2000 & 0921-263 & 2AC & 1.56 & 256 &  1.3 & 1418.470                  & \nodata & 16 \\
\\
Antlia          & BnA & AO215 & 2007-10-06  & 10:04:04.1 & -27:19:52 & 2000 & 0921-263 & 2AD & 0.78 & 256 & 0.6  & 1418.740                  & \nodata & 12 \\
Antlia          & BnA & AO215 & 2007-10-13  & 10:04:04.1 & -27:19:52 & 2000 & 0921-263 & 2AD & 0.78 & 256 & 0.6  & 1418.755                  & \nodata & 12 \\
Antlia          & CnB & AA232 & 1998-11-02  & 10:01:47.5 & -27:05:15 & 1950 & 1015-314 & 2AD & 0.78 & 128 &  1.3 & \nodata                   & 360.0   &  0 \\
Antlia          & CnB & AA232 & 1998-11-13  & 10:01:47.5 & -27:05:15 & 1950 & 1015-314 & 2AD & 0.78 & 128 &  1.3 & \nodata                   & 360.0   &  0 \\
Antlia          & DnC & AO215 & 2008-06-15  & 10:04:04.1 & -27:19:52 & 2000 & 0921-263 & 2AC & 0.78 & 256 & 0.6  & 1418.578                  & \nodata & 15 \\
Antlia          & DnC & AO215 & 2008-07-26  & 10:04:04.1 & -27:19:52 & 2000 & 0921-263 & 2AC & 0.78 & 256 & 0.6  & 1418.614                  & \nodata & 16 \\
\\
KDG\,63         & B  &  AO215 & 2007-11-29  & 10:05:06.4 & +66:33:32 & 2000 & 1035+564 & 2AD & 0.78 & 256 & 0.6  & 1421.093\tablenotemark{a} & \nodata & 12 \\ 
KDG\,63         & C  &  AO215 & 2008-04-05  & 10:05:06.4 & +66:33:32 & 2000 & 1035+564 & 2AC & 0.78 & 256 & 0.6  & 1420.930\tablenotemark{a} & \nodata & 14 \\ 
KDG\,63         & D  &  AO215 & 2008-08-08  & 10:05:06.4 & +66:33:32 & 2000 & 1035+564 & 2AC & 0.78 & 256 & 0.6  & 1421.010\tablenotemark{a} & \nodata & 16 \\ 
KDG\,63         & D  &  AO215 & 2008-08-17  & 10:05:06.4 & +66:33:32 & 2000 & 1035+564 & 2AC & 0.78 & 256 & 0.6  & 1421.052\tablenotemark{a} & \nodata & 17 \\ 
\\                   
Sextans\,A      & B  &  AO215 & 2007-11-21  & 10:11:00.8 & -04:41:34 & 2000 & 1024-008 & 2AD & 1.56 & 256 &  1.3 & 1418.981                  & \nodata & 12 \\
Sextans\,A      & C  &  AO215 & 2008-03-16  & 10:11:00.8 & -04:41:34 & 2000 & 1024-008 & 2AC & 1.56 & 256 &  1.3 & 1418.799                  & \nodata & 14 \\
Sextans\,A      & D  &  AO215 & 2008-04-12  & 10:11:00.8 & -04:41:34 & 2000 & 1024-008 & 2AC & 1.56 & 256 &  1.3 & 1418.770                  & \nodata & 16 \\
Sextans\,A      & D  &  AO215 & 2008-08-17  & 10:11:00.8 & -04:41:34 & 2000 & 0943-083 & 2AC & 1.56 & 256 &  1.3 & 1418.819                  & \nodata & 17 \\
\\                   
HS\,117         & B  &  AO215 & 2007-11-28  & 10:21:25.2 & +71:06:51 & 2000 & 1035+564 & 2AD & 0.78 & 256 & 0.6  & 1420.649\tablenotemark{a} & \nodata & 12 \\
HS\,117         & C  &  AO215 & 2008-04-11  & 10:21:25.2 & +71:06:51 & 2000 & 1035+564 & 2AC & 0.78 & 256 & 0.6  & 1420.500\tablenotemark{a} & \nodata & 15 \\
HS\,117         & D  &  AO215 & 2008-07-14  & 10:21:25.2 & +71:06:51 & 2000 & 1035+564 & 2AC & 0.78 & 256 & 0.6  & 1420.575\tablenotemark{a} & \nodata & 16 \\
HS\,117         & D  &  AO215 & 2008-08-11  & 10:21:25.2 & +71:06:51 & 2000 & 1035+564 & 2AC & 0.78 & 256 & 0.6  & 1420.608\tablenotemark{a} & \nodata & 16 \\
NG\\                   
DDO\,82         & B  &  AO215 & 2007-12-13  & 10:30:35.0 & +70:37:07 & 2000 & 1035+564 & 2AD & 1.56 & 256 &  1.3 & 1420.194\tablenotemark{a} & \nodata & 12 \\ 
DDO\,82         & C  &  AO215 & 2008-04-08  & 10:30:35.0 & +70:37:07 & 2000 & 1035+564 & 2AC & 1.56 & 256 &  1.3 & 1420.061\tablenotemark{a} & \nodata & 14 \\ 
DDO\,82         & D  &  AO215 & 2008-08-11  & 10:30:35.0 & +70:37:07 & 2000 & 1035+564 & 2AC & 1.56 & 256 &  1.3 & 1420.165\tablenotemark{a} & \nodata & 16 \\ 
DDO\,82         & D  &  AO215 & 2008-08-16  & 10:30:35.0 & +70:37:07 & 2000 & 1035+564 & 2AC & 1.56 & 256 &  1.3 & 1420.150\tablenotemark{a} & \nodata & 16 \\ 
\\                   
KDG\,73         & B  &  AO215 & 2007-11-27  & 10:52:57.1 & +69:32:58 & 2000 & 1313+675 & 2AD & 0.78 & 256 & 0.6  & 1419.928\tablenotemark{a} & \nodata & 12 \\
KDG\,73         & C  &  AO215 & 2008-04-11  & 10:52:57.1 & +69:32:58 & 2000 & 1313+675 & 2AC & 0.78 & 256 & 0.6  & 1419.778\tablenotemark{a} & \nodata & 15 \\
KDG\,73         & D  &  AO215 & 2008-07-14  & 10:52:57.1 & +69:32:58 & 2000 & 1313+675 & 2AC & 0.78 & 256 & 0.6  & 1419.845\tablenotemark{a} & \nodata & 16 \\
KDG\,73         & D  &  AO215 & 2008-08-16  & 10:52:57.1 & +69:32:58 & 2000 & 1313+675 & 2AC & 0.78 & 256 & 0.6  & 1419.883\tablenotemark{a} & \nodata & 17 \\
\\                   
NGC\,3741       & B  &  AO215 & 2007-11-06  & 11:36:06.2 & +45:17:01 & 2000 & 1146+399 & 2AD & 1.56 & 256 &  1.3 & 1419.434\tablenotemark{a} & \nodata & 12 \\
NGC\,3741       & C  &  AO215 & 2008-05-05  & 11:36:06.2 & +45:17:01 & 2000 & 1146+399 & 2AC & 1.56 & 256 &  1.3 & 1419.218\tablenotemark{a} & \nodata & 15 \\
NGC\,3741       & D  &  AO215 & 2008-08-04  & 11:36:06.2 & +45:17:01 & 2000 & 1146+399 & 2AC & 1.56 & 256 &  1.3 & 1419.295\tablenotemark{a} & \nodata & 17 \\
\\                   
DDO\,99         & B  &  AO215 & 2007-12-04  & 11:50:53.0 & +38:52:49 & 2000 & 1146+399 & 2AD & 1.56 & 256 &  1.3 & 1419.363                  & \nodata & 12 \\
DDO\,99         & C  &  AO215 & 2008-04-05  & 11:50:53.0 & +38:52:49 & 2000 & 1146+399 & 2AC & 1.56 & 256 &  1.3 & 1419.180                  & \nodata & 14 \\
DDO\,99         & D  &  AO215 & 2008-08-07  & 11:50:53.0 & +38:52:49 & 2000 & 1146+399 & 2AC & 1.56 & 256 &  1.3 & 1419.180                  & \nodata & 16 \\
DDO\,99         & D  &  AO215 & 2008-08-10  & 11:50:53.0 & +38:52:49 & 2000 & 1146+399 & 2AC & 1.56 & 256 &  1.3 & 1419.204                  & \nodata & 16 \\
\\                   
NGC\,4163       & B  &  AO215 & 2007-11-23  & 12:12:09.1 & +36:10:09 & 2000 & 1227+365 & 2AD & 0.78 & 256 & 0.6  & 1419.740                  & \nodata & 12 \\
NGC\,4163       & B  &  AH927\tablenotemark{b} & 2008-02-12  & 12:12:09.1 & +36:10:09 & 2000 & 1227+365 & 2AC & 0.78 & 256 & 0.6  & 1419.719\tablenotemark{a} & \nodata & 13 \\
NGC\,4163       & C  &  AO215 & 2008-04-08  & 12:12:09.1 & +36:10:09 & 2000 & 1227+365 & 2AC & 0.78 & 256 & 0.6  & 1419.563                  & \nodata & 14 \\
NGC\,4163       & C  &  AH927\tablenotemark{b} & 2008-06-01  & 12:12:09.1 & +36:10:09 & 2000 & 1227+365 & 2AC & 0.78 & 256 & 0.6  & 1419.546\tablenotemark{a} & \nodata & 15 \\
NGC\,4163       & D  &  AO215 & 2008-08-06  & 12:12:09.1 & +36:10:09 & 2000 & 1227+365 & 2AC & 0.78 & 256 & 0.6  & 1419.540                  & \nodata & 16 \\
\\                   
NGC\,4190       & B  &  AO215 & 2007-24-24  & 12:13:44.8 & +36:38:03 & 2000 & 1227+365 & 2AD & 1.56 & 256 &  1.3 & 1419.441\tablenotemark{a} & \nodata & 13 \\
NGC\,4190       & C  &  AO215 & 2008-03-09  & 12:13:44.8 & +36:38:03 & 2000 & 1227+365 & 2AC & 1.56 & 256 &  1.3 & 1419.323\tablenotemark{a} & \nodata & 14 \\
NGC\,4190       & D  &  AO215 & 2008-08-11  & 12:13:44.8 & +36:38:03 & 2000 & 1227+365 & 2AC & 1.56 & 256 &  1.3 & 1419.275\tablenotemark{a} & \nodata & 16 \\
\\                   
DDO\,113        & B  &  AO215 & 2007-12-01  & 12:14:57.9 & +36:13:08 & 2000 & 1227+365 & 2AD & 1.56 & 256 &  1.3 & 1419.179                  & \nodata & 12 \\ 
DDO\,113        & C  &  AO215 & 2008-04-04  & 12:14:57.9 & +36:13:08 & 2000 & 1227+365 & 2AC & 1.56 & 256 &  1.3 & 1419.007                  & \nodata & 14 \\ 
DDO\,113        & D  &  AO215 & 2008-08-15  & 12:14:57.9 & +36:13:08 & 2000 & 1227+365 & 2AC & 1.56 & 256 &  1.3 & 1419.040                  & \nodata & 16 \\ 
DDO\,113        & D  &  AO215 & 2008-08-17  & 12:14:57.9 & +36:13:08 & 2000 & 1227+365 & 2AC & 1.56 & 256 &  1.3 & 1419.020                  & \nodata & 17 \\ 
\\                   
MCG\,+09-20-131 & B  & AO215 & 2007-11-30  & 12:15:46.8 & +52:23:17 & 2000 & 1219+484 & 2AD & 1.56 & 256 &  1.3 & 1419.749\tablenotemark{a} & \nodata & 12 \\ 
MCG\,+09-20-131 & C  & AO215 & 2008-05-05  & 12:15:46.8 & +52:23:17 & 2000 & 1219+484 & 2AC & 1.56 & 256 &  1.3 & 1419.566\tablenotemark{a} & \nodata & 15 \\ 
MCG\,+09-20-131 & D  & AO215 & 2008-08-10  & 12:15:46.8 & +52:23:17 & 2000 & 1219+484 & 2AC & 1.56 & 256 &  1.3 & 1419.626\tablenotemark{a} & \nodata & 16 \\ 
\\                   
DDO\,125        & B  & AO215 & 2007-11-25  & 12:27:40.9 & +43:29:44 & 2000 & 1227+365 & 2AD & 0.78 & 256 & 0.6  & 1419.588                  & \nodata & 12 \\
DDO\,125        & C  & AO215 & 2008-03-09  & 12:27:40.9 & +43:29:44 & 2000 & 1227+365 & 2AC & 0.78 & 256 & 0.6  & 1419.478                  & \nodata & 14 \\
DDO\,125        & D  & AO215 & 2008-08-06  & 12:27:40.9 & +43:29:44 & 2000 & 1227+365 & 2AC & 0.78 & 256 & 0.6  & 1419.410                  & \nodata & 16 \\
DDO\,125        & D  & AO215 & 2008-08-15  & 12:27:40.9 & +43:29:44 & 2000 & 1227+365 & 2AC & 0.78 & 256 & 0.6  & 1419.446                  & \nodata & 17 \\
\\                   
UGCA\,292       & B  & AO215 & 2007-12-03  & 12:38:40.0 & +32:46:01 & 2000 & 1227+365 & 2AD & 0.78 & 256 & 0.6  & 1419.062                  & \nodata & 12 \\
UGCA\,292       & C  & AH927\tablenotemark{b} & 2008-02-06  & 12:38:40.0 & +32:46:01 & 2000 & 1227+365 & 2AC & 0.78 & 256 & 0.6  & 1419.053                  & \nodata & 13 \\
UGCA\,292       & D  & AO215 & 2008-07-11  & 12:38:40.0 & +32:46:01 & 2000 & 1227+365 & 2AC & 0.78 & 256 & 0.6  & 1418.865                  & \nodata & 16 \\
UGCA\,292       & D  & AO215 & 2008-08-16  & 12:38:40.0 & +32:46:01 & 2000 & 1227+365 & 2AC & 0.78 & 256 & 0.6  & 1418.892                  & \nodata & 17 \\
UGCA\,292       & D  & AH927\tablenotemark{b} & 2008-07-21  & 12:38:40.0 & +32:46:01 & 2000 & 1227+365 & 2AC & 0.78 & 256 & 0.6  & 1418.874                  & \nodata & 16 \\
\\                   
GR\,8           & B  & AO215 & 2007-11-12  & 12:58:40.4 & +14:13:03 & 2000 & 1254+116 & 2AD & 0.78 & 256 & 0.6  & 1419.485                  & \nodata & 11 \\
GR\,8           & C  & AH927\tablenotemark{b} & 2008-02-10  & 12:58:40.4 & +14:13:03 & 2000 & 1347+122 & 2AC & 0.78 & 256 & 0.6  & 1419.524                  & \nodata & 13 \\
GR\,8           & D  & AO215 & 2008-08-02  & 12:58:40.4 & +14:13:03 & 2000 & 1254+116 & 2AC & 0.78 & 256 & 0.6  & 1419.285                  & \nodata & 16 \\
GR\,8           & D  & AO215 & 2008-08-17  & 12:58:40.4 & +14:13:03 & 2000 & 1254+116 & 2AC & 0.78 & 256 & 0.6  & 1419.304                  & \nodata & 17 \\
GR\,8           & D  & AH927\tablenotemark{b} & 2008-08-02  & 12:58:40.4 & +14:13:03 & 2000 & 1347+122 & 2AC & 0.78 & 256 & 0.6  & 1419.287                  & \nodata & 16 \\
\\                   
UGC\,8508       & B  & AO215 & 2007-12-10  & 13:30:44.4 & +54:54:36 & 2000 & 1400+621 & 2AD & 0.78 & 256 & 0.6  & 1420.190\tablenotemark{a} & \nodata & 12 \\
UGC\,8508       & B  & AH927\tablenotemark{b} & 2008-02-09  & 13:30:44.4 & +54:54:36 & 2000 & 1400+621 & 2AC & 0.78 & 256 & 0.6  & 1420.175\tablenotemark{a} & \nodata & 13 \\
UGC\,8508       & C  & AO215 & 2008-03-15  & 13:30:44.4 & +54:54:36 & 2000 & 1400+621 & 2AC & 0.78 & 256 & 0.6  & 1420.102\tablenotemark{a} & \nodata & 14 \\
UGC\,8508       & C  & AH927\tablenotemark{b} & 2008-05-31  & 13:30:44.4 & +54:54:36 & 2000 & 1400+621 & 2AC & 0.78 & 256 & 0.6  & 1420.062\tablenotemark{a} & \nodata & 15 \\
UGC\,8508       & D  & AO215 & 2008-07-31  & 13:30:44.4 & +54:54:36 & 2000 & 1400+621 & 2AC & 0.78 & 256 & 0.6  & 1420.060\tablenotemark{a} & \nodata & 16 \\
UGC\,8508       & D  & AH927\tablenotemark{b} & 2008-08-03  & 13:30:44.4 & +54:54:36 & 2000 & 1400+621 & 2AC & 0.78 & 256 & 0.6  & 1420.071\tablenotemark{a} & \nodata & 16 \\
UGC\,8508       & D  & AO215 & 2008-08-17  & 13:30:44.4 & +54:54:36 & 2000 & 1400+621 & 2AC & 0.78 & 256 & 0.6  & 1420.084\tablenotemark{a} & \nodata & 17 \\
\\                   
DDO\,181        & B  & AO215 & 2007-12-06  & 13:39:53.8 & +40:44:21 & 2000 & 1331+305 & 2AD & 1.56 & 256 &  1.3 & 1419.543\tablenotemark{a} & \nodata & 12 \\
DDO\,181        & C  & AO215 & 2008-03-09  & 13:39:53.8 & +40:44:21 & 2000 & 1331+305 & 2AC & 1.56 & 256 &  1.3 & 1419.474\tablenotemark{a} & \nodata & 14 \\
DDO\,181        & D  & AO215 & 2008-08-16  & 13:39:53.8 & +40:44:21 & 2000 & 1331+305 & 2AC & 1.56 & 256 &  1.3 & 1419.388\tablenotemark{a} & \nodata & 17 \\
DDO\,181        & D  & AO215 & 2008-08-18  & 13:39:53.8 & +40:44:21 & 2000 & 1331+305 & 2AC & 1.56 & 256 &  1.3 & 1419.390\tablenotemark{a} & \nodata & 17 \\
\\                   
DDO\,183        & B  & AO215 & 2007-12-08  & 13:50:50.6 & +38:01:09 & 2000 & 1331+305 & 2AD & 1.56 & 256 &  1.3 & 1419.591\tablenotemark{a} & \nodata & 12 \\
DDO\,183        & C  & AO215 & 2008-03-15  & 13:50:50.6 & +38:01:09 & 2000 & 1331+305 & 2AC & 1.56 & 256 &  1.3 & 1419.518\tablenotemark{a} & \nodata & 14 \\
DDO\,183        & D  & AO215 & 2008-08-11  & 13:50:50.6 & +38:01:09 & 2000 & 1331+305 & 2AC & 1.56 & 256 &  1.3 & 1419.433\tablenotemark{a} & \nodata & 16 \\
DDO\,183        & D  & AO215 & 2008-08-11  & 13:50:50.6 & +38:01:09 & 2000 & 1331+305 & 2AC & 1.56 & 256 &  1.3 & 1419.424\tablenotemark{a} & \nodata & 16 \\
\\                   
KKH\,86         & B  & AO215 & 2007-11-11  & 13:54:33.5 & +04:14:35 & 2000 & 1347+122 & 2AD & 0.78 & 256 & 0.6  & 1419.105                  & \nodata & 11 \\
KKH\,86         & C  & AO215 & 2008-03-28  & 13:54:33.5 & +04:14:35 & 2000 & 1347+122 & 2AC & 0.78 & 256 & 0.6  & 1419.074                  & \nodata & 14 \\
KKH\,86         & D  & AO215 & 2008-08-08  & 13:54:33.5 & +04:14:35 & 2000 & 1347+122 & 2AC & 0.78 & 256 & 0.6  & 1418.944                  & \nodata & 16 \\
\\                   
UGC\,8833       & B  & AO215 & 2007-11-18  & 13:54:48.7 & +35:50:15 & 2000 & 1331+305 & 2AD & 0.78 & 256 & 0.6  & 1419.406                  & \nodata & 12 \\
UGC\,8833       & C  & AZ121 & 2000-04-15  & 13:52:38.2 & +36:04:60 & 1950 & 1413+349 & 2AD & 1.56 & 128 &  2.6 & \nodata                   & 225.0   &  0 \\
UGC\,8833       & D  & AO215 & 2008-08-05  & 13:54:48.7 & +35:50:15 & 2000 & 1331+305 & 2AC & 0.78 & 256 & 0.6  & 1419.265                  & \nodata & 16 \\
UGC\,8833       & D  & AO215 & 2008-08-15  & 13:54:48.7 & +35:50:15 & 2000 & 1331+305 & 2AC & 0.78 & 256 & 0.6  & 1419.259                  & \nodata & 17 \\
\\                   
KK\,230         & B\tablenotemark{c} & AO215 & 2007-11-10 & 14:07:10.5 & +35:03:37 & 2000 & 1331+305 & 2AD & 0.78 & 256 & 0.6  & 1420.171\tablenotemark{a} & \nodata & 12 \\
KK\,230         & C  & AO215 & 2008-04-03  & 14:07:10.5 & +35:03:37 & 2000 & 1331+305 & 2AC & 0.78 & 256 & 0.6  & 1420.114\tablenotemark{a} & \nodata & 13 \\
KK\,230         & D  & AO215 & 2008-08-15  & 14:07:10.5 & +35:03:37 & 2000 & 1331+305 & 2AC & 0.78 & 256 & 0.65 & 1420.036\tablenotemark{a} & \nodata & 17 \\
\\                   
DDO\,187        & B  & AO215 & 2007-11-17  & 14:15:56.5 & +23:03:19 & 2000 & 1330+251 & 2AD & 1.56 & 256 &  1.3 & 1419.751\tablenotemark{a} & \nodata & 13 \\
DDO\,187        & B  & AH927\tablenotemark{b} & 2008-02-10  & 14:15:56.5 & +23:03:19 & 2000 & 1330+251 & 2AC & 1.56 & 256 &  1.3 & 1419.805\tablenotemark{a} & \nodata & 13 \\
DDO\,187        & B  & AH927\tablenotemark{b} & 2008-02-12  & 14:15:56.5 & +23:03:19 & 2000 & 1330+251 & 2AC & 1.56 & 256 &  1.3 & 1419.805\tablenotemark{a} & \nodata & 13 \\
DDO\,187        & C  & AO215 & 2008-03-28  & 14:15:56.5 & +23:03:19 & 2000 & 1330+251 & 2AC & 1.56 & 256 &  1.3 & 1419.704\tablenotemark{a} & \nodata & 14 \\
DDO\,187        & C  & AH927\tablenotemark{b} & 2008-05-30  & 14:15:56.5 & +23:03:19 & 2000 & 1330+251 & 2AC & 1.56 & 256 &  1.3 & 1419.674\tablenotemark{a} & \nodata & 14 \\
DDO\,187        & D  & AH927\tablenotemark{b} & 2008-08-05  & 14:15:56.5 & +23:03:19 & 2000 & 1330+251 & 2AC & 1.56 & 256 &  1.3 & 1419.580\tablenotemark{a} & \nodata & 16 \\ 
DDO\,187        & D  & AO215 & 2008-08-06  & 14:15:56.5 & +23:03:19 & 2000 & 1330+251 & 2AC & 1.56 & 256 &  1.3 & 1419.598\tablenotemark{a} & \nodata & 16 \\
DDO\,187        & D  & AO215 & 2008-08-16  & 14:15:56.5 & +23:03:19 & 2000 & 1330+251 & 2AC & 1.56 & 256 &  1.3 & 1419.588\tablenotemark{a} & \nodata & 17 \\
\\                   
DDO\,190        & B  & AO215 & 2007-12-14  & 14:24:43.4 & +44:31:33 & 2000 & 1506+375 & 2AD & 0.78 & 256 & 0.6  & 1419.776                  & \nodata & 12 \\
DDO\,190        & C  & AZ121 & 2000-04-20  & 14:22:48.8 & +44:44:60 & 1950 & 1413+349 & 2AD & 1.56 & 128 &  2.6 & \nodata                   & 160.0   &  0 \\
DDO\,190        & D  & AO215 & 2008-08-16  & 14:24:43.4 & +44:31:33 & 2000 & 1506+375 & 2AC & 0.78 & 256 & 0.6  & 1419.637                  & \nodata & 17 \\
\\                   
KKR\,25         & B  & AO215 & 2008-01-31  & 16:13:47.9 & +54:22:16 & 2000 & 1634+627 & 2AD & 0.78 & 256 & 0.6  & 1421.112\tablenotemark{a} & \nodata & 13 \\
KKR\,25         & C  & AO215 & 2008-04-03  & 16:13:47.9 & +54:22:16 & 2000 & 1634+627 & 2AC & 0.78 & 256 & 0.6  & 1421.083\tablenotemark{a} & \nodata & 13 \\
KKR\,25         & D  & AO215 & 2008-08-07  & 16:13:47.9 & +54:22:16 & 2000 & 1634+627 & 2AC & 0.78 & 256 & 0.6  & 1421.035\tablenotemark{a} & \nodata & 16 \\
KKR\,25         & D  & AO215 & 2008-08-16  & 16:13:47.9 & +54:22:16 & 2000 & 1634+627 & 2AC & 0.78 & 256 & 0.6  & 1421.030\tablenotemark{a} & \nodata & 17 \\
\\                   
KKH\,98         & B  & AO215 & 2007-12-07  & 23:45:34.0 & +38:43:04 & 2000 & 0029+349 & 2AD & 0.78 & 256 & 0.6  & 1420.954\tablenotemark{a} & \nodata & 12 \\
KKH\,98         & C  & AO215 & 2008-03-15  & 23:45:34.0 & +38:43:04 & 2000 & 0029+349 & 2AC & 0.78 & 256 & 0.6  & 1421.016\tablenotemark{a} & \nodata & 14 \\
KKH\,98         & D  & AO215 & 2008-07-10  & 23:45:34.0 & +38:43:04 & 2000 & 0029+349 & 2AC & 0.78 & 256 & 0.6  & 1421.150\tablenotemark{a} & \nodata & 16 \\
KKH\,98         & D  & AO215 & 2008-08-21  & 23:45:34.0 & +38:43:04 & 2000 & 0029+349 & 2AC & 0.78 & 256 & 0.6  & 1421.135\tablenotemark{a} & \nodata & 17 

\enddata
\tablenotetext{a}{Potential MW Interference; offset flux calibrators}
\tablenotetext{b}{additional data from the LITTLE THINGS survey}
\tablenotetext{c}{Source is named KKR 25 but is actually KK 230}

\end{deluxetable}


\clearpage

\begin{deluxetable}{llrrrcccc}
\tabletypesize{\scriptsize}
\tablecaption{Properties of the VLA-ANGST Data Cubes. \label{tab:dataproperties}}

\tablehead{
\colhead{(1)}    & 
\colhead{(2)}    &
\colhead{(3)}    & 
\colhead{(4)}    &
\colhead{(5)}    &
\colhead{(6)}    &
\colhead{(7)}    &
\colhead{(8)}    &
\colhead{(9)}     \\

\colhead{Galaxy}            & 
\colhead{Weighting}         &
\colhead{B$_{major}$}       & 
\colhead{B$_{minor}$}      &
\colhead{BPA}               &
\colhead{Noise}             &
\colhead{Channel Width}     &
\colhead{N$_{pixels}$}      &
\colhead{Pixel scale}      \\

\colhead{}                  & 
\colhead{}                  &
\colhead{[$\arcsec$]}       & 
\colhead{[$\arcsec$]}       &
\colhead{[$\degr$]}         &
\colhead{[mJy\,beam$^{-1}$]} &
\colhead{[km s$^{-1}$]}     &
                  &
\colhead{[$\arcsec$]}      
}
\startdata
NGC\,247            & Natural &  9.0 &  6.2 &  10.5 & 0.9 &  2.6 & 2048$^{2}$ & 1.0 \\
                    & Robust  &  6.5 &  4.8 &  12.7 & 0.9 &      &            &  \\
DDO\,6              & Natural & 12.3 & 10.3 &  52.0 & 1.9 & 0.65 & 1024$^{2}$ & 1.5 \\
                    & Robust  &  7.2 &  6.3 &  49.0 & 2.1 &      &           &      \\
NGC\,404            & Natural & 13.7 & 12.4 & -34.7 & 0.9 &  2.6 & 1024$^{2}$ & 1.5 \\
                    & Robust  &  7.1 &  6.1 & -32.6 & 0.9 &      &            &      \\
KKH\,37\tablenotemark{a}             & Natural &  9.7 &  8.1 & -86.2 & 1.6 & 0.65 & 1024$^{2}$ & 1.5 \\
                    & Robust  &  6.5 &  5.8 & -66.9 & 1.8 &      &           &      \\
UGC\,4483           & Natural & 12.2 &  9.8 &  61.3 & 0.5 &  2.6 & 1024$^{2}$ & 1.5 \\
                    & Robust  &  7.6 &  5.7 &  57.1 & 0.6 &      &           &      \\
KK\,77\tablenotemark{a}                & Natural & 12.2 &  8.1 & -79.0 & 0.9 &  2.6 & 1024$^{2}$ & 1.5 \\
                    & Robust  &  6.1 &  5.8 & -66.6 & 0.7 &      &           &      \\
BK3N                & Natural & 12.0 &  8.1 & -85.1 & 1.8 & 0.65 & 1024$^{2}$ & 1.5 \\
                    & Robust  &  6.3 &  5.8 &  61.5 & 1.8 &      &           &      \\
AO\,0952+69         & Natural & 10.1 &  8.8 &  73.5 & 1.3 &  1.3 & 1024$^{2}$ & 1.5 \\
                    & Robust  &  6.4 &  5.9 & -71.0 & 1.2 &      &           &      \\
Sextans~B           & Natural & 15.0 & 14.1 &  10.5 & 0.8 &  1.3 & 1024$^{2}$ & 1.5 \\
                    & Robust  &  9.5 &  7.5 &  41.6 & 1.0 &      &           &      \\
NGC~3109            & Natural & 10.3 &  8.8 &  22.0 & 1.6 &  1.3 & 2048$^{2}$ & 1.0 \\
                    & Robust  &  7.6 &  5.0 &   8.8 & 1.7 &      &           &      \\
Antlia              & Natural & 14.1 & 13.9 & -81.3 & 1.0 &  1.3 & 1024$^{2}$ & 1.5 \\
                    & Robust  & 10.5 &  9.6 &  71.1 & 1.2 &      &           &      \\
KDG\,63\tablenotemark{a}               & Natural & 10.8 &  9.2 &  85.5 & 1.4 & 0.65 & 1024$^{2}$ & 1.5 \\
                    & Robust  &  6.2 &  6.0 &  77.0 & 1.6 &       &           &      \\
Sextans\,A          & Natural & 11.8 & 10.1 &  38.5 & 1.2 &  1.3 & 1024$^{2}$ & 1.5 \\
                    & Robust  &  7.3 &  6.0 &  35.1 & 1.3 &      &           &      \\
HS\,117\tablenotemark{a}               & Natural & 13.2 &  8.5 & -59.6 & 1.6 & 0.65 & 1024$^{2}$ & 1.5 \\
                    & Robust  &  8.6 &  6.1 & -77.8 & 1.7 &      &           &      \\
DDO\,82             & Natural &  9.3 &  7.7 & -81.0 & 1.3 &  1.3 & 1024$^{2}$ & 1.5 \\
                    & Robust  &  5.8 &  5.7 &  65.0 & 1.4 &      &           &      \\
KDG\,73             & Natural & 10.0 &  7.6 &  84.3 & 1.6 & 0.65 & 1024$^{2}$ & 1.5 \\
                    & Robust  &  6.9 &  5.6 &  65.2 & 1.7 &      &           &      \\
NGC\,3741           & Natural &  7.6 &  6.2 &  81.1 & 1.0 &  1.3 & 1024$^{2}$ & 1.5 \\
                    & Robust  &  5.5 &  4.8 &  75.4 & 1.1 &      &           &      \\
DDO\,99             & Natural & 12.4 &  7.6 & -86.7 & 1.0 &  1.3 & 1024$^{2}$ & 1.5 \\
                    & Robust  &  7.7 &  5.2 &  72.5 & 1.1 &      &           &      \\
NGC\,4163           & Natural & 12.3 & 10.4 & -89.6 & 1.3 & 0.65 & 1024$^{2}$ & 1.5 \\
                    & Robust  &  7.6 &  5.4 & -85.2 & 1.4 &      &           &      \\
NGC\,4190           & Natural & 10.5 &  8.9 &  83.8 & 0.9 &  1.3 & 1024$^{2}$ & 1.5 \\
                    & Robust  &  6.1 &  5.3 &  81.4 & 1.0 &      &           &      \\
DDO\,113\tablenotemark{a}              & Natural & 15.0 & 14.0 & -55.2 & 1.4 &  1.3 & 1024$^{2}$ & 1.5 \\
                    & Robust  &  9.9 &  7.7 &  82.8 & 1.5 &      &           &      \\
MCG\,+09$-$20$-$131 & Natural &  9.7 &  7.4 &  69.7 & 1.0 &  1.3 & 1024$^{2}$ & 1.5 \\
                    & Robust  &  6.1 &  5.3 &  69.1 & 1.1 &      &           &      \\
DDO\,125            & Natural & 11.5 & 10.6 & -68.2 & 1.5 & 0.65 & 1024$^{2}$ & 1.5 \\
                    & Robust  &  6.3 &  5.4 & -80.1 & 1.5 &      &           &      \\
UGCA\,292           & Natural &  9.7 &  6.9 &  69.4 & 1.5 & 0.65 & 1024$^{2}$ & 1.5 \\
                    & Robust  &  7.0 &  5.0 &  65.2 & 1.6 &      &           &      \\
GR\,8               & Natural &  7.6 &  7.3 & -55.9 & 1.5 & 0.65 & 1024$^{2}$ & 1.5 \\
                    & Robust  &  5.8 &  5.4 & -28.8 & 1.6 &      &           &      \\
UGC\,8508           & Natural & 13.9 & 12.1 &  83.6 & 1.3 & 0.65 & 1024$^{2}$ & 1.5 \\
                    & Robust  &  8.2 &  6.4 &  86.1 & 1.5 &      &           &      \\
DDO\,181            & Natural & 12.8 &  9.5 & -75.7 & 1.0 &  1.3 & 1024$^{2}$ & 1.5 \\
                    & Robust  &  7.6 &  5.5 & -80.4 & 1.1 &      &           &      \\
DDO\,183            & Natural & 12.7 & 10.9 & -76.7 & 1.1 &  1.3 & 1024$^{2}$ & 1.5 \\
                    & Robust  &  7.6 &  6.2 &  88.2 & 1.2 &      &           &      \\
KKH\,86             & Natural & 11.0 &  9.9 &  -8.2 & 1.5 & 0.65 & 1024$^{2}$ & 1.5 \\
                    & Robust  &  7.5 &  5.8 & -21.2 & 1.7 &      &           &      \\
UGC\,8833           & Natural & 16.4 & 15.4 & -87.4 & 0.6 &  2.6 & 1024$^{2}$ & 1.5 \\
                    & Robust  & 12.4 & 11.2 &  81.4 & 0.6 &      &           &      \\
KK\,230             & Natural &  8.2 &  7.3 & -56.5 & 1.4 & 0.65 & 1024$^{2}$ & 1.5 \\
                    & Robust  &  5.9 &  5.2 & -41.6 & 1.5 &      &           &      \\
DDO\,187            & Natural & 12.2 & 10.4 & -82.4 & 0.9 &  1.3 & 1024$^{2}$ & 1.5 \\
                    & Robust  &  7.1 &  5.7 &  88.5 & 1.0 &      &           &      \\
DDO\,190            & Natural & 15.6 & 14.2 &  88.1 & 0.6 &  2.6 & 1024$^{2}$ & 1.5 \\
                    & Robust  & 10.8 &  9.9 &  84.1 & 0.6 &      &           &      \\
KKR\,25\tablenotemark{a}               & Natural &  8.5 &  5.0 &  63.8 & 0.4 & 0.65 & 1024$^{2}$ & 1.5 \\
                    & Robust  &  5.5 &  4.4 &  65.0 & 0.4 &      &           &      \\
KKH\,98             & Natural &  9.9 &  7.4 &  82.2 & 1.3 & 0.65 & 1024$^{2}$ & 1.5 \\
                    & Robust  &  6.2 &  5.2 &  80.4 & 1.4 &      &           &      \\
\enddata
\tablenotetext{a}{Non-detection}

\end{deluxetable}


\clearpage

\begin{deluxetable}{lrrccccc}
\tabletypesize{\scriptsize}
\rotate
\tablecaption{Galaxy \hi\ Properties. \label{tab:hiproperties}}
\tablehead{

  \colhead{(1)} &
  \colhead{(2)} &
  \colhead{(3)} &
  \colhead{(4)} &
  \colhead{(5)} &
  \colhead{(6)} &
  \colhead{(7)} &
  \colhead{(8)}\\

  \colhead{Galaxy} &
  \colhead{$S_{\textsc{Hi}}$} &
  \colhead{$M_{\textsc{Hi}}$} &
  \colhead{$S_{\rm HI}^{\rm SD}$} &
  \colhead{$w_{\rm 20}$} &
  \colhead{$w_{\rm 50}$} &
  \colhead{$v_{\textrm{cen}}$} &
  \colhead{Peak $N_{\textsc{Hi}}$} \\

  \colhead{} & 
  \colhead{[Jy~km~s$^{-1}$]} &
  \colhead{[$10^6 M_{\odot}$]} &
  \colhead{[Jy~km~s$^{-1}$]} &
  \colhead{[\kms]} &
  \colhead{[\kms]} &
  \colhead{[\kms]} &
  \colhead{[$10^{21}$\,cm$^{-2}$]}

}
\startdata
NGC\,247          &  382.6  &  1106.2  & 608\tablenotemark{a}  &  201.3  &  193.9  &  163.7  &   5.4  \\
DDO\,6            &   1.2  &  3.2  &  3.7\tablenotemark{b}  &  20.9  &  13.7  &  292.5  &   0.9  \\
NGC\,404        \tablenotemark{c}  &  66.7  &  146.4  &  76\tablenotemark{d}  &  80.5  &  63.2  &  -54.0  &   0.5  \\
KKH\,37\tablenotemark{e}   &  $<3.4$   &  $<0.8$ & 1.8\tablenotemark{f,g} & \nodata & \nodata & \nodata
& $<0.08$ \\      
UGC\,4483         &  12.0  &  32.8  &  13.6\tablenotemark{h}  &  51.2  &  34.3  &  153.9  &   3.2  \\
KK\,77\tablenotemark{e}   & $<4.4$ & $<2.3$ & $<5.5$\tablenotemark{i} & \nodata & \nodata & \nodata & $<0.06$ \\
BK3N             &   6.3  &  22.0  &  $<0.75$\tablenotemark{j}  &  44.4  &  20.0  &  -42.5  &   0.7  \\
AO\,0952+69     \tablenotemark{k}  &  61.3  &  206.6  & $<0.59$\tablenotemark{j} &  56.0  &  45.6  &  112.8  &  1.3  \\
Sextans\,B        &  91.0  &  41.5  &  72.9\tablenotemark{h}  &  58.1  &  40.6  &  302.2  &   2.6  \\
NGC\,3109         &  720.9  &  270.1  &  1148\tablenotemark{a}  &  127.7  &  116.0  &  405.1  &   6.6  \\
Antlia           &   1.4  &  0.5  &  1.7\tablenotemark{l}  &  23.4  &  13.4  &  363.4  &   0.3  \\
KDG\,63\tablenotemark{e} & $<4.2$ & $<1.1$ & $<0.2$\tablenotemark{m} & \nodata & \nodata & \nodata & $<0.06$ \\
Sextans\,A        &  138.1  &  62.1  &  169\tablenotemark{a}  &  59.8  &  46.2  &  324.8  &   6.1  \\
HS\,117\tablenotemark{e} & $<1.7$ & $<0.6$ & \nodata  & \nodata & \nodata & \nodata & $<0.03$ \\
DDO\,82           &   0.8  &  2.8  &   $<0.62$\tablenotemark{j}  &  35.8  &  26.7  &  56.2  &   0.9  \\
KDG\,73           &   0.1  &  0.5  &  1.0\tablenotemark{h}  &   9.2  &   8.5  &  116.3  &   0.1  \\
NGC\,3741         &  32.8  &  81.1  &  44.6\tablenotemark{o}  &  85.4  &  70.6  &  229.1  &   3.4  \\
DDO\,99          &  29.7  &  46.9  &  47.1\tablenotemark{o}  &  51.6&  28.6  &  242.1  &   2.6  \\
NGC\,4163        &   4.8  &  9.3  &  9.6\tablenotemark{p}  &  33.5  &  22.7  &  161.6  &   2.1  \\
NGC 4190         &  15.5  &  44.8  &  23.2\tablenotemark{p}  &  73.2  &  52.8  &  227.0  &   3.5  \\
DDO\,113\tablenotemark{e} &  $<1.6$ & $<0.4$ & 23.6\tablenotemark{p}&  \nodata & \nodata & \nodata & $<0.04$ \\         
MCG\,+09-20-131   &   3.1  &  11.9  &  5.2\tablenotemark{q}  &  39.0  &  26.1  &  157.6  &   3.3  \\
DDO\,125          &  18.3  &  28.7  &  21.8\tablenotemark{h}  &  39.7  &  27.0  &  196.1  &   2.1  \\
UGCA\,292         &  12.9  &  40.0  &  14.3\tablenotemark{h}  &  37.1  &  25.2  &  308.3  &   4.2  \\
GR\,8             &   5.8  &  5.9  &  7.8\tablenotemark{h}  &  30.7  &  21.4  &  213.7  &   1.7  \\
UGC\,8508         &  12.3  &  19.3  &  14.8\tablenotemark{p}  &  62.7  &  48.1  &  59.8  &   2.9  \\
DDO\,181          &  10.5  &  24.4  &  12.5\tablenotemark{o}  &  52.1  &  40.8  &  201.4  &   1.7  \\
DDO\,183          &   8.2  &  20.1  &  9.6\tablenotemark{p}  &  42.2  &  26.4  &  191.2  &   2.2  \\
KKH\,86           &   0.1  &  0.1  &  0.5\tablenotemark{h}  &   7.7  &   6.9  &  285.5  &   0.2  \\
UGC\,8833         &   5.9  &  13.1  &  6.0\tablenotemark{h}  &  41.0  &  29.4  &  225.9  &   2.2  \\
KK\,230           &   0.8  &  0.7  &  2.6\tablenotemark{h}  &  17.4  &  11.5  &  60.6  &   0.6  \\
DDO\,187          &  10.1  &  11.6  &  12.0\tablenotemark{h}  &  46.0  &  31.8  &  152.2  &   3.2  \\
DDO\,190          &  22.5  &  41.3  &  8.5\tablenotemark{o}  &  62.3  &  45.2  &  148.8  &   3.6  \\
KKR\,25\tablenotemark{e} & $<1.0$ & $<0.1$ &
2.2\tablenotemark{r}\tablenotemark{g} &\nodata & \nodata & \nodata & $<0.03$ \\
KKH\,98           &   2.2  &  3.3  &  4.1\tablenotemark{h} &  25.5  &  17.0  &  -137.8  &   0.8  \\

\enddata
\tablenotetext{a}{\citet{huc98}}
\tablenotetext{b}{\citet{mey04}}
\tablenotetext{c}{NGC~404 is contaminated by foreground Milky Way \hi\ emission.}
\tablenotetext{d}{\citet{baa76}}
\tablenotetext{e}{VLA-ANGST non-detection. Limits based on a width of 20\,\kms\ and the optical diameter $D_{25}$.}
\tablenotetext{f}{\citet{kar01}}
\tablenotetext{g}{might be Galactic \hi}
\tablenotetext{h}{\citet{huc03}}
\tablenotetext{i}{\citet{huc00a}}
\tablenotetext{j}{\citet{vdr98}}
\tablenotetext{k}{AO~0952+62 is contaminated by M81 tidal HI
  emission.}
\tablenotetext{l}{\citet{bar01}}
\tablenotetext{m}{\citet{sch90}}
\tablenotetext{n}{\citet{bar04}}
\tablenotetext{o}{\citet{spr05}}
\tablenotetext{p}{\citet{kor04}}
\tablenotetext{q}{\citet{pus07}}
\tablenotemark{r}{\citet{huc00b}}
\end{deluxetable}


\acknowledgments

We thank the National Radio Astronomy Observatory for their generous
time allocation, observing, and data reduction support for this Large
Project. The National Radio Astronomy Observatory is a facility of the
National Science Foundation operated under cooperative agreement by
Associated Universities, Inc. Support for this work was provided by
the National Science Foundation collaborative research grant `Star
Formation, Feedback, and the ISM: Time Resolved Constraints from a
Large VLA Survey of Nearby Galaxies,' grant number AST-0807710. This
material is based on work supported by the National Science Foundation
under Grant No. DGE-0718124. SRW is grateful for support from a
Penrose Fellowship, a University of Minnesota Degree Dissertation
Fellowship, and a NRAO Research Fellowship number 807515. We would
like to thank the LITTLE THINGS and THINGS teams for collaboration on
the calibration and imaging pipeline. We have made use of the
NASA/IPAC Extragalactic Database (NED), which is operated by the Jet
Propulsion Laboratory, California Institute of Technology, under
contract with NASA. This research has also made use of NASA's
Astrophysics Data System (ADS).


{\it Facilities:} \facility{VLA}, \facility{HST}



\clearpage

\clearpage

\clearpage
\begin{figure}
\includegraphics[angle=0,scale=0.27]{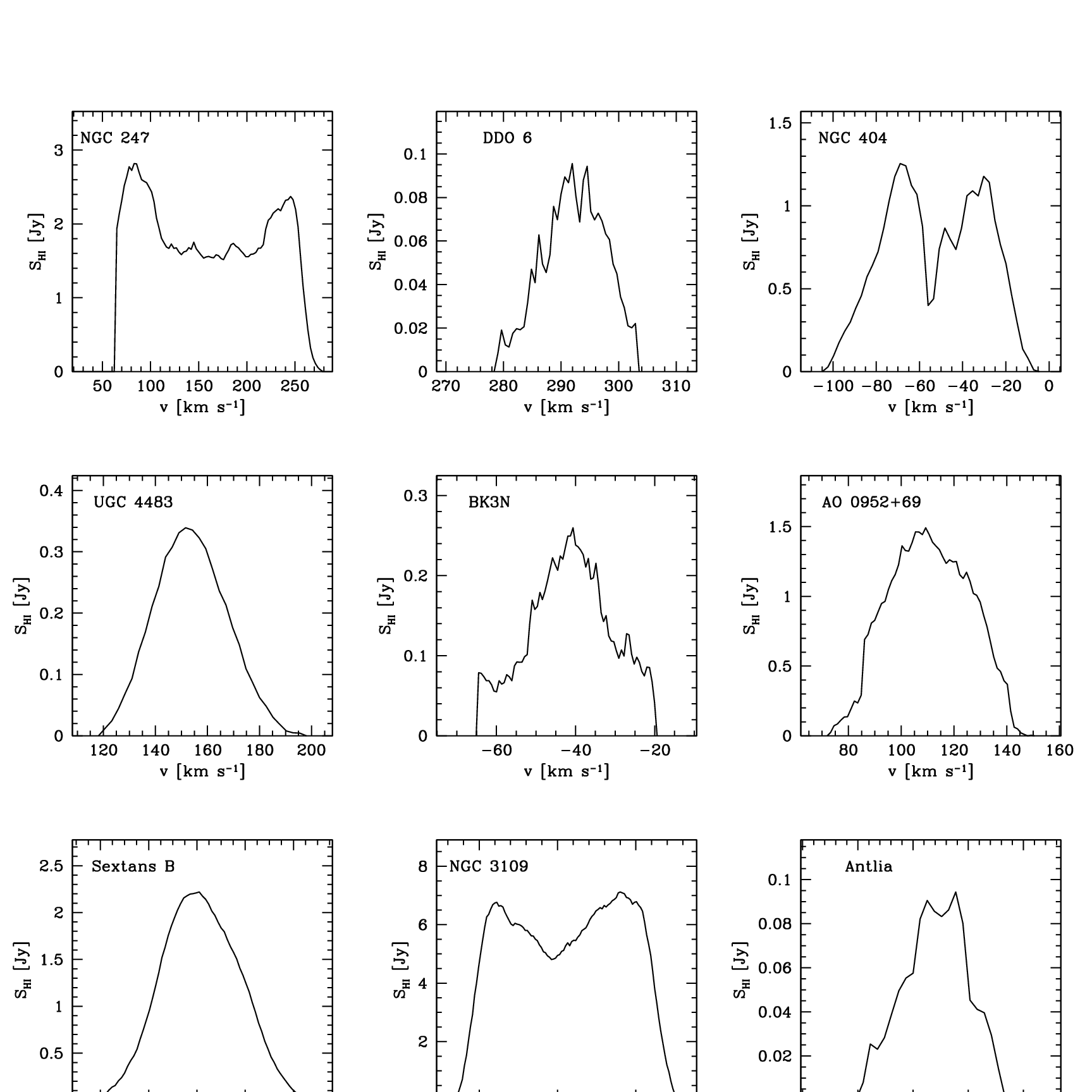}
\caption{Spatially integrated \hi\ spectra of the VLA-ANGST galaxies.\label{fig:spectra}}
\end{figure}

\clearpage
\addtocounter{figure}{-1}
\begin{figure}
\includegraphics[angle=0,scale=0.27]{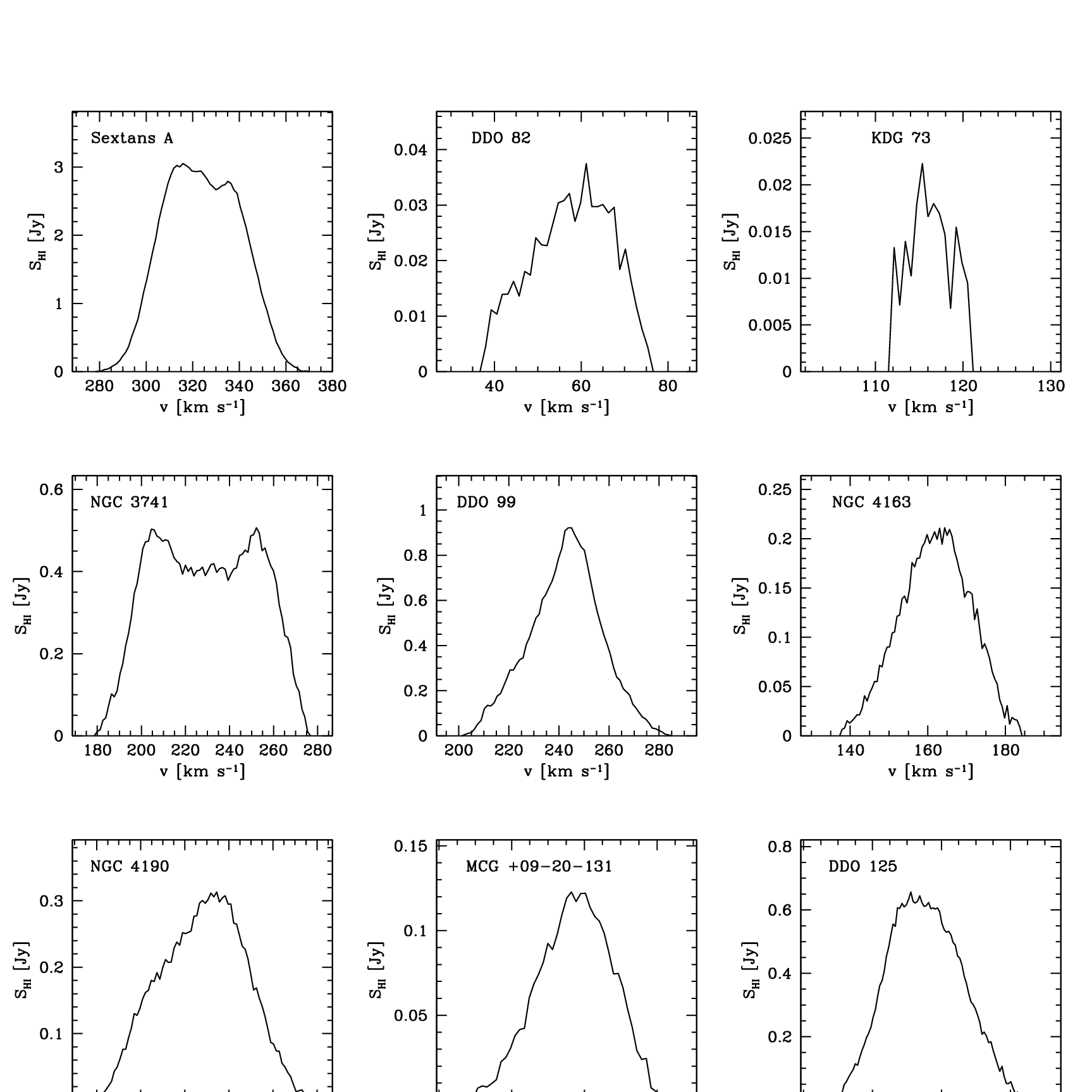}
\caption{continued.}
\end{figure}

\clearpage
\addtocounter{figure}{-1}
\begin{figure}
\includegraphics[angle=0,scale=0.27]{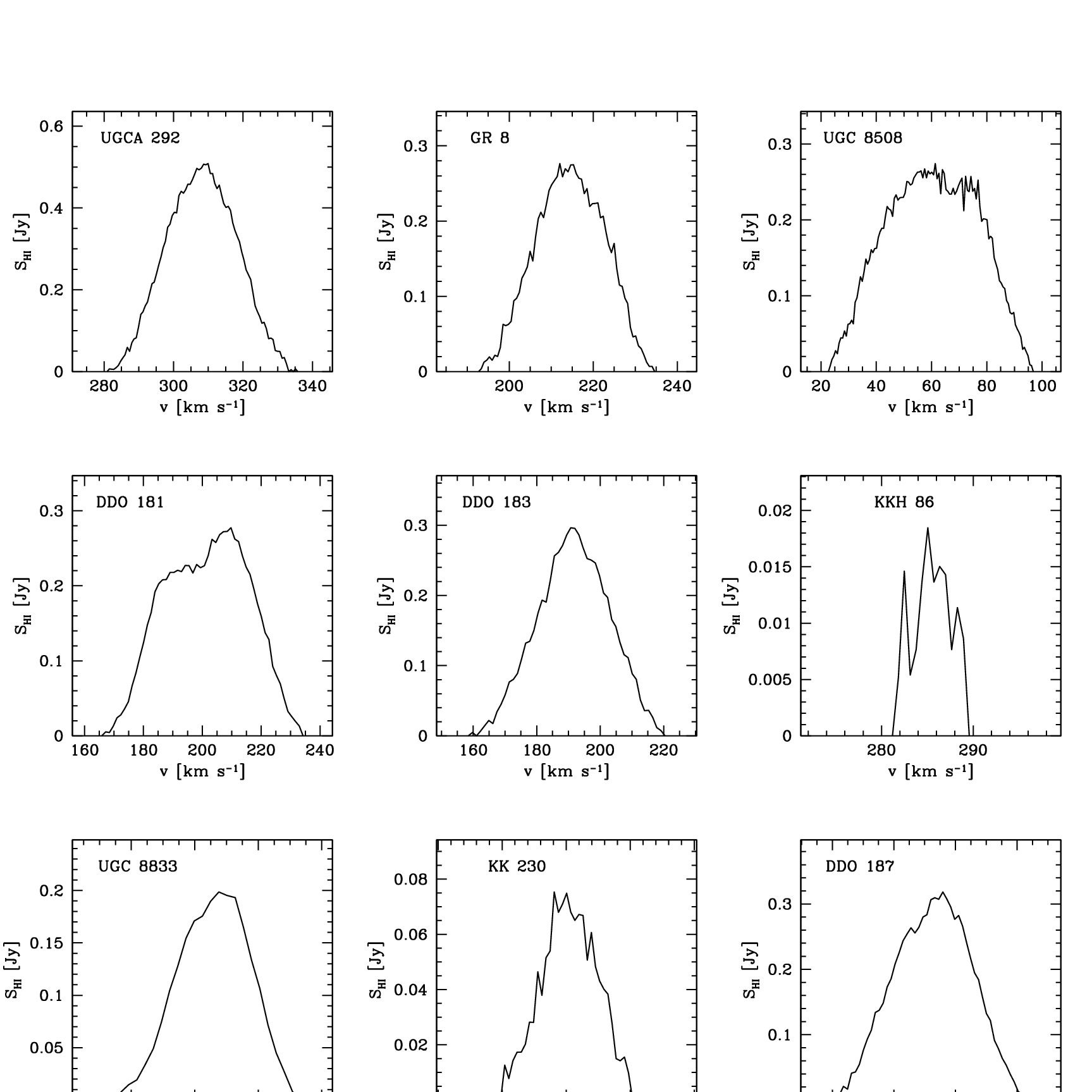}
\caption{continued.}
\end{figure}

\clearpage
\addtocounter{figure}{-1}
\begin{figure}
\includegraphics[angle=0,scale=0.27]{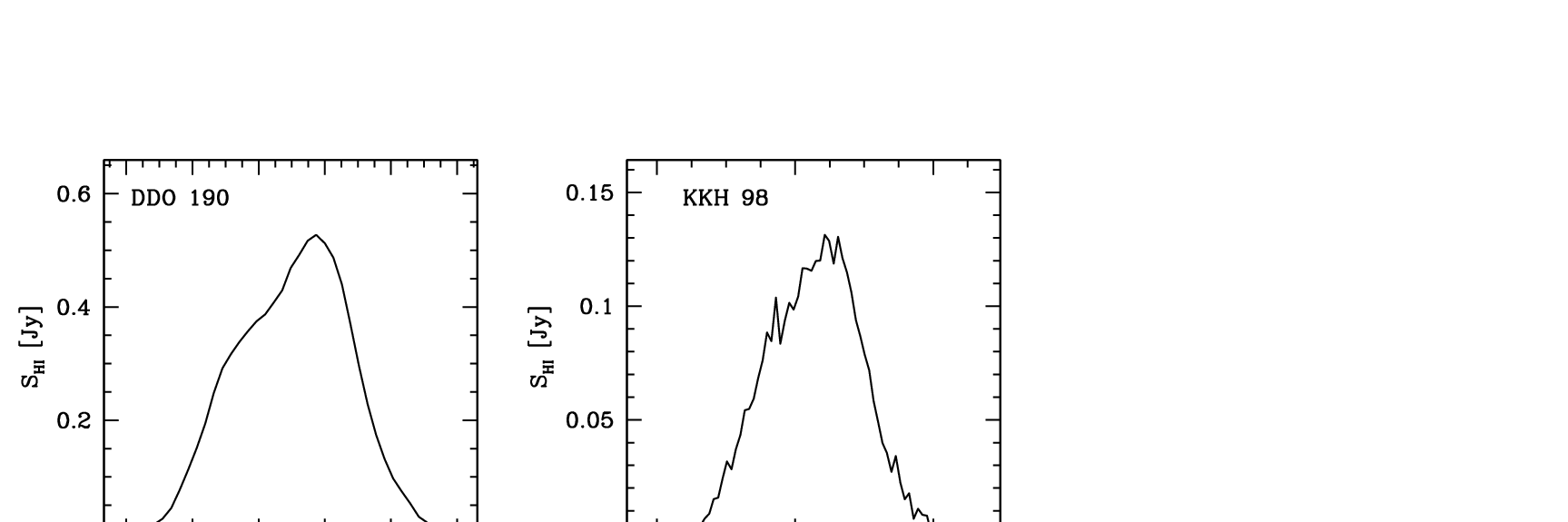}
\caption{continued.}
\end{figure}

\clearpage
\begin{figure}
\includegraphics[angle=-90,scale=1]{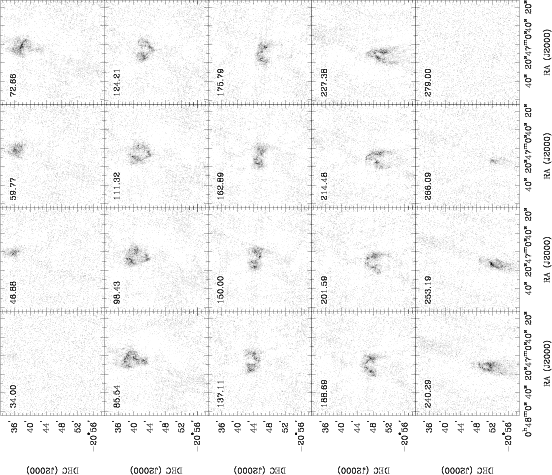}
\caption{ \textbf{NGC\,247:} Channel maps based on the natural-weighted cube
  (grayscale range: $-0.02$ to 12.2\,mJy\,beam$^{-1}$). Every fourth
  channel is shown (channel width 2.6\,\kms{}) and each map has
  the same size as the moment maps in the following panels. \label{fig:n247}}
\end{figure}

\clearpage
\addtocounter{figure}{-1}
\begin{figure}
\includegraphics[angle=0,scale=1.2]{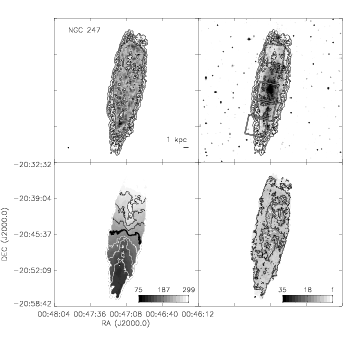}
\caption{continued. {\it Top left:} The integrated \hi\ intensity map
  for NGC\,247.  The grayscale covers a range from 1$\times$10$^{19}$
  to $5.4\times10^{21}$\,cm$^{-2}$with contours of 1$\times$10$^{20}$,
  5$\times$10$^{20}$, 1$\times$10$^{21}$, and 5$\times$10$^{21}$
  cm$^{-2}$.  {\it Top Right:} An optical 4680\,\AA\ image from the DSS with the
  same column density contours overlaid.  The HST ACS footprints are
  the fields covered by the ANGST survey.  {\it Bottom Left:} The \hi\
  velocity field.  Black contours (lighter grayscale) indicate
  approaching emission, white contours (darker grayscale) receding
  emission.  The thick black contour is the central velocity
  (v$_{cen}$ = 163.7\,km\,s$^{-1}$) and the isovelocity contours are
  spaced by $\Delta$v = 25\,km\,s$^{-1}$.  {\it Bottom Right:} The
  \hi\ velocity dispersion.  Contours are plotted at 5, 10, 15, and
  20\,km\,s$^{-1}$. Colorbars are in units of \kms.}
\end{figure}


\clearpage
\begin{figure}
\includegraphics[angle=-90,scale=1]{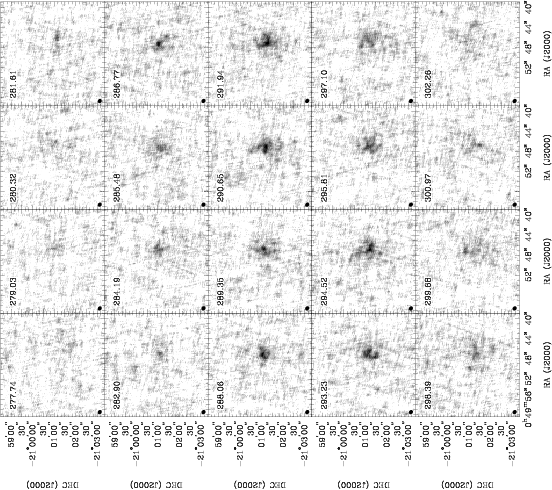}
\caption{\textbf{DDO\,6:} Channel maps based on the natural-weighted cube
  (grayscale range: $-0.02$ to 13.3\,mJy\,beam$^{-1}$). Every channel is
  shown (channel width 0.6\,\kms{}) and each map has
  the same size as the moment maps in the following panels. \label{fig:ddo6}}
\end{figure}

\clearpage
\addtocounter{figure}{-1}
\begin{figure}
\includegraphics[angle=0,scale=1.2]{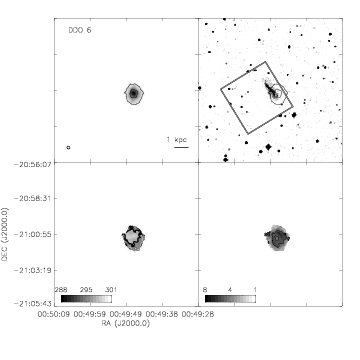}
\caption{continued. {\it Top left:} The integrated \hi\ intensity map
  for DDO\,6.  The grayscale covers a range from 1$\times$10$^{19}$ to
  $9.4\times10^{20}$\,cm$^{-2}$ with contours of 1$\times$10$^{20}$
  and 5$\times$10$^{20}$\,cm$^{-2}$.  {\it Top Right:} An optical
  4680\,\AA\ image from the DSS with the same column density contours
  overlaid.  The HST ACS footprint is the field covered by the ANGST
  survey.  {\it Bottom Left:} The \hi\ velocity field.  Black contours
  (lighter gray scale) indicate approaching emission, white contours
  (darker gray scale) receding emission.  The thick black contour is
  the central velocity (v$_{cen}$ = 292.5\,km\,s$^{-1}$) and the
  isovelocity contours are spaced by $\Delta$v = 3\,km\,s$^{-1}$.
  {\it Bottom Right:} The \hi\ velocity dispersion.  A contour is
  plotted at 5\,km\,s$^{-1}$. Colorbars are in units of \kms.}
\end{figure}


\clearpage
\begin{figure}
\includegraphics[angle=-90,scale=1]{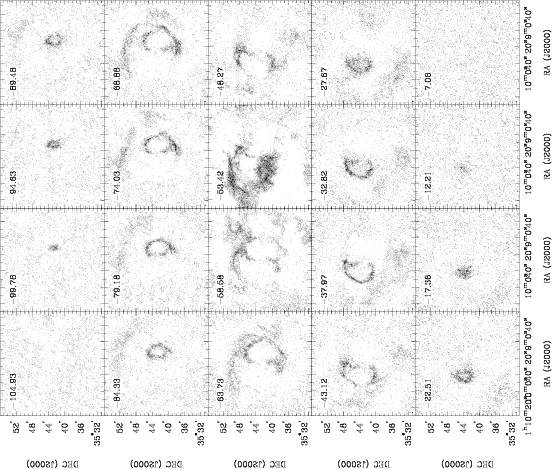}
\caption{ \textbf{NGC\,404:} Channel maps based on the natural-weighted cube
  (grayscale range: $-0.02$ to 8.5\,mJy beam$^{-1}$). Every channel is
  shown (channel width 2.6\,\kms{}) and each map has
  the same size as the moment maps in the following panels. Confusion with \hi{}
  emission from the Milky Way is present in this galaxy between
  velocities $-58$ to $-50$\,\kms{} and can be seen in two of the above
  channels.\label{fig:n404}}
\end{figure}

\clearpage
\addtocounter{figure}{-1}
\begin{figure}
\includegraphics[angle=0,scale=1.2]{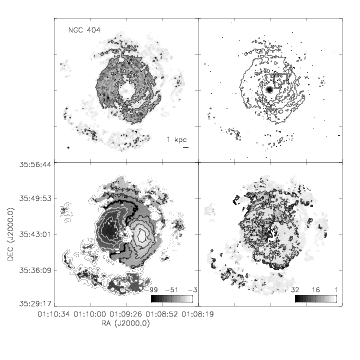}
\caption{continued. {\it Top left:} The integrated \hi\ intensity map
  for NGC\,404.  The grayscale covers a range from 1$\times$10$^{19}$
  to $5.0\times10^{20}$\,cm$^{-2}$ with contours of
  1$\times$10$^{20}$\,cm$^{-2}$ and 5$\times$10$^{20}$ cm$^{-2}$.
  {\it Top Right:} An optical 6450\,\AA\ image from the DSS with the
  same column density contours overlaid.  The HST WFPC2 footprint is
  the field covered by the ANGST survey. The bright, large disk in the
  lower right is a foreground star. {\it Bottom Left:} The \hi\
  velocity field.  Black contours (lighter gray scale) indicate
  approaching emission, white contours (darker gray scale) receding
  emission.  The thick black contour is the central velocity
  (v$_{cen}$ = $-54.0$\,km\,s$^{-1}$) and the isovelocity contours are
  spaced by $\Delta$v = 10\,km\,s$^{-1}$.  {\it Bottom Right:} The
  \hi\ velocity dispersion.  Contours are plotted at 5, 10, 15, and
  20\,km\,s$^{-1}$. Colorbars are in units of \kms.}
\end{figure}

\clearpage
\begin{figure}
\includegraphics[angle=-90,scale=1]{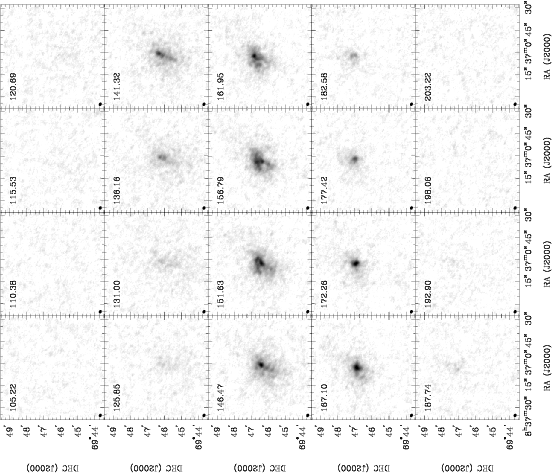}
\caption{\textbf{UGC\,4483:} Channel maps based on the natural-weighted
  cube (grayscale range: $-0.02$ to 14.8\,mJy\,beam$^{-1}$). Every channel
  is shown (channel width 2.6\,\kms{}) and each map has
  the same size as the moment maps in the following panels. \label{fig:u4483}}
\end{figure}

\clearpage
\addtocounter{figure}{-1}
\begin{figure}
\includegraphics[angle=0,scale=1.2]{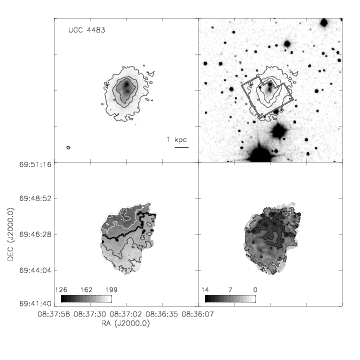}
\caption{continued. {\it Top left:} The integrated \hi\ intensity map
  for UGC\,4483.  The grayscale covers a range from 1$\times$10$^{19}$
  to $3.2\times10^{21}$\,cm$^{-2}$ with contours of
  1$\times$10$^{20}$, 5$\times$10$^{20}$, and
  1$\times$10$^{21}$\,cm$^{-2}$.  {\it Top Right:} An optical
  6450\,\AA\ image from the DSS with the same column density contours
  overlaid.  The HST WFPC2 footprint is the field covered by the ANGST
  survey.  {\it Bottom Left:} The \hi\ velocity field.  Black contours
  (lighter gray scale) indicate approaching emission, white contours
  (darker gray scale) receding emission.  The thick black contour is
  the central velocity (v$_{cen}$ = 153.9\,km\,s$^{-1}$) and the
  isovelocity contours are spaced by $\Delta$v = 10\,km\,s$^{-1}$.
  {\it Bottom Right:} The \hi\ velocity dispersion.  Contours are
  plotted at 5 and 10\,km\,s$^{-1}$. Colorbars are in units of \kms.}
\end{figure}

\clearpage
\begin{figure}
\includegraphics[angle=-90,scale=1]{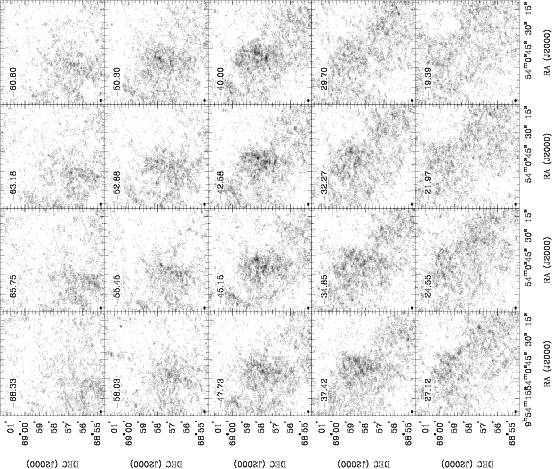}
\caption{\textbf{BK3N:} Channel maps based on the natural-weighted cube (grayscale
  range: $-0.02$ to 12.9\,mJy\,beam$^{-1}$). Every third channel is shown
  (channel width 0.6\,\kms{}) and each map has
  the same size as the moment maps in the following panels. \hi{} emission from
  M81 is present in every channel of the data cube. \label{fig:bk3n}}
\end{figure}

\clearpage
\addtocounter{figure}{-1}
\begin{figure}
\includegraphics[angle=0,scale=1.2]{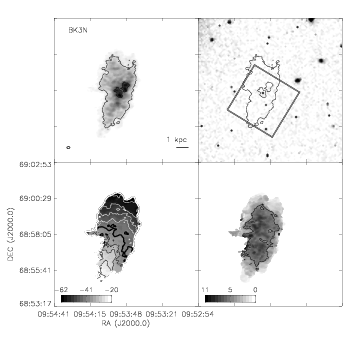}
\caption{continued. {\it Top left:} The integrated \hi\ intensity map
  for BK3N.  The grayscale covers a range from 1$\times$10$^{19}$ to
  $7.1\times10^{20}$\,cm$^{-2}$ withcontours of 1$\times$10$^{20}$ and
  5$\times$10$^{20}$ cm$^{-2}$.  {\it Top Right:} An optical g-band
  image from the SDSS with the same column density contours overlaid.
  The HST ACS footprint is the field covered by the ANGST survey.
  {\it Bottom Left:} The \hi\ velocity field.  Black contours (lighter
  gray scale) indicate approaching emission, white contours (darker
  gray scale) receding emission.  The thick black contour is the
  central velocity (v$_{cen}$ = $-42.5$\,km\,s$^{-1}$) and the
  isovelocity contours are spaced by $\Delta$v = $-5$\,km\,s$^{-1}$.
  {\it Bottom Right:} The \hi\ velocity dispersion.  Contours are
  plotted at 5 and 10\,km\,s$^{-1}$. Colorbars are in units of \kms.}
\end{figure}

\clearpage
\begin{figure}
\includegraphics[angle=-90,scale=1]{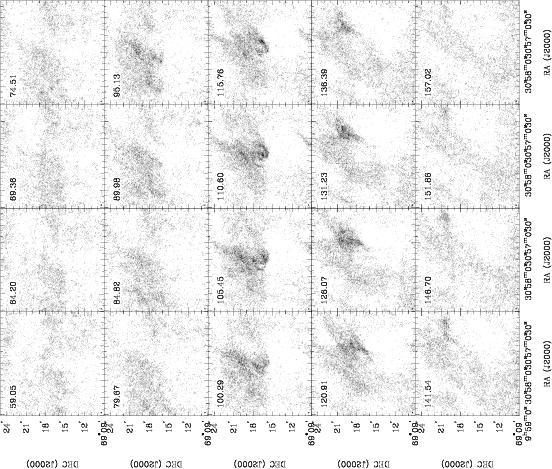}
\caption{ \textbf{AO\,0952+69:} Channel maps based on the natural-weighted cube
  (grayscale range: $-0.02$ to 10.5\,mJy\,beam$^{-1}$). Every third
  channel is shown (channel width 1.3\,\kms{}) and each map has
  the same size as the moment maps in the following panels.  This field
  is in the M81 group and therefore tidal \hi{} from member
  interactions is also visible. \label{fig:a09}}
\end{figure}

\clearpage
\addtocounter{figure}{-1}
\begin{figure}
\includegraphics[angle=0,scale=1.2]{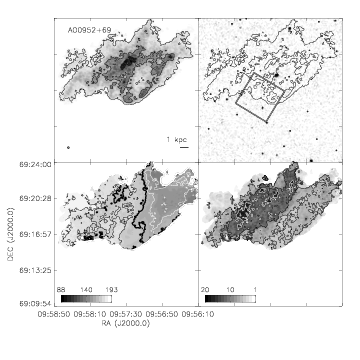}
\caption{continued. {\it Top left:} The integrated \hi\ intensity map
  for AO\,0952+69.  The grayscale covers a range from
  1$\times$10$^{19}$ to $1.13\times10^{21}$\,cm$^{-2}$ with contours
  of 1$\times$10$^{20}$, 5$\times$10$^{20}$, and
  1$\times$10$^{21}$\,cm$^{-2}$.  {\it Top Right:} An optical g-band
  image from the SDSS with the same column density contours overlaid.
  The HST ACS footprint is the field covered by the ANGST survey.
  {\it Bottom Left:} The \hi\ velocity field.  Black contours (lighter
  gray scale) indicate approaching emission, white contours (darker
  gray scale) receding emission.  The thick black contour is the
  central velocity (v$_{cen}$ = 112.8\,km\,s$^{-1}$) and the
  isovelocity contours are spaced by $\Delta$v = 10\,km\,s$^{-1}$.
  {\it Bottom Right:} The \hi\ velocity dispersion.  Contours are
  plotted at 5, 10, and 15\,km\,s$^{-1}$. Colorbars are in units of \kms. }
\end{figure}

\clearpage
\begin{figure}
\includegraphics[angle=-90,scale=1]{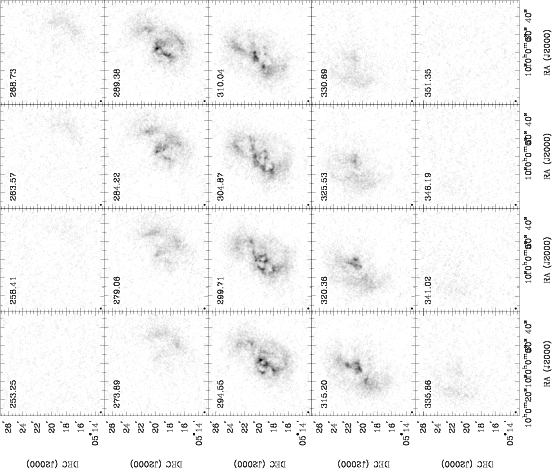}
\caption{\textbf{Sextans\,B:} Channel maps based on the natural-weighted
  cube (grayscale range: $-0.02$ to 25.3\,mJy\,beam$^{-1}$). Every third
  channel is shown (channel width 1.3\,\kms{}) and each map has
  the same size as the moment maps in the following panels. \label{fig:sexb}}
\end{figure}

\clearpage
\addtocounter{figure}{-1}
\begin{figure}
\includegraphics[angle=0,scale=1.2]{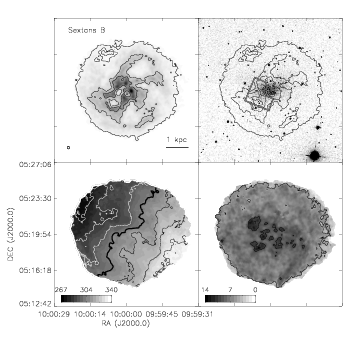}
\caption{continued. {\it Top left:} The integrated \hi\ intensity map
  for Sextans\,B.  The grayscale covers a range from
  1$\times$10$^{19}$ to $2.6\times10^{21}$\,cm$^{-2}$ with contours of
  1$\times$10$^{20}$, 5$\times$10$^{20}$, and 1$\times$10$^{21}$
  cm$^{-2}$. {\it Top Right:} An optical g-band image from the SDSS
  with the same column density contours overlaid.  The HST WFPC2
  footprint is the field covered by the ANGST survey.  {\it Bottom
    Left:} The \hi\ velocity field.  Black contours (lighter gray
  scale) indicate approaching emission, white contours (darker gray
  scale) receding emission.  The thick black contour is the central
  velocity (v$_{cen}$ = 302.2\,km\,s$^{-1}$) and the isovelocity
  contours are spaced by $\Delta$v = 10\,km\,s$^{-1}$.  {\it Bottom
    Right:} The \hi\ velocity dispersion.  Contours are plotted at 5
  and 10\,km\,s$^{-1}$. Colorbars are in units of \kms.}
\end{figure}

\clearpage
\begin{figure}
\includegraphics[angle=-90,scale=1]{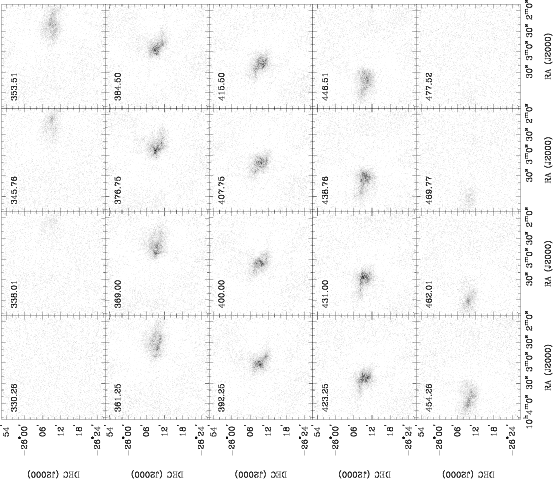}
\caption{\textbf{NGC\,3109:} Channel maps based on the natural-weighted
  cube (grayscale range: $-0.02$ to 31.7\,mJy\,beam$^{-1}$). Every fifth
  channel is shown (channel width 1.3\,\kms{}) and each map has
  the same size as the moment maps in the following panels. \label{fig:n3109}}
\end{figure}

\clearpage
\addtocounter{figure}{-1}
\begin{figure}
\includegraphics[angle=0,scale=1.2]{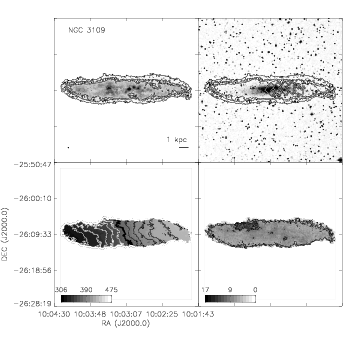}
\caption{continued. {\it Top left:} The integrated \hi\ intensity map
  for NGC\,3109.  The grayscale covers a range from 1$\times$10$^{19}$
  to $6.6\times10^{21}$\,cm$^{-2}$ with contours of
  1$\times$10$^{20}$, 5$\times$10$^{20}$, 1$\times$10$^{21}$, and
  5$\times$10$^{21}$ cm$^{-2}$.  {\it Top Right:} An optical
  4680\,\AA\ image from the DSS with the same column density contours
  overlaid.  The HST WFPC2 footprints are the fields covered by the
  ANGST survey.  {\it Bottom Left:} The \hi\ velocity field.  Black
  contours (lighter grayscale) indicate approaching emission, white
  contours (darker grayscale) receding emission.  The thick black
  contour is the central velocity (v$_{cen}$ = 405.1\,km\,s$^{-1}$)
  and the isovelocity contours are spaced by $\Delta$v =
  10\,km\,s$^{-1}$.  {\it Bottom Right:} The \hi\ velocity dispersion.
  Contours are plotted at 5, 10, and 15\,km\,s$^{-1}$. Colorbars are in units of \kms.}
\end{figure}

\clearpage
\begin{figure}
\includegraphics[angle=-90,scale=1]{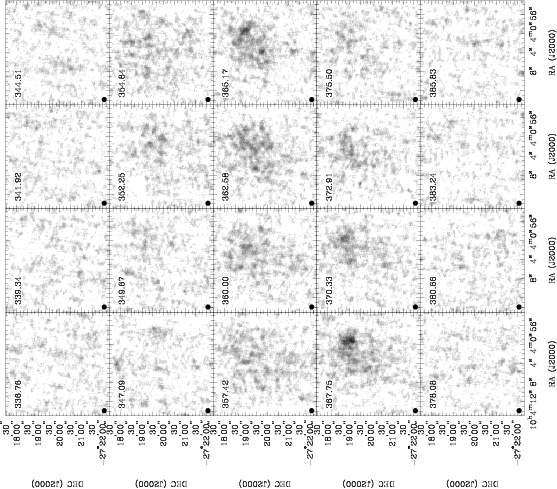}
\caption{ \textbf{Antlia:} Channel maps based on the natural-weighted cube
  (grayscale range: $-0.02$ to 8.8\,mJy\,beam$^{-1}$) and each map has
  the same size as the moment maps in the following panels.\label{fig:antlia}}
\end{figure}

\clearpage
\addtocounter{figure}{-1}
\begin{figure}
\includegraphics[angle=0,scale=1.2]{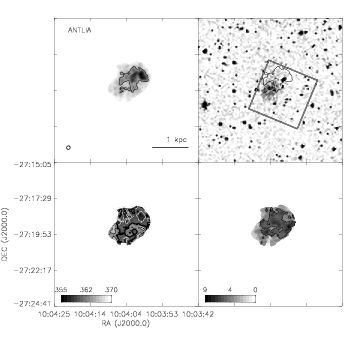}
\caption{continued. {\it Top left:} The integrated \hi\ intensity map
  for Antlia.  The grayscale covers a range from 1$\times$10$^{19}$ to
  $2.9\times10^{20}$\,cm$^{-2}$ with a contour of
  1$\times$10$^{20}$\,cm$^{-2}$.  {\it Top Right:} An optical
  4680\,\AA\ image from the DSS with the same column density contours
  overlaid.  The HST ACS footprint is the field covered by the ANGST
  survey.  {\it Bottom Left:} The \hi\ velocity field.  Black contours
  (lighter grayscale) indicate approaching emission, white contours
  (darker grayscale) receding emission.  The thick black contour is
  the central velocity (v$_{cen}$ = 363.0\,km\,s$^{-1}$) and the
  isovelocity contours are spaced by $\Delta$v = 3\,km\,s$^{-1}$.
  {\it Bottom Right:} The \hi\ velocity dispersion.  A contour is
  plotted at 5\,km\,s$^{-1}$. Colorbars are in units of \kms.}
\end{figure}

\clearpage
\begin{figure}
\includegraphics[angle=-90,scale=1]{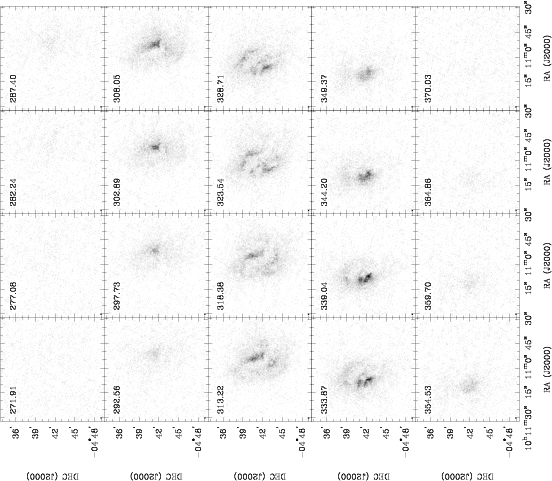}
\caption{\textbf{Sextans\,A:} Channel maps based on the natural-weighted
  cube (grayscale range: $-0.02$ to 35.4\,mJy\,beam$^{-1}$). Every third
  channel is shown (channel width 1.3\,\kms{}) and each map has
  the same size as the moment maps in the following panels.\label{fig:sexa}}
\end{figure}

\clearpage
\addtocounter{figure}{-1}
\begin{figure}
\includegraphics[angle=0,scale=1.2]{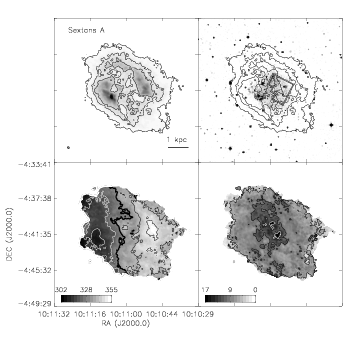}
\caption{continued. {\it Top left:} The integrated \hi\ intensity map
  for Sextans\,A.  The grayscale covers a range from
  1$\times$10$^{19}$ to $6.1\times10^{21}$\,cm$^{-2}$ with contours of
  1$\times$10$^{20}$, 5$\times$10$^{20}$, 1$\times$10$^{21}$, and
  5$\times$10$^{21}$ cm$^{-2}$.  {\it Top Right:} An optical
  4680\,\AA\ image from the DSS with the same column density contours
  overlaid.  The HST WFPC2 footprint is the field covered by the ANGST
  survey.  {\it Bottom Left:} The \hi\ velocity field.  Black contours
  (lighter gray scale) indicate approaching emission, white contours
  (darker gray scale) receding emission.  The thick black contour is
  the central velocity (v$_{cen}$ = 324.8\,km\,s$^{-1}$) and the
  isovelocity contours are spaced by $\Delta$v = 10\,km\,s$^{-1}$.
  {\it Bottom Right:} The \hi\ velocity dispersion.  Contours are
  plotted at 5, 10, and 15\,km\,s$^{-1}$. Colorbars are in units of \kms. }
\end{figure}

\clearpage
\begin{figure}
\includegraphics[angle=-90,scale=1]{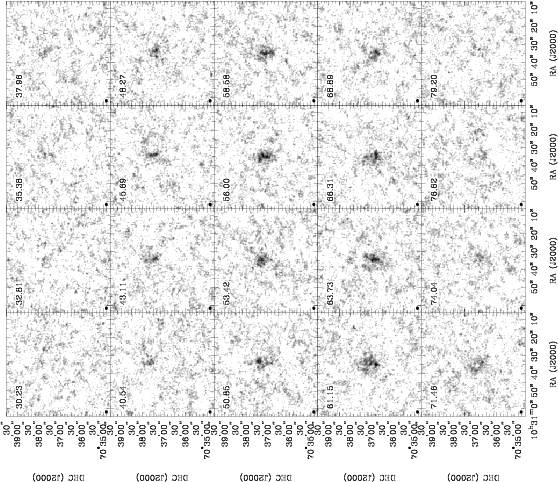}
\caption{\textbf{DDO\,82:} Channel maps based on the natural-weighted cube
  (grayscale range: $-0.02$ to 6.7\,mJy\,beam$^{-1}$). Every channel is
  shown (channel width 1.3\,\kms{}) and each map has
  the same size as the moment maps in the following panels.\label{fig:ddo82}}
\end{figure}

\clearpage
\addtocounter{figure}{-1}
\begin{figure}
\includegraphics[angle=0,scale=1.2]{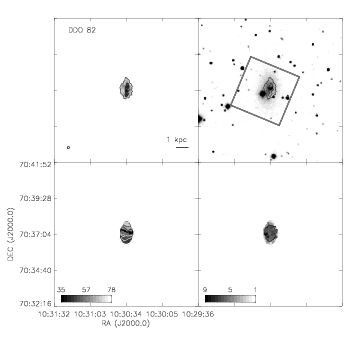}
\caption{continued. {\it Top left:} The integrated \hi\ intensity map
  for DDO\,82.  The grayscale covers a range from 1$\times$10$^{19}$
  to $9.3\times10^{20}$\,cm$^{-2}$ with contours of 1$\times$10$^{20}$
  and 5$\times$10$^{20}$\,cm$^{-2}$.  {\it Top Right:} An optical
  6450\,\AA\ image from the DSS with the same column density contours
  overlaid.  The HST ACS footprint is the field covered by the ANGST
  survey.  {\it Bottom Left:} The \hi\ velocity field.  Black contours
  (lighter gray scale) indicate approaching emission, white contours
  (darker gray scale) receding emission.  The thick black contour is
  the central velocity (v$_{cen}$ = 56.2\,km\,s$^{-1}$) and the
  isovelocity contours are spaced by $\Delta$v = 5\,km\,s$^{-1}$.
  {\it Bottom Right:} The \hi\ velocity dispersion.  A contour is
  plotted at 5\,km\,s$^{-1}$. Colorbars are in units of \kms.}
\end{figure}

\clearpage
\begin{figure}
\includegraphics[angle=-90,scale=1]{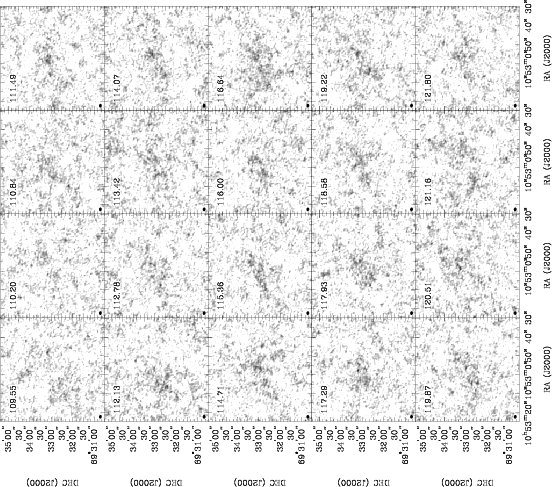}
\caption{\textbf{KDG\,73:} Channel maps based on the natural-weighted cube
  (grayscale range: $-0.02$ to 8.6\,mJy\,beam$^{-1}$). Every third channel
  is shown (channel width 0.6\,\kms{}) and each map has
  the same size as the moment maps in the following panels.\label{fig:k73}}
\end{figure}

\clearpage
\addtocounter{figure}{-1}
\begin{figure}
\includegraphics[angle=0,scale=1.2]{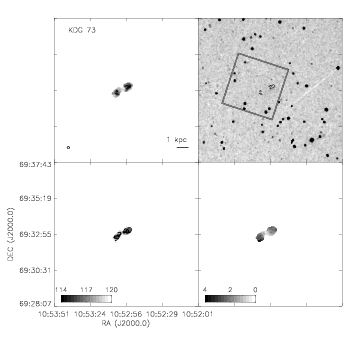}
\caption{continued. {\it Top left:} The integrated \hi\ intensity map
  for KDG\,73.  The grayscale covers a range from 1$\times$10$^{19}$
  to $1.4\times10^{20}$\,cm$^{-2}$ with a contour of
  1$\times$10$^{20}$\,cm$^{-2}$.  {\it Top Right:} An optical
  6450\,\AA\ image from the DSS with the same column density contours
  overlaid.  The HST ACS footprint is the field covered by the ANGST
  survey.  {\it Bottom Left:} The \hi\ velocity field.  Black contours
  (lighter gray scale) indicate approaching emission, white contours
  (darker gray scale) receding emission.  The thick black contour is
  the central velocity (v$_{cen}$ = 116.3\,km\,s$^{-1}$) and the
  isovelocity contours are spaced by $\Delta$v = 10\,km\,s$^{-1}$.
  {\it Bottom Right:} The \hi\ velocity dispersion. Colorbars are in units of \kms.}
\end{figure}

\clearpage
\begin{figure}
\includegraphics[angle=-90,scale=1]{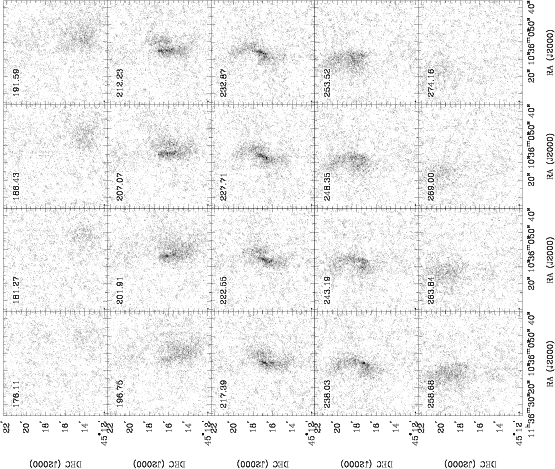}
\caption{\textbf{NGC\,3741:} Channel maps based on the natural-weighted
  cube (grayscale range: $-0.02$ to 9.6\,mJy\,beam$^{-1}$). Every third
  channel is shown (channel width 1.3\,\kms{}) and each map has
  the same size as the moment maps in the following panels.\label{fig:n3741}}
\end{figure}

\clearpage
\addtocounter{figure}{-1}
\begin{figure}
\includegraphics[angle=0,scale=1.2]{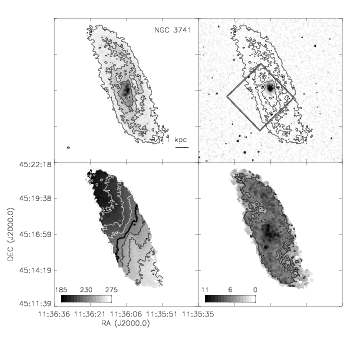}
\caption{continued. {\it Top left:} The integrated \hi\ intensity map
  for NGC\,3741.  The grayscale covers a range from 1$\times$10$^{19}$
  to $3.4\times10^{21}$\,cm$^{-2}$ with contours of
  1$\times$10$^{20}$, 5$\times$10$^{20}$, and
  1$\times$10$^{21}$\,cm$^{-2}$.  {\it Top Right:} An optical g-band
  image from the SDSS with the same column density contours overlaid.
  The HST ACS footprint is the field covered by the ANGST survey.
  {\it Bottom Left:} The \hi\ velocity field.  Black contours (lighter
  gray scale) indicate approaching emission, white contours (darker
  gray scale) receding emission.  The thick black contour is the
  central velocity (v$_{cen}$ = 229.1\,km\,s$^{-1}$) and the
  isovelocity contours are spaced by $\Delta$v = 10\,km\,s$^{-1}$.
  {\it Bottom Right:} The \hi\ velocity dispersion.  Contours are
  plotted at 5 and 10\,km\,s$^{-1}$. Colorbars are in units of \kms.}
\end{figure}

\clearpage
\begin{figure}
\includegraphics[angle=-90,scale=1]{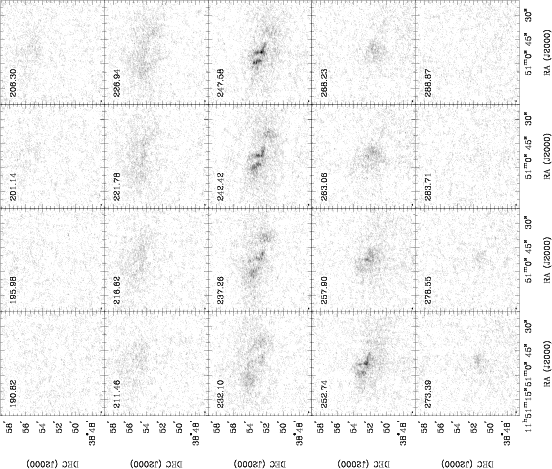}
\caption{\textbf{DDO\,99:} Channel maps based on the natural-weighted cube
  (grayscale range: $-0.02$ to 14.7\,mJy\,beam$^{-1}$). Every third
  channel is shown (channel width 1.3\,\kms{}) and each map has
  the same size as the moment maps in the following panels.\label{fig:ddo99}}
\end{figure}

\clearpage
\addtocounter{figure}{-1}
\begin{figure}
\includegraphics[angle=0,scale=1.2]{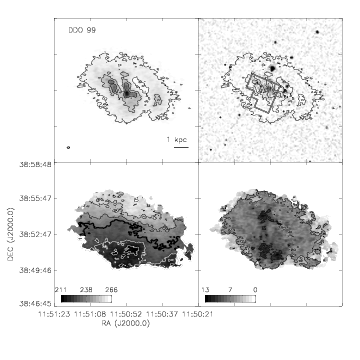}
\caption{continued. {\it Top left:} The integrated \hi\ intensity map
  for DDO\,99.  The grayscale covers a range from 1$\times$10$^{19}$
  to $2.6\times10^{21}$\,cm$^{-2}$ with contours of
  1$\times$10$^{20}$, 5$\times$10$^{20}$, and
  1$\times$10$^{21}$\,cm$^{-2}$.  {\it Top Right:} An optical g-band
  image from the SDSS with the same column density contours overlaid.
  The HST WFPC2 footprint is the field covered by the ANGST survey.
  {\it Bottom Left:} The \hi\ velocity field.  Black contours (lighter
  gray scale) indicate approaching emission, white contours (darker
  gray scale) receding emission.  The thick black contour is the
  central velocity (v$_{cen}$ = 242.1\,km\,s$^{-1}$) and the
  isovelocity contours are spaced by $\Delta$v = 10\,km\,s$^{-1}$.
  {\it Bottom Right:} The \hi\ velocity dispersion.  Contours are
  plotted at 5 and 10\,km\,s$^{-1}$. Colorbars are in units of \kms.}
\end{figure}

\clearpage
\begin{figure}
\includegraphics[angle=-90,scale=1]{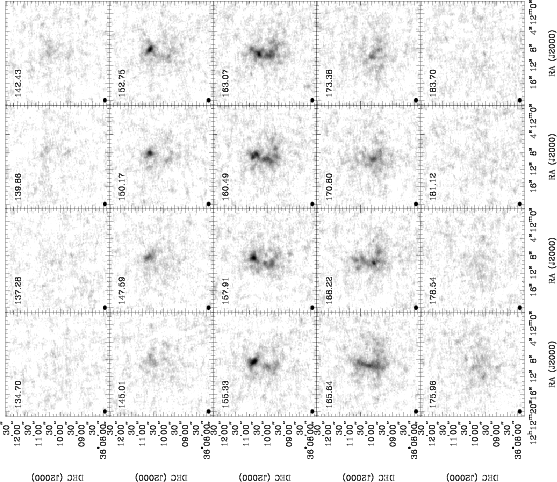}
\caption{\textbf{NGC\,4163:} Channel maps based on the natural-weighted
  cube (grayscale range: $-0.02$ to 16.1\,mJy\,beam$^{-1}$). Every third
  channel is shown (channel width 0.6\,\kms{}) and each map has
  the same size as the moment maps in the following panels.\label{fig:n4163}}
\end{figure}

\clearpage
\addtocounter{figure}{-1}
\begin{figure}
\includegraphics[angle=0,scale=1.2]{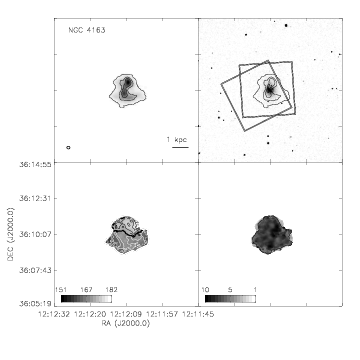}
\caption{continued. {\it Top left:} The integrated \hi\ intensity map
  for NGC\,4163.  The grayscale covers a range from 1$\times$10$^{19}$
  to $2.1\times10^{21}$\,cm$^{-2}$ with contours of
  1$\times$10$^{20}$, 5$\times$10$^{20}$, and
  1$\times$10$^{21}$\,cm$^{-2}$.  {\it Top Right:} An optical g-band
  image from the SDSS with the same column density contours overlaid.
  The HST ACS footprints are the fields covered by the ANGST survey.
  {\it Bottom Left:} The \hi\ velocity field.  Black contours (lighter
  gray scale) indicate approaching emission, white contours (darker
  gray scale) receding emission.  The thick black contour is the
  central velocity (v$_{cen}$ = 161.6\,km\,s$^{-1}$) and the
  isovelocity contours are spaced by $\Delta$v = 3\,km\,s$^{-1}$.
  {\it Bottom Right:} The \hi\ velocity dispersion.  Contours are
  plotted at 5 and 10\,km\,s$^{-1}$. Colorbars are in units of \kms.}
\end{figure}

\clearpage
\begin{figure}
\includegraphics[angle=-90,scale=1]{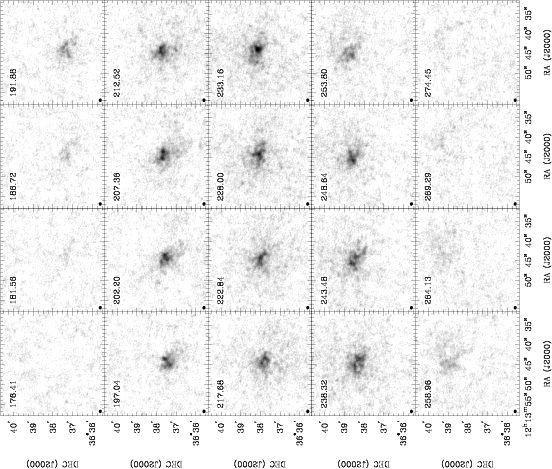}
\caption{\textbf{NGC\,4190:} Channel maps based on the natural-weighted
  cube (grayscale range: $-0.02$ to 14.7\,mJy\,beam$^{-1}$). Every third
  channel is shown (channel width 1.3\,\kms{}) and each map has
  the same size as the moment maps in the following panels.\label{fig:n4190}}
\end{figure}

\clearpage
\addtocounter{figure}{-1}
\begin{figure}
\includegraphics[angle=0,scale=1.2]{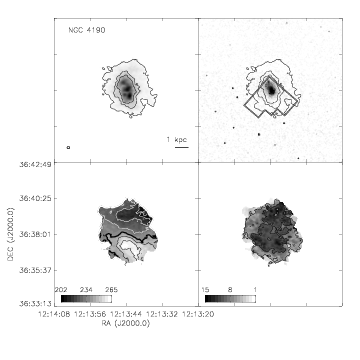}
\caption{continued. {\it Top left:} The integrated \hi\ intensity map
  for NGC\,4190.  The grayscale covers a range from 1$\times$10$^{19}$
  to $3.5\times10^{21}$\,cm$^{-2}$ with contours of
  1$\times$10$^{20}$, 5$\times$10$^{20}$, and
  1$\times$10$^{21}$\,cm$^{-2}$.  {\it Top Right:} An optical g-band
  image from the SDSS with the same column density contours overlaid.
  The HST WFPC2 footprint is the field covered by the ANGST survey.
  {\it Bottom Left:} The \hi\ velocity field.  Black contours (lighter
  gray scale) indicate approaching emission, white contours (darker
  gray scale) receding emission.  The thick black contour is the
  central velocity (v$_{cen}$ = 227.0\,km\,s$^{-1}$) and the
  isovelocity contours are spaced by $\Delta$v = 10\,km\,s$^{-1}$.
  {\it Bottom Right:} The \hi\ velocity dispersion.  Contours are
  plotted at 5, 10, and 15\,km\,s$^{-1}$. Colorbars are in units of \kms. }
\end{figure}

\clearpage
\begin{figure}
\includegraphics[angle=-90,scale=1]{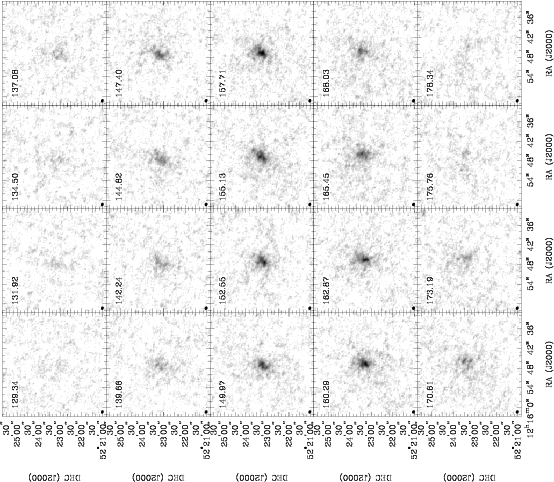}
\caption{\textbf{MCG\,+09-20-131:} Channel maps based on the
  natural-weighted cube (grayscale range: $-0.02$ to 12.3\,mJy\,beam$^{-1}$). Every
  channel is shown (channel width 1.3\,\kms{}) and each map has
  the same size as the moment maps in the following panels.\label{fig:mcg9}}
\end{figure}

\clearpage
\addtocounter{figure}{-1}
\begin{figure}
\includegraphics[angle=0,scale=1.2]{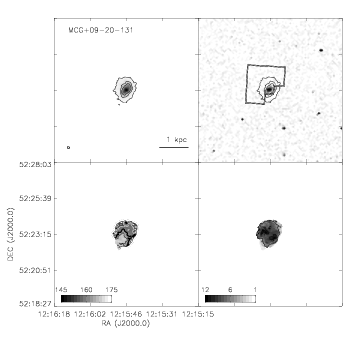}
\caption{continued. {\it Top left:} The integrated \hi\ intensity map
  for MCG\,+09-20-131.  The grayscale covers a range from
  1$\times$10$^{19}$ to $3.3\times10^{21}$\,cm$^{-2}$ with contours of
  1$\times$10$^{20}$, 5$\times$10$^{20}$, and
  1$\times$10$^{21}$\,cm$^{-2}$.  {\it Top Right:} An optical g-band
  image from the SDSS with the same column density contours overlaid.
  The HST WFPC2 footprint is the field covered by the ANGST survey.
  {\it Bottom Left:} The \hi\ velocity field.  Black contours (lighter
  gray scale) indicate approaching emission, white contours (darker
  gray scale) receding emission.  The thick black contour is the
  central velocity (v$_{cen}$ = 157.6\,km\,s$^{-1}$) and the
  isovelocity contours are spaced by $\Delta$v = 5\,km\,s$^{-1}$.
  {\it Bottom Right:} The \hi\ velocity dispersion.  Contours are
  plotted at 5 and 10\,km\,s$^{-1}$. Colorbars are in units of \kms. }
\end{figure}

\clearpage
\begin{figure}
\includegraphics[angle=-90,scale=1]{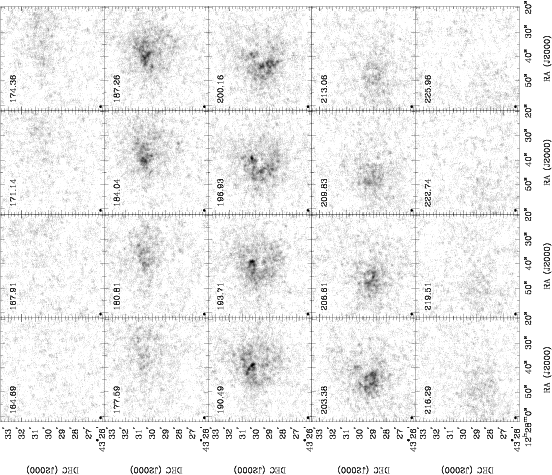}
\caption{\textbf{DDO\,125:} Channel maps based on the natural-weighted
  cube (grayscale range: $-0.02$ to 16.0\,mJy\,beam$^{-1}$). Every fourth
  channel is shown (channel width 0.6\,\kms{}) and each map has
  the same size as the moment maps in the following panels.\label{fig:ddo125}}
\end{figure}

\clearpage
\addtocounter{figure}{-1}
\begin{figure}
\includegraphics[angle=0,scale=1.2]{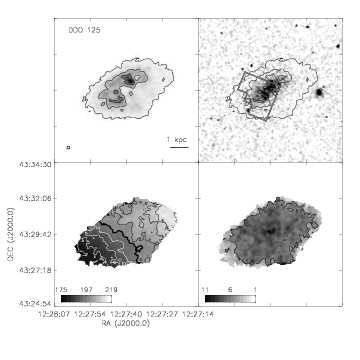}
\caption{continued. {\it Top left:} The integrated \hi\ intensity map
  for DDO\,125.  The grayscale covers a range from 1$\times$10$^{19}$
  to $2.1\times10^{21}$\,cm$^{-2}$ with contours of
  1$\times$10$^{20}$, 5$\times$10$^{20}$, and
  1$\times$10$^{21}$\,cm$^{-2}$.  {\it Top Right:} An optical g-band
  image from the SDSS with the same column density contours overlaid.
  The HST WFPC2 footprint is the field covered by the ANGST survey.
  {\it Bottom Left:} The \hi\ velocity field.  Black contours (lighter
  gray scale) indicate approaching emission, white contours (darker
  gray scale) receding emission.  The thick black contour is the
  central velocity (v$_{cen}$ = 196.1\,km\,s$^{-1}$) and the
  isovelocity contours are spaced by $\Delta$v = 5\,km\,s$^{-1}$.
  {\it Bottom Right:} The \hi\ velocity dispersion.  Contours are
  plotted at 5 and 10\,km\,s$^{-1}$. Colorbars are in units of \kms.}
\end{figure}

\clearpage
\begin{figure}
\includegraphics[angle=-90,scale=1]{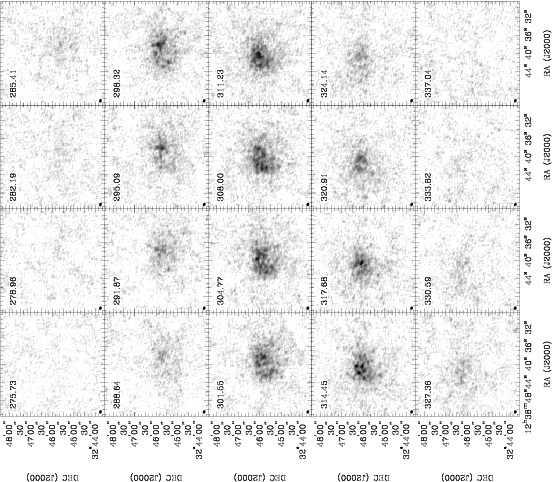}
\caption{\textbf{UGCA\,292:} Channel maps based on the natural-weighted
  cube (grayscale range: $-0.02$ to 15.8\,mJy\,beam$^{-1}$). Every fourth
  channel is shown (channel width 0.6\,\kms{}) and each map has
  the same size as the moment maps in the following panels.\label{fig:ua292}}
\end{figure}

\clearpage
\addtocounter{figure}{-1}
\begin{figure}
\includegraphics[angle=0,scale=1.2]{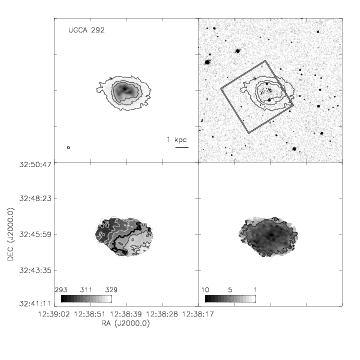}
\caption{continued. {\it Top left:} The integrated \hi\ intensity map
  for UGCA\,292.  The grayscale covers a range from 1$\times$10$^{19}$
  to $4.2\times10^{21}$\,cm$^{-2}$ with contours of
  1$\times$10$^{20}$, 5$\times$10$^{20}$, and
  1$\times$10$^{21}$\,cm$^{-2}$.  {\it Top Right:} An optical g-band
  image from the SDSS with the same column density contours overlaid.
  The HST ACS footprint is the field covered by the ANGST survey.
  {\it Bottom Left:} The \hi\ velocity field.  Black contours (lighter
  gray scale) indicate approaching emission, white contours (darker
  gray scale) receding emission.  The thick black contour is the
  central velocity (v$_{cen}$ = 308.3\,km\,s$^{-1}$) and the
  isovelocity contours are spaced by $\Delta$v = 5\,km\,s$^{-1}$.
  {\it Bottom Right:} The \hi\ velocity dispersion.  Contours are
  plotted at 5 and 10\,km\,s$^{-1}$. Colorbars are in units of \kms.}
\end{figure}

\clearpage
\begin{figure}
\includegraphics[angle=-90,scale=1]{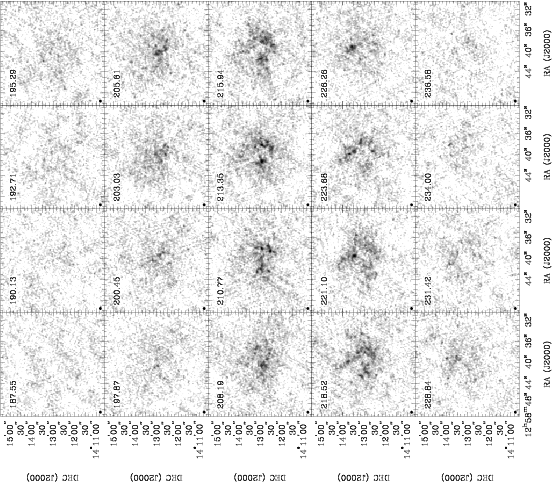}
\caption{\textbf{GR\,8:} Channel maps based on the natural-weighted cube
  (grayscale range: $-0.02$ to 10.6\,mJy\,beam$^{-1}$). Every third
  channel is shown (channel width 0.6\,\kms{}) and each map has
  the same size as the moment maps in the following panels.\label{fig:gr8}}
\end{figure}

\clearpage
\addtocounter{figure}{-1}
\begin{figure}
\includegraphics[angle=0,scale=1.2]{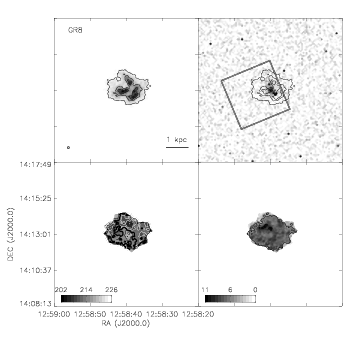}
\caption{continued. {\it Top left:} The integrated \hi\ intensity map
  for GR\,8.  The grayscale covers a range from 1$\times$10$^{19}$ to
  $1.7\times10^{21}$\,cm$^{-2}$ with contours of 1$\times$10$^{20}$,
  5$\times$10$^{20}$, and 1$\times$10$^{21}$\,cm$^{-2}$.  {\it Top
    Right:} An optical g-band image from the SDSS with the same column
  density contours overlaid.  The HST ACS footprint is the field
  covered by the ANGST survey.  {\it Bottom Left:} The \hi\ velocity
  field.  Black contours (lighter gray scale) indicate approaching
  emission, white contours (darker gray scale) receding emission.  The
  thick black contour is the central velocity (v$_{cen}$ =
  213.7\,km\,s$^{-1}$) and the isovelocity contours are spaced by
  $\Delta$v = 3\,km\,s$^{-1}$.  {\it Bottom Right:} The \hi\ velocity
  dispersion.  Contours are plotted at 5 and 10\,km\,s$^{-1}$. Colorbars are in units of \kms.}
\end{figure}

\clearpage
\begin{figure}
\includegraphics[angle=-90,scale=1]{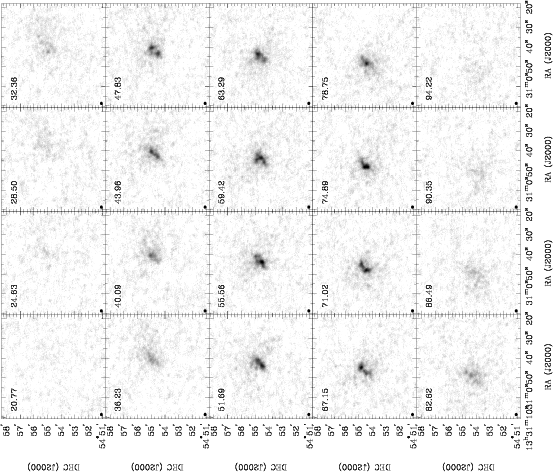}
\caption{\textbf{UGC\,8508:} Channel maps based on the natural-weighted
  cube (grayscale range: $-0.02$ to 23.0\,mJy\,beam$^{-1}$). Every fifth
  channel is shown (channel width 0.6\,\kms{}) and each map has
  the same size as the moment maps in the following panels.\label{fig:u8508}}
\end{figure}

\clearpage
\addtocounter{figure}{-1}
\begin{figure}
\includegraphics[angle=0,scale=1.2]{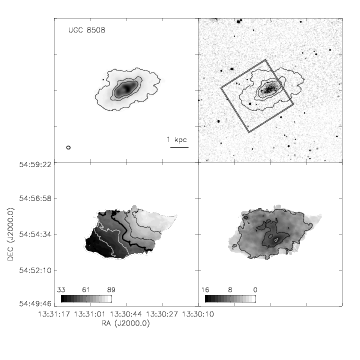}
\caption{continued. {\it Top left:} The integrated \hi\ intensity map
  for UGC\,8508.  The grayscale covers a range from 1$\times$10$^{19}$
  to $2.9\times10^{21}$\,cm$^{-2}$ with contours of
  1$\times$10$^{20}$, 5$\times$10$^{20}$, and
  1$\times$10$^{21}$\,cm$^{-2}$.  {\it Top Right:} An optical g-band
  image from the SDSS with the same column density contours overlaid.
  The HST ACS footprint is the field covered by the ANGST survey.
  {\it Bottom Left:} The \hi\ velocity field.  Black contours (lighter
  gray scale) indicate approaching emission, white contours (darker
  gray scale) receding emission.  The thick black contour is the
  central velocity (v$_{cen}$ = 62.0\,km\,s$^{-1}$) and the
  isovelocity contours are spaced by $\Delta$v = 10\,km\,s$^{-1}$.
  {\it Bottom Right:} The \hi\ velocity dispersion.  Contours are
  plotted at 5, 10, and 15\,km\,s$^{-1}$. Colorbars are in units of \kms.}
\end{figure}

\clearpage
\begin{figure}
\includegraphics[angle=-90,scale=1]{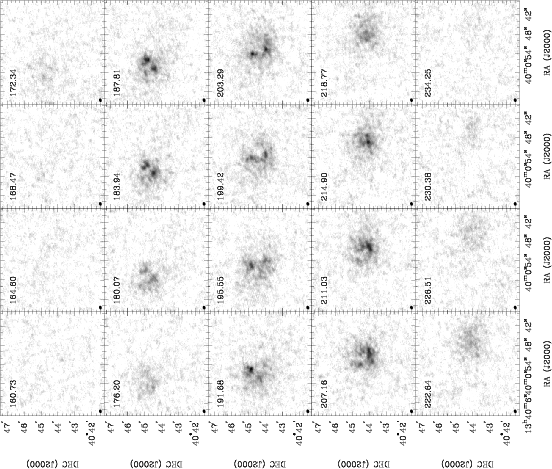}
\caption{\textbf{DDO\,181:} Channel maps based on the natural-weighted
  cube (grayscale range: $-0.02$ to 13.9\,mJy\,beam$^{-1}$). Every second
  channel is shown (channel width 1.3\,\kms{}) and each map has
  the same size as the moment maps in the following panels.\label{fig:ddo181}}
\end{figure}

\clearpage
\addtocounter{figure}{-1}
\begin{figure}
\includegraphics[angle=0,scale=1.2]{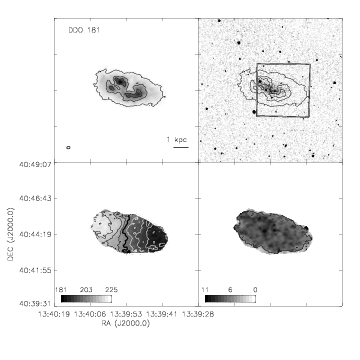}
\caption{continued. {\it Top left:} The integrated \hi\ intensity map
  for DDO\,181.  The grayscale covers a range from 1$\times$10$^{19}$
  to $1.7\times10^{21}$\,cm$^{-2}$ with contours of
  1$\times$10$^{20}$, 5$\times$10$^{20}$, and
  1$\times$10$^{21}$\,cm$^{-2}$.  {\it Top Right:} An optical g-band
  image from the SDSS with the same column density contours overlaid.
  The HST ACS footprint is the field covered by the ANGST survey.
  {\it Bottom Left:} The \hi\ velocity field.  Black contours (lighter
  gray scale) indicate approaching emission, white contours (darker
  gray scale) receding emission.  The thick black contour is the
  central velocity (v$_{cen}$ = 201.4\,km\,s$^{-1}$) and the
  isovelocity contours are spaced by $\Delta$v = 5\,km\,s$^{-1}$.
  {\it Bottom Right:} The \hi\ velocity dispersion.  Contours are
  plotted at 5 and 10\,km\,s$^{-1}$. Colorbars are in units of \kms.}
\end{figure}

\clearpage
\begin{figure}
\includegraphics[angle=-90,scale=1]{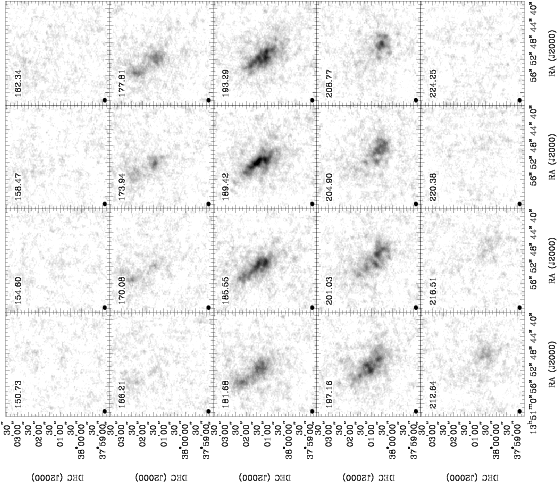}
\caption{\textbf{DDO\,183:} Channel maps based on the natural-weighted
  cube (grayscale range: $-0.02$ to 15.7\,mJy\,beam$^{-1}$). Every second
  channel is shown (channel width 1.3\,\kms{}) and each map has
  the same size as the moment maps in the following panels.\label{fig:ddo183}}
\end{figure}

\clearpage
\addtocounter{figure}{-1}
\begin{figure}
\includegraphics[angle=0,scale=1.2]{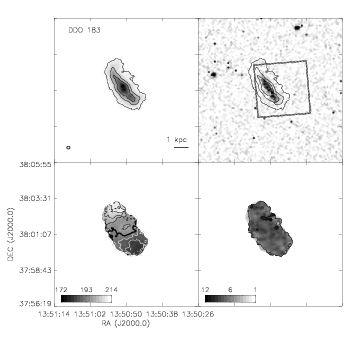}
\caption{continued. {\it Top left:} The integrated \hi\ intensity map
  for DDO\,183.  The grayscale covers a range from 1$\times$10$^{19}$
  to $2.2\times10^{21}$\,cm$^{-2}$ with contours of
  1$\times$10$^{20}$, 5$\times$10$^{20}$, and
  1$\times$10$^{21}$\,cm$^{-2}$.  {\it Top Right:} An optical g-band
  image from the SDSS with the same column density contours overlaid.
  The HST ACS footprint is the field covered by the ANGST survey.
  {\it Bottom Left:} The \hi\ velocity field.  Black contours (lighter
  gray scale) indicate approaching emission, white contours (darker
  gray scale) receding emission.  The thick black contour is the
  central velocity (v$_{cen}$ = 191.2\,km\,s$^{-1}$) and the
  isovelocity contours are spaced by $\Delta$v = 5\,km\,s$^{-1}$.
  {\it Bottom Right:} The \hi\ velocity dispersion.  Contours are
  plotted at 5 and 10\,km\,s$^{-1}$. Colorbars are in units of \kms.}
\end{figure}

\clearpage
\begin{figure}
\includegraphics[angle=-90,scale=1]{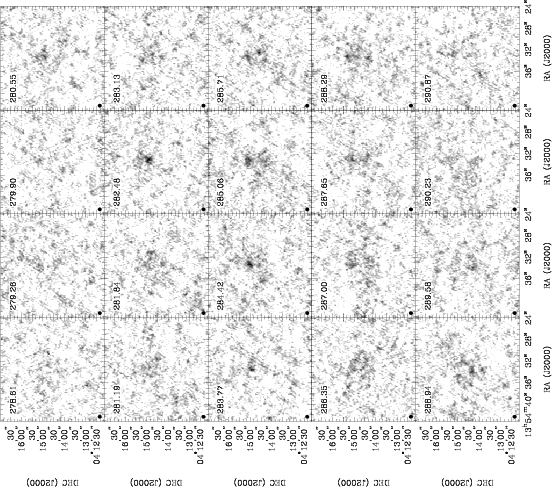}
\caption{\textbf{KKH\,86:} Channel maps based on the natural-weightedcube
  (grayscale range: $-0.02$ to 7.7\,mJy\,beam$^{-1}$). Every channel is
  shown (channel width 0.6\,\kms{}) and each map has
  the same size as the moment maps in the following panels.\label{fig:kkh86}}
\end{figure}

\clearpage
\addtocounter{figure}{-1}
\begin{figure}
\includegraphics[angle=0,scale=1.2]{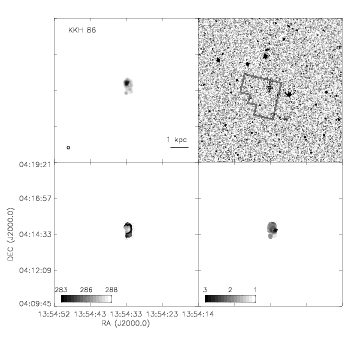}
\caption{continued. {\it Top left:} The integrated \hi\ intensity map
  for KKH\,86.  The grayscale covers a range from 1$\times$10$^{19}$
  to $1.5\times10^{20}$\,cm$^{-2}$ with a contour of
  1$\times$10$^{20}$\,cm$^{-2}$.  {\it Top Right:} An optical g-band
  image from the SDSS with the same column density contours overlaid.
  The HST WFPC2 footprint is the field covered by the ANGST survey.
  {\it Bottom Left:} The \hi\ velocity field.  Black contours (lighter
  gray scale) indicate approaching emission, white contours (darker
  gray scale) receding emission.  The thick black contour is the
  central velocity (v$_{cen}$ = 285.5\,km\,s$^{-1}$) and the
  isovelocity contours are spaced by $\Delta$v = 10\,km\,s$^{-1}$.
  {\it Bottom Right:} The \hi\ velocity dispersion. Colorbars are in units of \kms.}
\end{figure}

\clearpage
\begin{figure}
\includegraphics[angle=-90,scale=1]{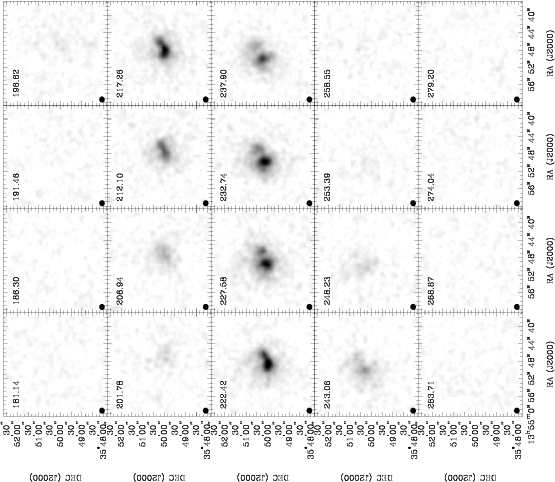}
\caption{\textbf{UGC\,8833:} Channel maps based on the natural-weighted
  cube (grayscale range: $-0.02$ to 20.8\,mJy\,beam$^{-1}$). Every channel
  is shown (channel width 2.6\,\kms{}) and each map has
  the same size as the moment maps in the following panels.\label{fig:u8833}}
\end{figure}

\clearpage
\addtocounter{figure}{-1}
\begin{figure}
\includegraphics[angle=0,scale=1.2]{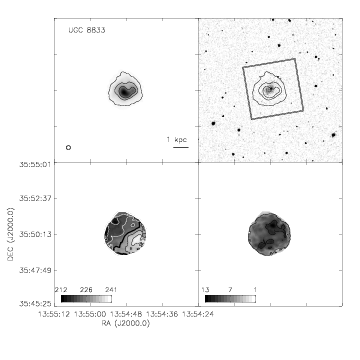}
\caption{continued. {\it Top left:} The integrated \hi\ intensity map
  for UGC\,8833.  The grayscale covers a range from 1$\times$10$^{19}$
  to $2.2\times10^{21}$\,cm$^{-2}$ with contours of
  1$\times$10$^{20}$, 5$\times$10$^{20}$, and
  1$\times$10$^{21}$\,cm$^{-2}$.  {\it Top Right:} An optical g-band
  image from the SDSS with the same column density contours overlaid.
  The HST ACS footprint is the field covered by the ANGST survey.
  {\it Bottom Left:} The \hi\ velocity field.  Black contours (lighter
  gray scale) indicate approaching emission, white contours (darker
  gray scale) receding emission.  The thick black contour is the
  central velocity (v$_{cen}$ = 225.9\,km\,s$^{-1}$) and the
  isovelocity contours are spaced by $\Delta$v = 5\,km\,s$^{-1}$.
  {\it Bottom Right:} The \hi\ velocity dispersion.  Contours are
  plotted at 5 and 10\,km\,s$^{-1}$. Colorbars are in units of \kms.}
\end{figure}

\clearpage
\begin{figure}
\includegraphics[angle=-90,scale=1]{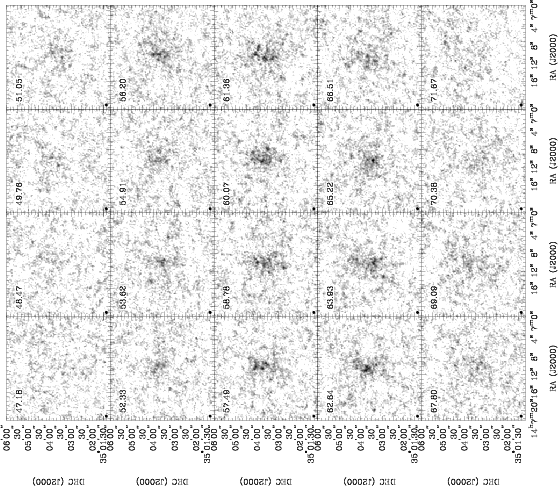}
\caption{\textbf{KK\,230:} Channel maps based on the natural-weighted cube
  (grayscale range: $-0.02$ to 8.2\,mJy\,beam$^{-1}$). Every channel is
  shown (channel width 0.6\,\kms{}) and each map has
  the same size as the moment maps in the following panels.\label{fig:kk230}}
\end{figure}

\clearpage
\addtocounter{figure}{-1}
\begin{figure}
\includegraphics[angle=0,scale=1.2]{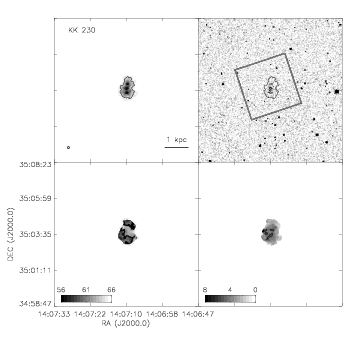}
\caption{continued. {\it Top left:} The integrated \hi\ intensity map
  for KK\,230.  The grayscale covers a range from 1$\times$10$^{19}$
  to $6.1\times10^{20}$\,cm$^{-2}$ with contours of 1$\times$10$^{20}$
  and 5$\times$10$^{20}$\,cm$^{-2}$.  {\it Top Right:} An optical
  g-band image from the SDSS with the same column density contours
  overlaid.  The HST ACS footprint is the field covered by the ANGST
  survey.  {\it Bottom Left:} The \hi\ velocity field.  Black contours
  (lighter gray scale) indicate approaching emission, white contours
  (darker gray scale) receding emission.  The thick black contour is
  the central velocity (v$_{cen}$ = 60.6\,km\,s$^{-1}$) and the
  isovelocity contours are spaced by $\Delta$v = 5\,km\,s$^{-1}$.
  {\it Bottom Right:} The \hi\ velocity dispersion.  A contour is
  plotted at 5\,km\,s$^{-1}$. Colorbars are in units of \kms.}
\end{figure}

\clearpage
\begin{figure}
\includegraphics[angle=-90,scale=1]{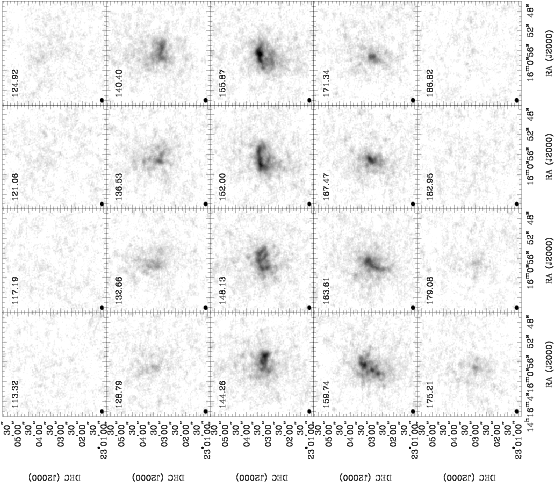}
\caption{\textbf{DDO\,187:} Channel maps based on the natural-weighted
  cube (grayscale range: $-0.02$ to 17.1\,mJy\,beam$^{-1}$). Every second
  channel is shown (channel width 1.3\,\kms{}) and each map has
  the same size as the moment maps in the following panels.\label{fig:ddo187}}
\end{figure}

\clearpage
\addtocounter{figure}{-1}
\begin{figure}
\includegraphics[angle=0,scale=1.2]{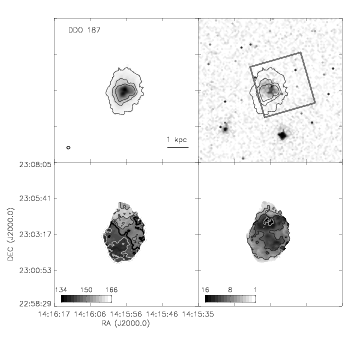}
\caption{continued. {\it Top left:} The integrated \hi\ intensity map
  for DDO\,187.  The grayscale covers a range from 1$\times$10$^{19}$
  to $3.2\times10^{21}$\,cm$^{-2}$ with contours of
  1$\times$10$^{20}$, 5$\times$10$^{20}$, and
  1$\times$10$^{21}$\,cm$^{-2}$.  {\it Top Right:} An optical g-band
  image from the SDSS with the same column density contours overlaid.
  The HST ACS footprint is the field covered by the ANGST survey.
  {\it Bottom Left:} The \hi\ velocity field.  Black contours (lighter
  gray scale) indicate approaching emission, white contours (darker
  gray scale) receding emission.  The thick black contour is the
  central velocity (v$_{cen}$ = 152.2\,km\,s$^{-1}$) and the
  isovelocity contours are spaced by $\Delta$v = 5\,km\,s$^{-1}$.
  {\it Bottom Right:} The \hi\ velocity dispersion.  Contours are
  plotted at 5, 10, and 15\,km\,s$^{-1}$. Colorbars are in units of \kms.}
\end{figure}

\clearpage
\begin{figure}
\includegraphics[angle=-90,scale=1]{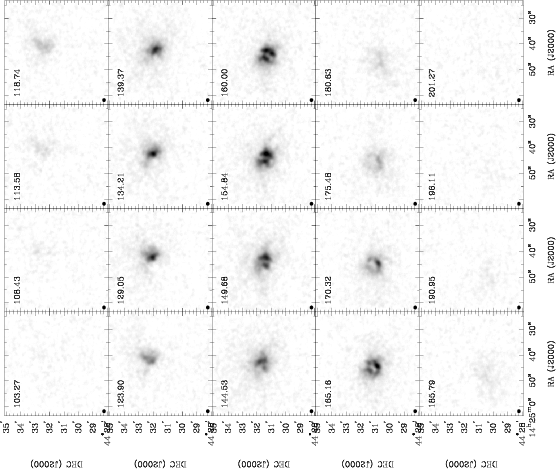}
\caption{\textbf{DDO\,190:} Channel maps based on the natural-weighted
  cube (grayscale range: $-0.02$ to 27.5\,mJy\,beam$^{-1}$). Every channel
  is shown (channel width 2.6\,\kms{}) and each map has
  the same size as the moment maps in the following panels.\label{fig:ddo190}}
\end{figure}

\clearpage
\addtocounter{figure}{-1}
\begin{figure}
\includegraphics[angle=0,scale=1.2]{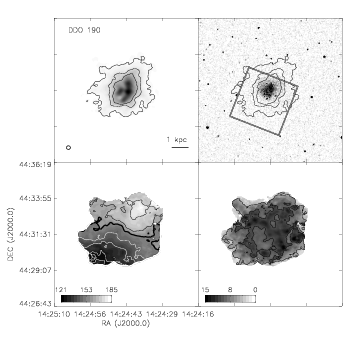}
\caption{continued. {\it Top left:} The integrated \hi\ intensity map
  for DDO\,190.  The grayscale covers a range from 1$\times$10$^{19}$
  to $3.6\times10^{21}$\,cm$^{-2}$ with contours of
  1$\times$10$^{20}$, 5$\times$10$^{20}$, and
  1$\times$10$^{21}$\,cm$^{-2}$.  {\it Top Right:} An optical g-band
  image from the SDSS with the same column density contours overlaid.
  The HST ACS footprint is the field covered by the ANGST survey.
  {\it Bottom Left:} The \hi\ velocity field.  Black contours (lighter
  gray scale) indicate approaching emission, white contours (darker
  gray scale) receding emission.  The thick black contour is the
  central velocity (v$_{cen}$ = 148.8\,km\,s$^{-1}$) and the
  isovelocity contours are spaced by $\Delta$v = 10\,km\,s$^{-1}$.
  {\it Bottom Right:} The \hi\ velocity dispersion.  Contours are
  plotted at 5 and 10\,km\,s$^{-1}$. Colorbars are in units of \kms.}
\end{figure}

\clearpage
\begin{figure}
\includegraphics[angle=-90,scale=1]{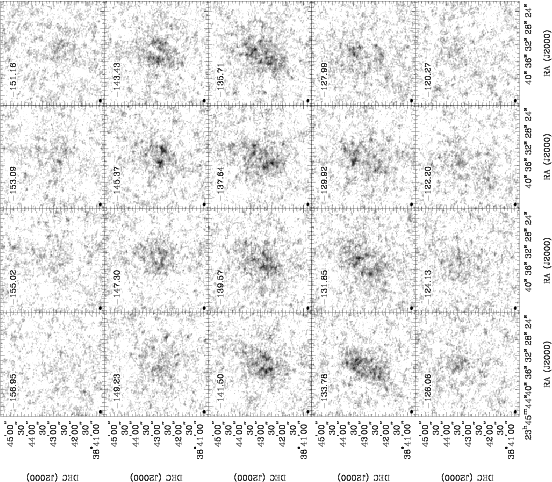}
\caption{\textbf{KKH\,98:} Channel maps based on the natural-weighted cube
  (grayscale range: $-0.02$ to 8.4\,mJy\,beam$^{-1}$). Every second
  channel is shown (channel width 0.6\,\kms{}) and each map has
  the same size as the moment maps in the following panels.\label{fig:kkh98}}
\end{figure}

\clearpage
\addtocounter{figure}{-1}
\begin{figure}
\includegraphics[angle=0,scale=1.2]{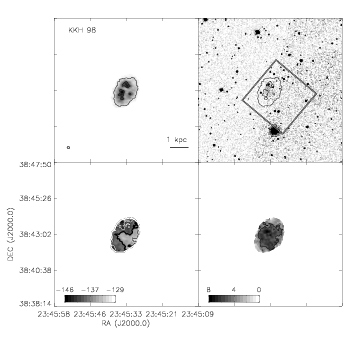}
\caption{continued. {\it Top left:} The integrated \hi\ intensity map
  for KKH\,98.  The grayscale covers a range from 1$\times$10$^{19}$
  to $7.7\times10^{20}$\,cm$^{-2}$ with contours of 1$\times$10$^{20}$
  and 5$\times$10$^{20}$\,cm$^{-2}$.  {\it Top Right:} An optical
  g-band image from the SDSS with the same column density contours
  overlaid.  The HST ACS footprint is the field covered by the ANGST
  survey.  {\it Bottom Left:} The \hi\ velocity field.  Black contours
  (lighter gray scale) indicate approaching emission, white contours
  (darker gray scale) receding emission.  The thick black contour is
  the central velocity (v$_{cen}$ = $-137.8$\,km\, s$^{-1}$) and the
  isovelocity contours are spaced by $\Delta$v = 5\,km\,s$^{-1}$.
  {\it Bottom Right:} The \hi\ velocity dispersion.  A contour is
  plotted at 5\,km\,s$^{-1}$. Colorbars are in units of \kms.}
\end{figure}

\end{document}